\RequirePackage{fix-cm}
\documentclass[fleqn,usenatbib]{mnras}

\usepackage[T1]{fontenc}
\usepackage{lmodern}
\rmfamily 
\DeclareFontShape{T1}{lmr}{b}{sc}{<->ssub*cmr/bx/sc}{}
\DeclareFontShape{T1}{lmr}{bx}{sc}{<->ssub*cmr/bx/sc}{}
\usepackage{slantsc}
\usepackage[margin=5pt]{subfig}
\usepackage{comment}
\usepackage{longtable}
\usepackage{caption}
\usepackage{placeins}
\usepackage{threeparttable}
\usepackage{booktabs}
\usepackage{multirow}
\usepackage[figuresright]{rotating}
\usepackage{textcase}
\newcommand{\be}{\begin{equation}}
\newcommand{\ee}{\end{equation}}
\newcommand{\bea}{\begin{eqnarray}}
\newcommand{\eea}{\end{eqnarray}}
\newcommand{\Om}{\Omega_{m0}}
\newcommand{\Ok}{\Omega_{k0}}
\newcommand{\wX}{w_{\rm X}}
\newcommand{\om}{$\Omega_{m0}$}
\newcommand{\ok}{$\Omega_{k0}$}
\newcommand{\wx}{$w_{\rm X}$}

\newcommand{\mq}{Mg\,\textsc{ii} QSO}
\newcommand{\mii}{Mg\,\textsc{ii}}
\newcommand{\cq}{C\,\textsc{iv} QSO}
\newcommand{\civ}{C\,\textsc{iv}}
\newcommand{\rfe}{${\cal R}_{\rm{Fe\,\textsc{ii}}}$}
\newcommand{\Feii}{Fe\,\textsc{ii}}
\newcommand{\obh}{\Omega_{b}h^2}
\newcommand{\och}{\Omega_{c}h^2}
\newcommand{\onh}{\Omega_{\nu}h^2}

\newcommand{\obhs}{$\Omega_{b}h^2$}
\newcommand{\ochs}{$\Omega_{c}h^2$}
\newcommand{\hunit}{$\rm{km \ s^{-1} \ Mpc^{-1}}$}
\newcommand{\lcdm}{$\Lambda$CDM}
\newcommand{\pcdm}{$\phi$CDM}
\usepackage{array}
\usepackage{threeparttable}
\usepackage{scalerel}
\usepackage{tikz}
\usetikzlibrary{svg.path}

\definecolor{orcidlogocol}{HTML}{A6CE39}
\tikzset{
  orcidlogo/.pic={
    \fill[orcidlogocol] svg{M256,128c0,70.7-57.3,128-128,128C57.3,256,0,198.7,0,128C0,57.3,57.3,0,128,0C198.7,0,256,57.3,256,128z};
    \fill[white] svg{M86.3,186.2H70.9V79.1h15.4v48.4V186.2z}
                 svg{M108.9,79.1h41.6c39.6,0,57,28.3,57,53.6c0,27.5-21.5,53.6-56.8,53.6h-41.8V79.1z M124.3,172.4h24.5c34.9,0,42.9-26.5,42.9-39.7c0-21.5-13.7-39.7-43.7-39.7h-23.7V172.4z}
                 svg{M88.7,56.8c0,5.5-4.5,10.1-10.1,10.1c-5.6,0-10.1-4.6-10.1-10.1c0-5.6,4.5-10.1,10.1-10.1C84.2,46.7,88.7,51.3,88.7,56.8z};
  }
}

\newcommand\orcidicon[1]{\href{https://orcid.org/#1}{\mbox{\scalerel*{
\begin{tikzpicture}[yscale=-1,transform shape]
\pic{orcidlogo};
\end{tikzpicture}
}{|}}}}

\usepackage{hyperref} 

\DeclareRobustCommand{\VAN}[3]{#2}
\let\VANthebibliography\thebibliography
\def\thebibliography{\DeclareRobustCommand{\VAN}[3]{##3}\VANthebibliography}

\usepackage{graphicx}	
\usepackage{amsmath}	
\usepackage{amssymb}	

\title[Heterogeneity \& time-lag in \mii\ \& \civ\ RM QSO cosmology]{Effects of heterogeneous data sets and time-lag measurement techniques on cosmological parameter constraints from \mii\ and \civ\ reverberation-mapped quasar data}

\author[Cao et al.]{
Shulei Cao$^{\orcidicon{0000-0003-2421-7071}{1,2}}$\thanks{E-mail: shuleicao@boisestate.edu},
Michal Zaja\v{c}ek$^{\orcidicon{0000-0001-6450-1187}{3}}$\thanks{E-mail: zajacek@physics.muni.cz},
Bo\.zena Czerny$^{\orcidicon{0000-0001-5848-4333}{4}}$\thanks{E-mail: bcz@cft.edu.pl},
\newauthor \hspace{0.1mm}
Swayamtrupta Panda$^{\orcidicon{0000-0002-5854-7426}{4,5}}$\thanks{E-mail: spanda@lna.br}\thanks{CNPq Fellow},
Bharat Ratra$^{\orcidicon{0000-0002-7307-0726}1}$\thanks{E-mail: ratra@phys.ksu.edu}
\\
$^{1}$Department of Physics, Kansas State University, 116 Cardwell Hall, Manhattan, KS 66506, USA\\
$^{2}$Department of Physics, Boise State University, 1910 University Drive, Boise, ID 83725, USA\\
$^{3}$Department of Theoretical Physics and Astrophysics, Faculty of Science, Masaryk University, Kotl\'a\v{r}sk\'a 2, 611 37 Brno, Czech Republic\\
$^{4}$Center for Theoretical Physics, Polish Academy of Sciences, Al.\ Lotnik\'{o}w 32/46, 02-668 Warsaw, Poland\\
$^{5}$Laborat\'orio Nacional de Astrof\'isica - MCTI, R. dos Estados Unidos, 154 - Na\c{c}\~oes, Itajub\'a - MG, 37504-364, Brazil\\
}

\date{Accepted XXX. Received YYY; in original form ZZZ}

\pubyear{2023}

\begin{document}
\label{firstpage}
\pagerange{\pageref{firstpage}--\pageref{lastpage}}
\maketitle

\begin{abstract}

Previously, we demonstrated that \mii\ and \civ\ reverberation-mapped quasars (RM QSOs) are standardizable and that the cosmological parameters inferred using the broad-line region radius-luminosity ($R-L$) relation are consistent with those determined from better-established cosmological probes. With more data expected from ongoing and future spectroscopic and photometric surveys, it is imperative to examine how new QSO data sets of varied quality, with their own specific luminosity and time-delay distributions, can be best used to determine more restrictive cosmological parameter constraints. In this study, we test the effect of adding 25 OzDES \mii\ RM QSOs as well as 25 lower-quality SDSS RM \civ\ QSOs, which increases the previous sample of RM QSOs by $\sim 36\%$. Although cosmological parameter constraints become tighter for some cosmological models after adding these new QSOs, the new combined data sets have increased differences between $R-L$ parameter values obtained in different cosmological models and thus a lower standardizability for the larger \mii\ + \civ\ compilation. Different time-delay methodologies, particularly the ICCF and CREAM methods used for inferring time delays of SDSS RM QSOs, slightly affect cosmological and $R-L$ relation parameter values, however, the effect is negligible for (smaller) compilations of robust time-delay detections. Our analysis indicates that increasing the sample size is not sufficient for tightening cosmological constraints and a quality cut is necessary to obtain a standardizable RM QSO sample.
\end{abstract}




\section{Introduction} \label{sec:intro}

There have been reports of inconsistencies between cosmological parameter values, in particular the Hubble constant ($H_0$) and the strength of the matter clustering $S_8$, which can be measured from local distance-ladder observations and weak lensing studies or galaxy clustering (current epoch data), respectively. On the other hand, they can also be inferred from model-dependent studies using cosmic microwave background (CMB) data (early epoch measurements), which typically make use of the current standard $\Lambda$CDM model \citep[where $\Lambda$ represents cosmological constant dark energy and CDM stands for cold dark matter,][]{peeb84}. These discrepancies motivate the search for alternate methods to measure the cosmological expansion rate \citep[see e.g.][for recent reviews]{PerivolaropoulosSkara2021, Morescoetal2022, Abdallaetal2022, HuWang2023}. Among the emerging probes, quasars (QSOs) hosting efficiently accreting supermassive black holes or bright active galactic nuclei \citep[hereafter AGN; see e.g.][for recent reviews]{2021bhns.confE...1K,2023Ap&SS.368....8C,2023FrASS..1030103P,2023arXiv230615082Z} stand out because they span a vast redshift range, from nearby sources to redshifts on the order of 7 or even higher. 

There are several ways in which QSOs can be used for cosmological studies. Gravitationally lensed QSOs are used for that purpose through the measurement of the time delays between different images of the same source \citep{Suyu2017,Shajib2020,shajib2023}. AGN jets are used for the angular size cosmological test \citep{Cao_et_al2017b, Ryanetal2019, CaoRyanRatra2020, CaoRyanRatra2021, CaoRyanRatra2022, Zhengetal2021, Lian_etal_2021}. The nonlinear relation between QSO X-ray and ultraviolet (UV) luminosities ($L_{X}-L_{UV}$ relation), discovered by \citet{tananbaum1979}, was used to try to standardize QSOs and to use these QSOs to derive cosmological constraints \citep{RisalitiLusso2015,RisalitiandLusso_2019} and initially appeared to work \citep{Khadka_2020a, Khadka_2020b, Yang_Banerjee_Colgain_2020}. However, a newer, larger $L_{X}-L_{UV}$ QSO data set \citep{Lusso_etal_2020}, when correctly analyzed,\footnote{By constraining simultaneously the $L_{X}-L_{UV}$ relation parameters and the cosmological parameters, thus avoiding the circularity problem, in a number of different cosmological models to verify whether the $L_{X}-L_{UV}$ relation is independent of a cosmological model and thus standardizable \citep{Khadka_2020c, Caoetal_2021}.} was shown to contain a significant number of non-standardizable QSOs, with possibly as many as 2/3 of the QSOs being un-standardizable \citep{KhadkaRatra2021,KhadkaRatra2022}.\footnote{For a rather limited sample of 58 X-ray detected reverberation-mapped (RM) QSOs it was possible to apply both the broad-line region radius-luminosity ($R-L$) relation (discussed in the main text next) and the $L_{X}-L_{UV}$ correlation \citep{Khadkaetal2023}. This revealed that the $L_{X}-L_{UV}$ relation prefers larger values of the current non-relativistic matter energy density parameter $\Omega_{m0}$ than does the $R-L$ relation, $\Omega_{m0}$ values larger than those measured by better-established cosmological probes. This results in a smaller source luminosity distance when the $L_{X}-L_{UV}$ relation is used, compared to the luminosity distance measured using the $R-L$ relation. \citet{2023arXiv230508179Z} have shown that this systematic effect can be attributed to local differential dust extinction of QSO X-ray and UV light in the QSO host galaxies. This suggests that the $L_{X}-L_{UV}$ relation cannot be used to standardize QSOs. For other discussions of the \citet{Lusso_etal_2020} QSO data see \citet{HuWang2022}, \citet{Colgainetal2022}, \citet{Petrosianetal2022}, \citet{dainotti2023}, and references therein.} 

Time-delay measurements between QSO continuum and broad-line region emission lines can also be used cosmologically, through the radius-luminosity ($R-L$) relation, as suggested by \citet{watson2011} and \citet{haas2011}, and implemented recently in practice in a number of works \citep{martinez2019, 2019FrASS...6...75P, Czernyetal2021, zajacek2021, Khadkaetal_2021a, Khadkaetal2022a, Caoetal_2022}. Our current paper focuses on this cosmological test. This is timely because of the increasing number of available time-delay measurements, as well as the prospects of much more such data from current and forthcoming surveys \citep[see e.g.,][for expectations from the planned Rubin Legacy Survey of Space and Time]{2019FrASS...6...75P,czerny2023}, raise questions about the optimal methodology for using these data. This is because measurements from extensive surveys might not always be of the highest quality needed for this cosmological test and because some of these measurements may be affected by systematic biases due to e.g. adopted survey strategies \citep[see][for the analysis of recovered BLR time delays from simulated optical photometric light curves, i.e. using the photometric reverberation mapping, adopting the planned Vera C. Rubin LSST survey strategies]{czerny2023}. This study indicates that the recovered time delays are expected to be mostly systematically smaller than the true intrinsic BLR time delays, depending on the source redshift.

One issue we address in this paper is whether an increasing number of largely heterogeneous QSO measurements leads to an increase in the cosmological parameter measurement precision. We showed in \citet{Hbeta_Khadka2022} that time-delay measurements of the H$\beta$ line compiled from the literature (and coming from several groups) were not satisfactory for performing cosmology. On the other hand, a similarly sized sample for \mii\ RM quasars \citep{Khadkaetal_2021a} can be used for cosmological purposes, and combining this sample with a sample of time-delay measurements of the \civ\ line improves the resulting cosmological parameter constraints \citep{Caoetal_2022}. Enlarging the sample of standardizable RM QSOs appears to be the only way to decrease the statistical uncertainties of both $R-L$ relation parameters and cosmological model parameters since the inclusion of the third parameter in the $R-L$ relation that is supposed to be correlated with the accretion rate (the ratio of the equivalent widths of UV \Feii\ and \mii, \rfe) did not result in a decrease in the intrinsic scatter of the $R-L$ relation \citep{Khadkaetal2022a}.\footnote{See \citet{Hbeta_Khadka2022} for a similar issue with H$\beta$ data.}

To this end, we compile the largest sample of QSO time delays for simultaneous determination of $R-L$ relation and cosmological model parameters, to test whether these data are standardizable and to derive cosmological parameter constraints. We compare results derived using the full data set, as well as those from various subsamples, to previous results from a smaller sample. In \citet{Caoetal_2022} we studied a combined sample of 78 \mii\ + 38 \civ\ measurements that were found to be standardizable. The derived cosmological constraints were weak but overall consistent with those from better-established cosmological probes, specifically a combination of Hubble parameter [$H(z)$] and baryon acoustic oscillation (BAO) data. Here we enlarge the sample of previously analyzed 69 \mii\ measurements \citep[see][]{zajacek2021} with 25 new OzDES \mii\ time-delay measurements \citep{Yuetal2023}. The \civ\ data set is enlarged by 25 SDSS lower-quality, albeit still significant, time-delay detections reported in \citet{2019ApJ...887...38G}. For the SDSS RM sources (altogether 41), we also have information on the time-delay inference methodology, and here we focus specifically on the standard Interpolated Cross-Correlation Function (ICCF) method and the newer CREAM inference code method (which incorporates the response function of the standard accretion disc). In comparison with the previous studies of \citet{Caoetal_2022} and \citet{Khadkaetal_2021a}, we also correct the redshift of nearby sources by taking into account peculiar velocities. In particular, we correct the redshift of two \mii\ time-delay measurements of NGC 4151 and six \civ\ QSOs. These modified redshifts propagate into luminosity-distance computations as well as the computation of rest-frame time delays.

Our manuscript is structured as follows. In Section~\ref{sec:model} we briefly summarize the cosmological models we use in this cosmological test. Different data sets and their statistical properties are presented in Section~\ref{sec:data}, and the analysis methodology is outlined in Section~\ref{sec:analysis}. We present constraints on the $R-L$ relation parameters in the standard flat $\Lambda$CDM model in Section~\ref{sec:results_fixedLCDM}. Simultaneous constraints on cosmological and $R-L$ relation parameters are presented in Section~\ref{sec:results_general}. Systematic effects across different data sets are discussed in Section~\ref{sec:discussion}. Finally, we conclude in Section~\ref{sec:conclusion}. The \civ\ and \mii\ data we use are tabulated in Appendices~\ref{sec:civdata} and \ref{sec:MgIIdata}.

\section{Cosmological models}
\label{sec:model}

In this paper, we use various combinations of data to constrain cosmological model parameters and \civ\ and \mq\ $R-L$ relation parameters in six spatially flat and non-flat relativistic dark energy cosmological models.\footnote{For recent observational constraints on spatial curvature, see \citet{Oobaetal2018b}, \citet{Yuetal2018}, \citet{ParkRatra2019b},  \citet{DiValentinoetal2021a}, \citet{Khadkaetal_2021b}, \citet{ArjonaNesseris2021}, \citet{Dhawanetal2021}, \citet{Renzietal2021}, \citet{Gengetal2022}, \citet{MukherjeeBanerjee2022}, \citet{Glanvilleetal2022}, \citet{Wuetal2023}, \citet{deCruzPerezetal2023}, \citet{DahiyaJain2022}, \cite{Stevensetal2023}, \citet{Favaleetal2023}, \citet{Qietal2023}, and references therein.} In each cosmological model we study we use the expansion rate $E(z, \textbf{p})$, as a function of redshift $z$ and the cosmological parameters $\textbf{p}$, to compute cosmological-parameter-dependent predictions. Here $E(z, \textbf{p})\equiv H(z, \textbf{p})/H_0$ with $H(z, \textbf{p})$ being the Hubble parameter and $H_0$ the Hubble constant. As in \cite{CaoRatra2023}, we assume one massive and two massless neutrino species, where the effective number of relativistic neutrino species and the total neutrino mass are set to be $N_{\rm eff} = 3.046$ and $\sum m_{\nu}=0.06$ eV, respectively. Consequently, the current value of the non-relativistic neutrino physical energy density parameter, $\onh=\sum m_{\nu}/(93.14\ \rm eV)$, is a constant, where $h$ is $H_0$ in units of 100 \hunit. The non-relativistic matter density parameter $\Om = (\onh + \obh + \och)/{h^2}$, with \obhs\ and \ochs\ being the current values of the baryonic and cold dark matter physical energy density parameters, respectively.

In the \lcdm\ models and XCDM parametrizations the dark energy equation of state parameter $w_{\rm DE}=p_{\rm DE}/\rho_{\rm DE}$ is $=-1$ and $\neq -1$, respectively, where $p_{\rm DE}$ and $\rho_{\rm DE}$ are the pressure and energy density of the dark energy, respectively. The expansion rate function is therefore 
\be
\label{eq:EzL}
\resizebox{0.475\textwidth}{!}{%
    $E(z, \textbf{p}) = \sqrt{\Om\left(1 + z\right)^3 + \Ok\left(1 + z\right)^2 + \Omega_{\rm DE}\left(1+z\right)^{1+w_{\rm DE}}},$%
    }
\ee
where \ok\ is the current value of the spatial curvature energy density parameter and $\Omega_{\rm DE} = 1 - \Om - \Ok$ is the current value of the dark energy density parameter. In the \lcdm\, dark energy is a cosmological constant $\Lambda$ and $\Omega_{\rm DE} =\Omega_{\Lambda}$, while in XCDM dynamical dark energy is an X-fluid and $\Omega_{\rm DE} =\Omega_{\rm X0}$. The cosmological parameters being constrained are $\textbf{p}=\{H_0, \obh\!, \och\!, \Ok\}$ and $\textbf{p}=\{H_0, \obh\!, \och\!, \wX, \Ok\}$ in the \lcdm\ models and XCDM parametrizations ($\Ok=0$ in the flat cases), respectively. Note that in analyses of \civ\ and \mii\ QSOs, $H_0=70$ \hunit\ and $\Omega_{b}=0.05$ are set with $\textbf{p}$ changing accordingly.

In the \pcdm\ models \citep{peebrat88,ratpeeb88,pavlov13}\footnote{For recent cosmological constraints on the \pcdm\ models, see \cite{ooba_etal_2018b, ooba_etal_2019}, \cite{park_ratra_2018, park_ratra_2019b, park_ratra_2020}, \citet{Singhetal2019}, \cite{UrenaLopezRoy2020}, \cite{SinhaBanerjee2021}, \cite{CaoKhadkaRatra2022}, \cite{deCruzetal2021}, \cite{Jesusetal2022}, \cite{Adiletal2023}, \cite{Dongetal2023}, \cite{VanRaamsdonkWaddell2023}, \cite{CaoRatra2023b}, and references therein.} dark energy is a dynamical scalar field $\phi$ with an inverse power-law scalar field potential energy density
\be
\label{PE}
V(\phi)=\frac{1}{2}\kappa m_p^2\phi^{-\alpha},
\ee
where $m_p$ is the Planck mass, $\alpha$ is a positive constant ($\alpha=0$ corresponding to \lcdm), and $\kappa$ is a constant that is determined by the shooting method implemented in the Cosmic Linear Anisotropy Solving System (\textsc{class}) code \citep{class}. The expansion rate is
\be
\label{eq:Ezp}
    E(z, \textbf{p}) = \sqrt{\Om\left(1 + z\right)^3 + \Ok\left(1 + z\right)^2 + \Omega_{\phi}(z,\alpha)},
\ee
where 
\be
\label{Op}
\Omega_{\phi}(z,\alpha)=\frac{1}{6H_0^2}\bigg[\frac{1}{2}\dot{\phi}^2+V(\phi)\bigg],
\ee
is the scalar field dynamical dark energy density parameter and is computed by numerically solving the Friedmann equation \eqref{eq:Ezp} and the equation of motion of the scalar field
\be
\label{em}
\ddot{\phi}+3H\dot{\phi}+V'(\phi)=0.
\ee 
Here an overdot and a prime denote a derivative to time and $\phi$, respectively. The cosmological parameters being constrained are $\textbf{p}=\{H_0, \obh\!, \och\!, \alpha, \Ok\}$ with $\Ok=0$ corresponding to the flat case.

\section{Data}
\label{sec:data}

In this work, we use updated and currently the most extended compilations of \civ\ and \mii\ QSO measurements and $H(z)$ + BAO data, as well as combinations of these data sets, to constrain cosmological-model and QSO $R-L$ relation parameters. These data sets are summarized next.

\begin{figure*}
    \centering
     \includegraphics[width=0.33\textwidth]{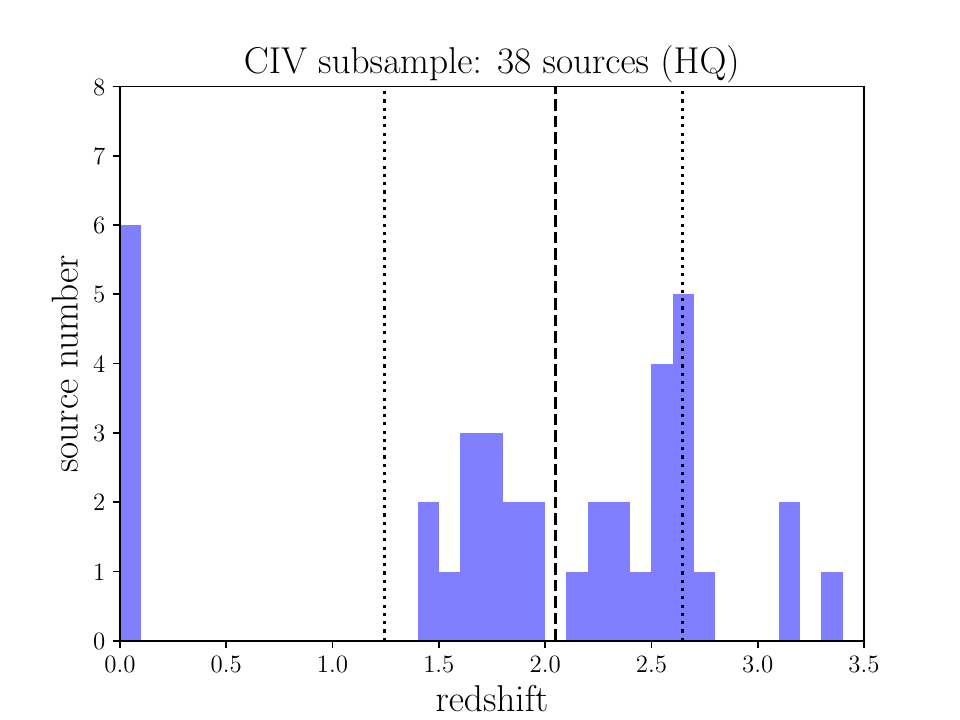}
       \includegraphics[width=0.33\textwidth]{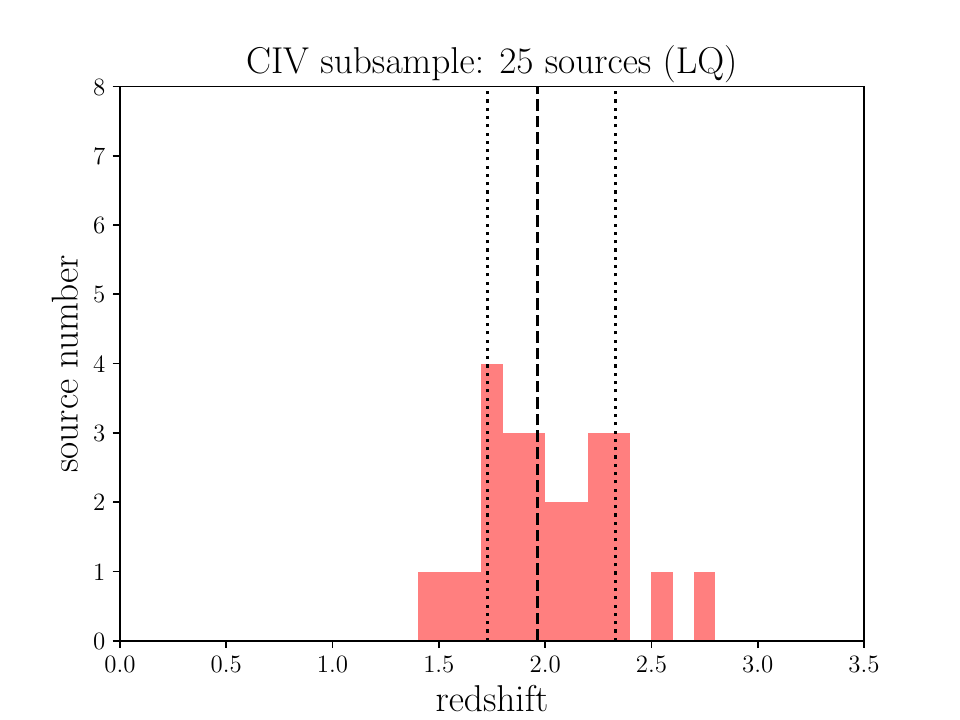}
      \includegraphics[width=0.33\textwidth]{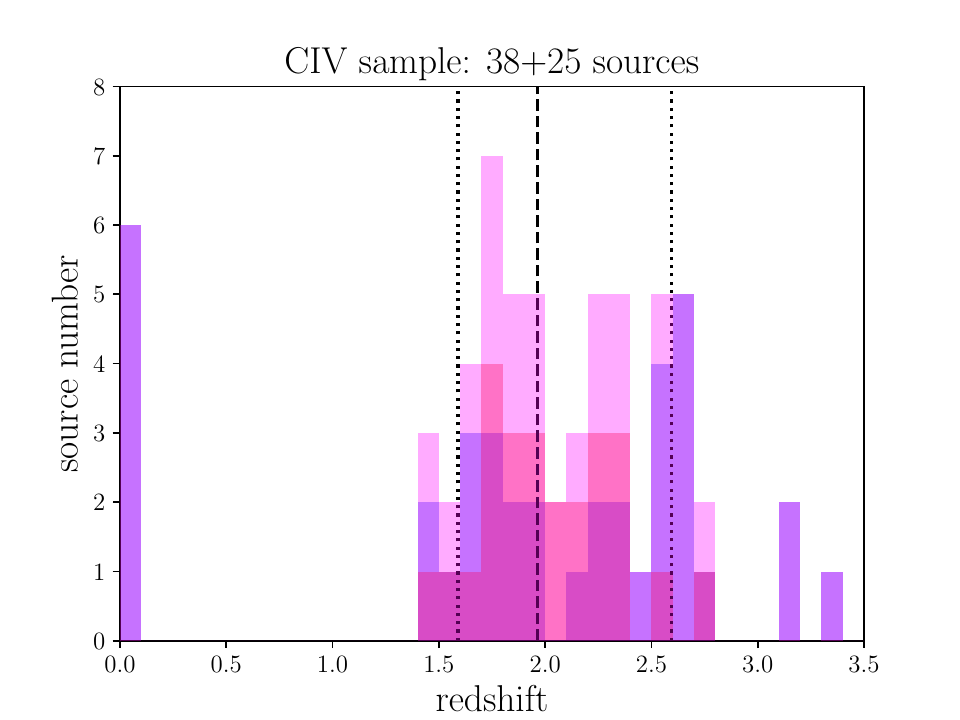}   
  \caption{Redshift distributions for the \civ\ QSOs. {\bf Left panel:} The sample of 38 higher-quality sources. {\bf Middle panel:} The sample of 25 lower-quality sources. {\bf Right panel:} The combined sample of $38+25 = 63$ sources. A vertical dashed line marks the distribution's median, while dotted lines stand for $16\%$ and $84\%$ percentiles. The bin width for all distributions is $\Delta z=0.1$.}
    \label{fig_CIV_redshift}
\end{figure*}

\begin{figure*}
    \centering
     \includegraphics[width=0.33\textwidth]{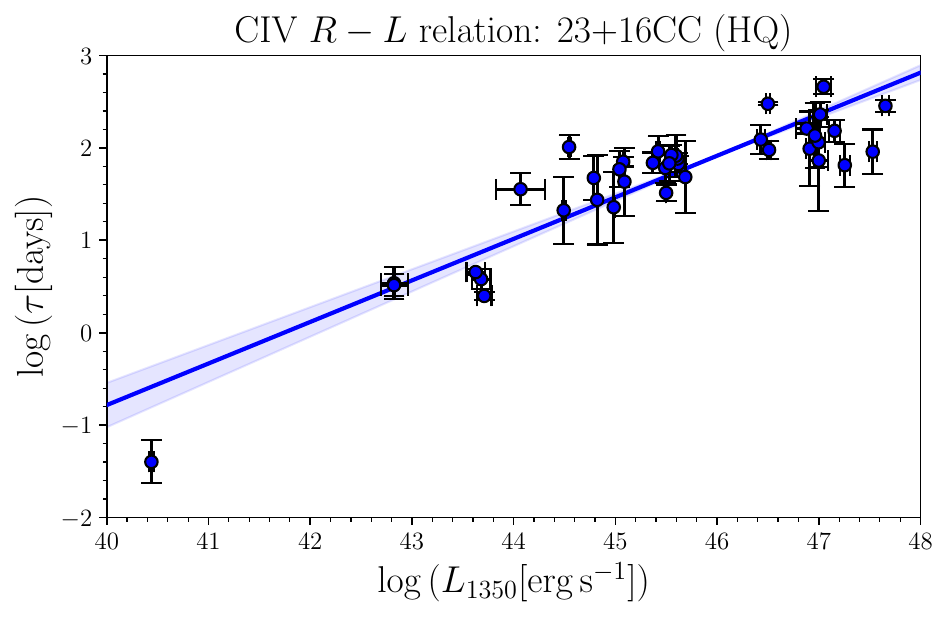}
       \includegraphics[width=0.33\textwidth]{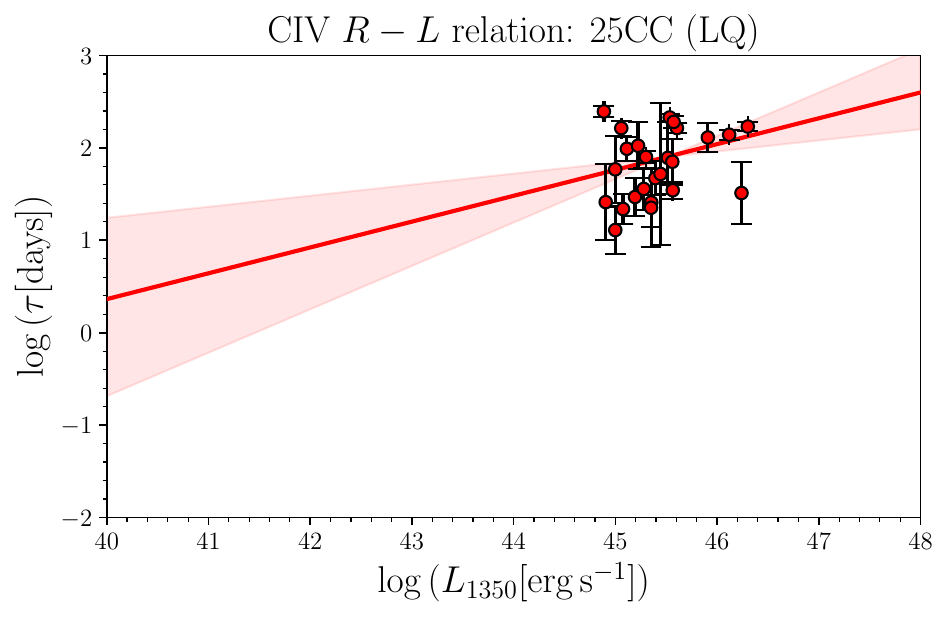}
      \includegraphics[width=0.33\textwidth]{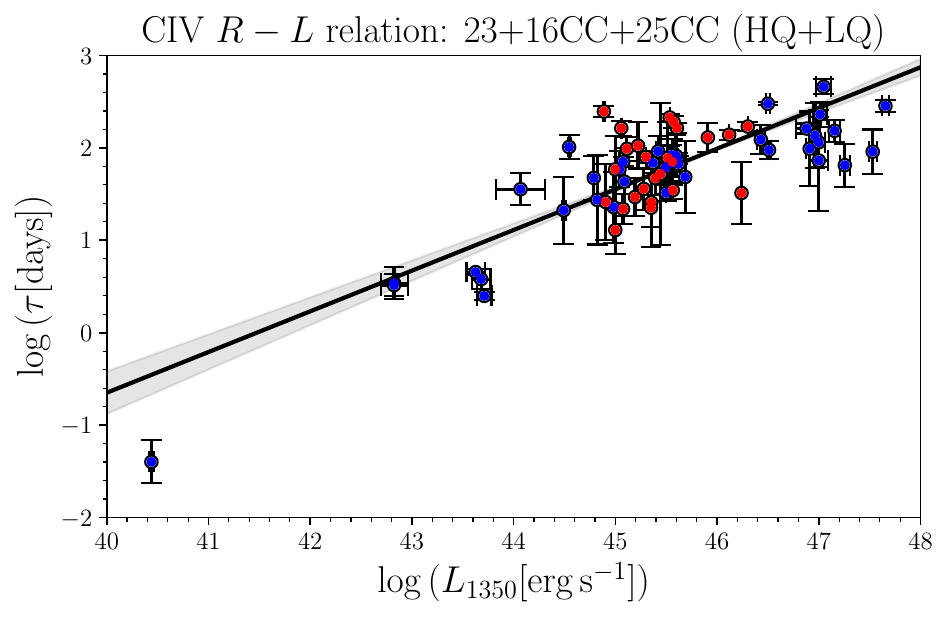}   
  \caption{\civ\ $R-L$ relations in the fixed flat $\Lambda$CDM cosmological model ($\Omega_{m0}=0.3$, $H_0=70\,{\rm km\,s^{-1}\,Mpc^{-1}}$). The time delays of SDSS QSOs were determined using the CC method. {\bf Left panel:} The subsample of 39 \civ\ measurements that have a clear detection of the time delay of \civ\ emission (the HQ subsample). The best-fit $R-L$ relation is $\log{\tau}=(0.45^{+0.04}_{-0.04})\log{L_{44}}+1.02^{+0.08}_{-0.08}$, with $\sigma=0.29^{+0.05}_{-0.04}$. {\bf Middle panel:} The subsample of \civ\ QSOs that have a quality class equal to or lower than 3 in the SDSS RM catalogue (the LQ subsample). The best-fit $R-L$ relation is $\log{\tau}=(0.28^{+0.19}_{-0.16})\log{L_{44}}+1.48^{+0.24}_{-0.29}$, with $\sigma=0.34^{+0.07}_{-0.06}$. {\bf Right panel:} The combined sample of 64 measurements. The best-fit $R-L$ relation is $\log{\tau}=(0.44^{+0.04}_{-0.04})\log{L_{44}}+1.11^{+0.07}_{-0.07}$, with $\sigma=0.33^{+0.04}_{-0.03}$. In each panel, we indicate the best-fit $R-L$ relation by a solid line. The shaded regions show the 1$\sigma$ confidence intervals. }
    \label{fig_RL_CIV}
\end{figure*}

\begin{figure*}
    \centering
     \includegraphics[width=0.33\textwidth]{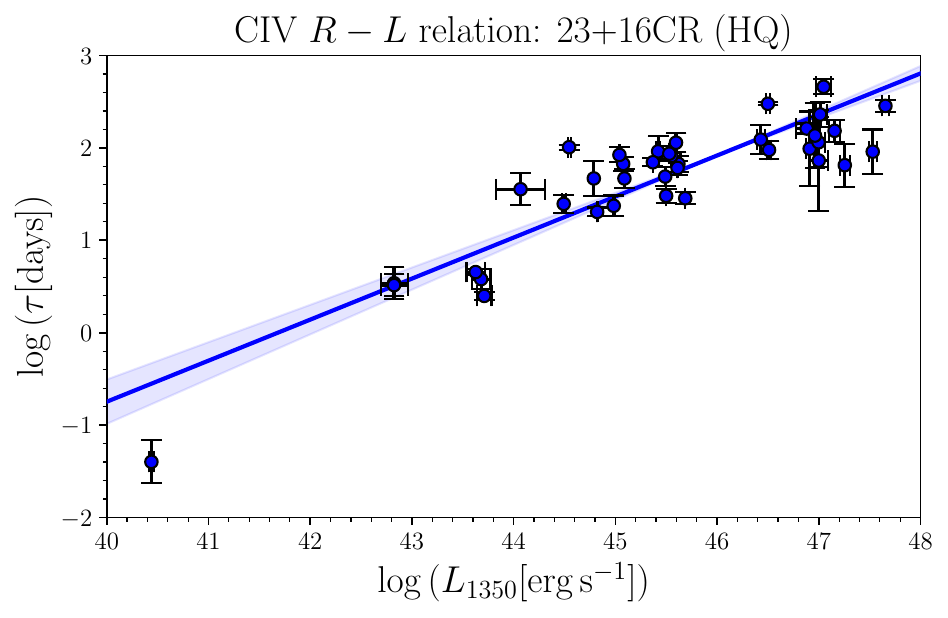}
       \includegraphics[width=0.33\textwidth]{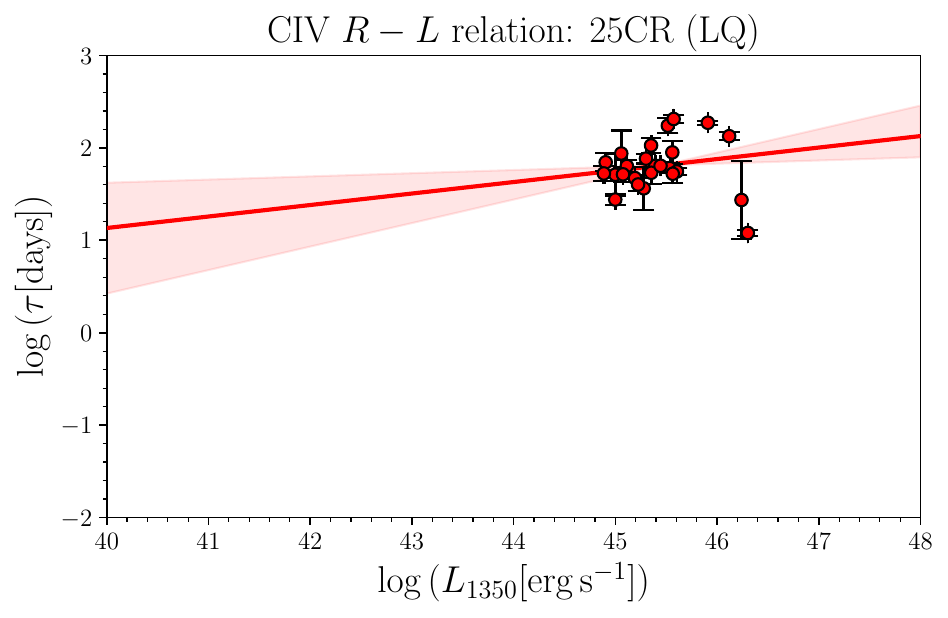}
      \includegraphics[width=0.33\textwidth]{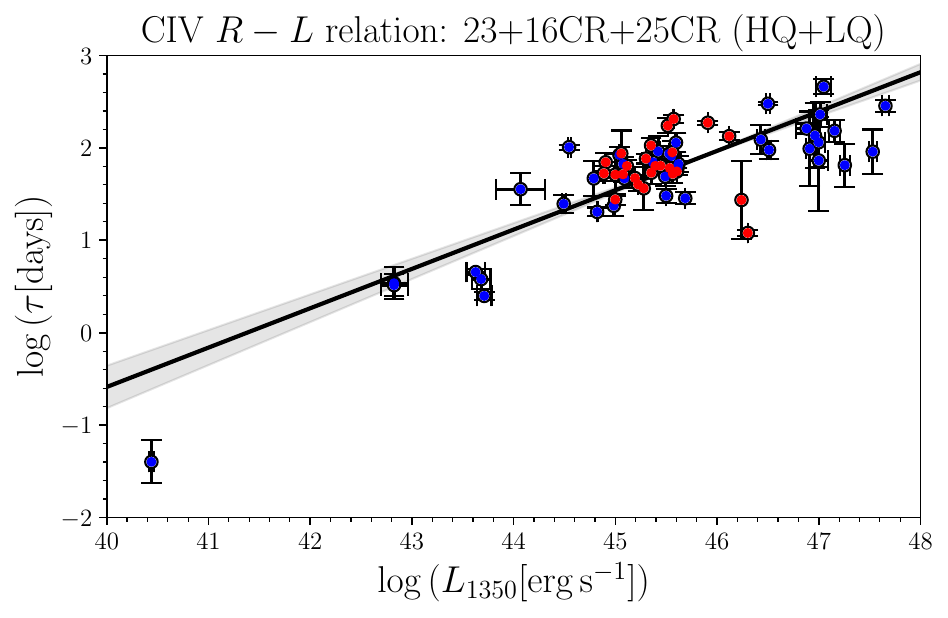}   
  \caption{\civ\ $R-L$ relations in the fixed flat $\Lambda$CDM cosmological model ($\Omega_{m0}=0.3$, $H_0=70\,{\rm km\,s^{-1}\,Mpc^{-1}}$) for the samples of 39 measurements (HQ subsample), 25 measurements (LQ subsample), and the combined sample of 64 measurements. These plots are analogous to Fig.~\ref{fig_RL_CIV} with the difference being that the time delays of RM SDSS sources are now based on the CREAM method (CR). \textbf{Left panel:} The best-fit $R-L$ relation is $\log{\tau}=(0.44^{+0.04}_{-0.04})\log{L_{44}}+1.03^{+0.08}_{-0.08}$ with $\sigma=0.30^{+0.05}_{-0.04}$. \textbf{Middle panel:} The best-fit $R-L$ relation is $\log{\tau}=(0.12^{+0.13}_{-0.09})\log{L_{44}}+1.63^{+0.13}_{-0.19}$ with $\sigma=0.29^{+0.05}_{-0.04}$. \textbf{Right panel:} The best-fit $R-L$ relation is $\log{\tau}=(0.43^{+0.04}_{-0.04})\log{L_{44}}+1.12^{+0.07}_{-0.07}$ with $\sigma=0.31^{+0.03}_{-0.03}$}.
    \label{fig_RL_CIV_cr}
\end{figure*}

\begin{figure*}
    \centering
     \includegraphics[width=0.33\textwidth]{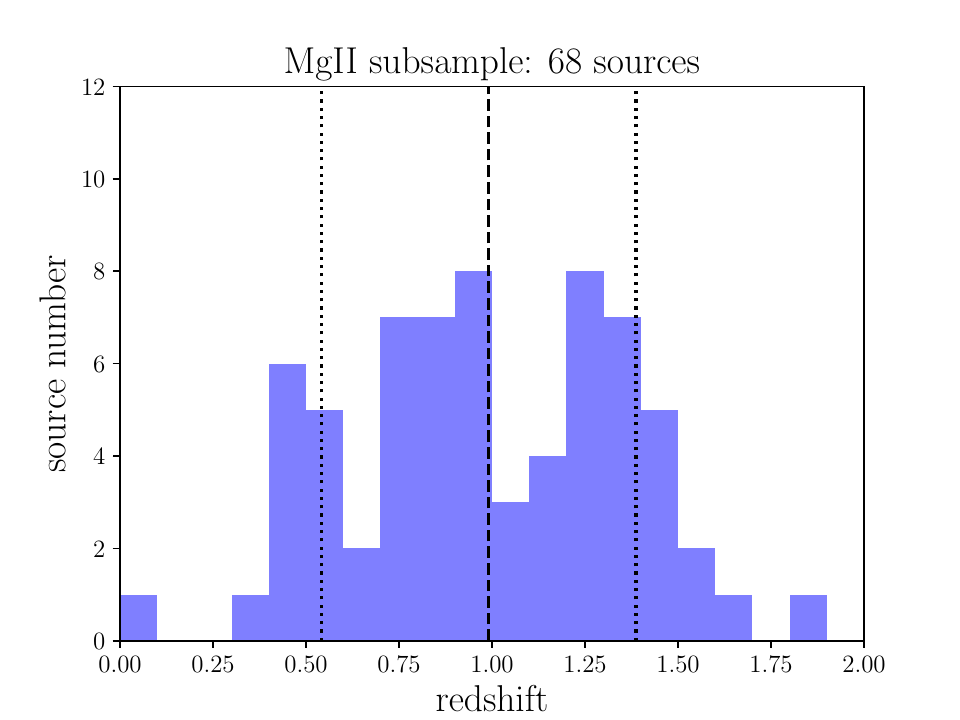}
       \includegraphics[width=0.33\textwidth]{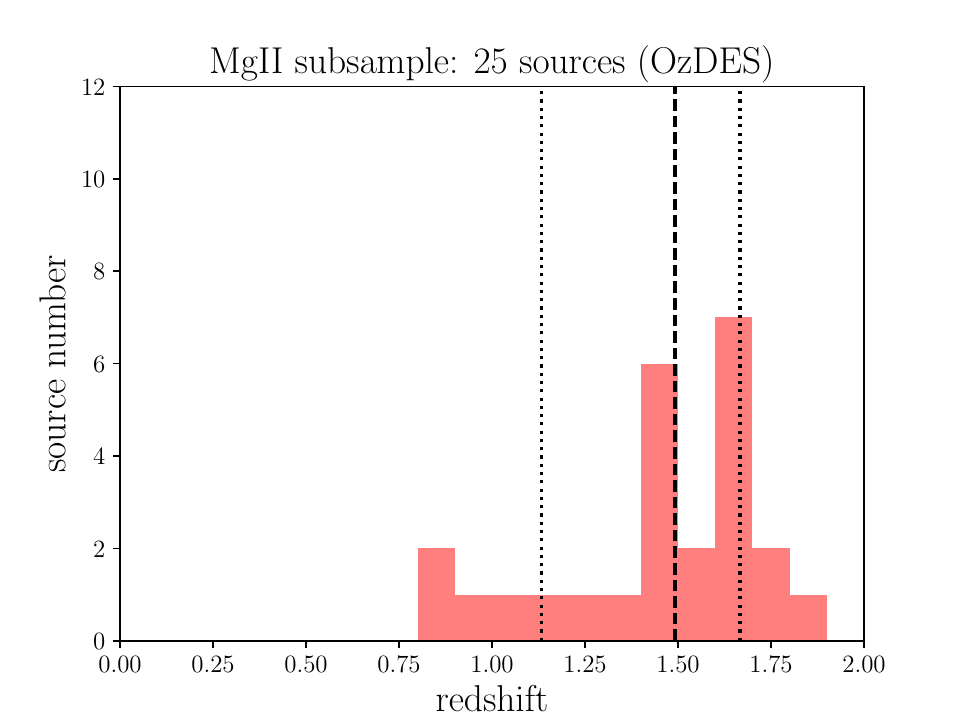}
      \includegraphics[width=0.33\textwidth]{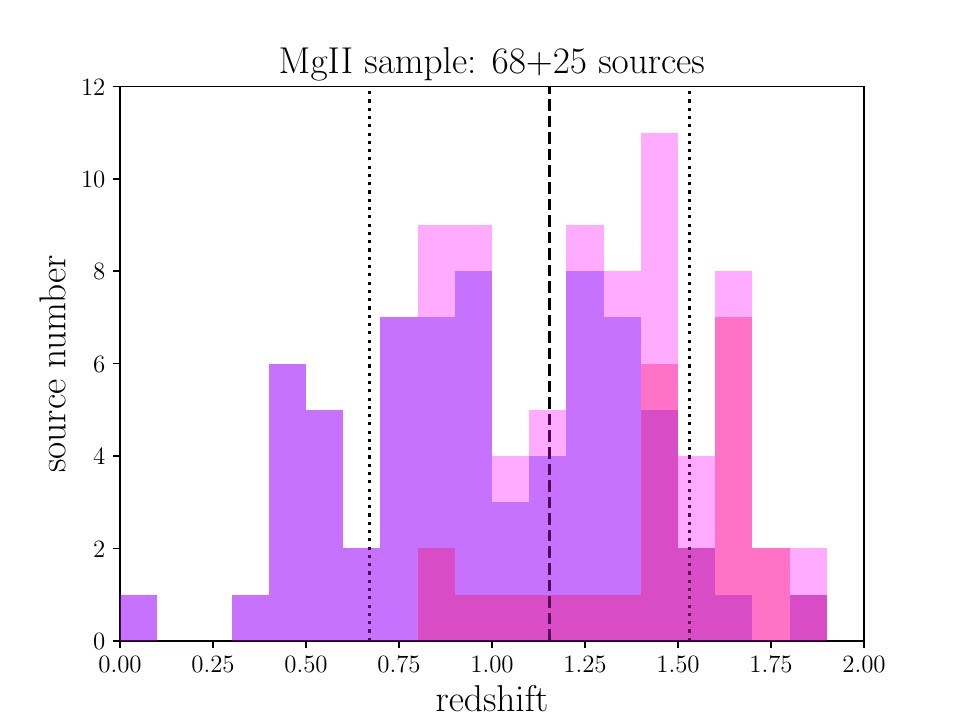}   
  \caption{Redshift distributions for the \mii\ QSOs. {\bf Left panel:} The sample of 68 sources. {\bf Middle panel:} The sample of 25 OzDES sources. {\bf Right panel:} The combined sample of $68+25=93$ sources. A vertical dashed line marks the distribution's median, while dotted lines stand for $16\%$ and $84\%$ percentiles. The bin width for all distributions is $\Delta z=0.1$.}
    \label{fig_MgII_redshift}
\end{figure*}

\begin{figure*}
    \centering
     \includegraphics[width=0.33\textwidth]{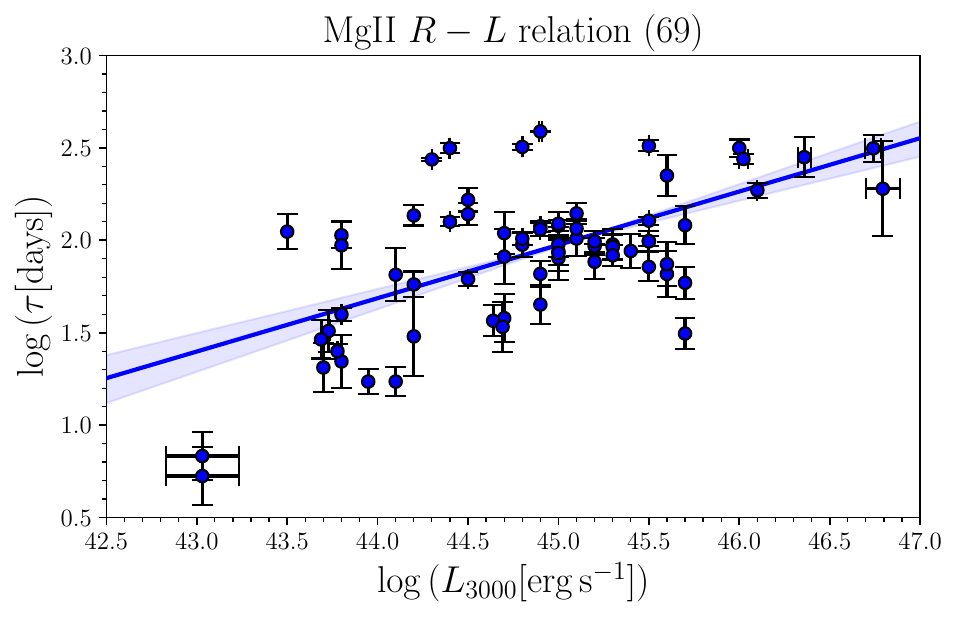}
       \includegraphics[width=0.33\textwidth]{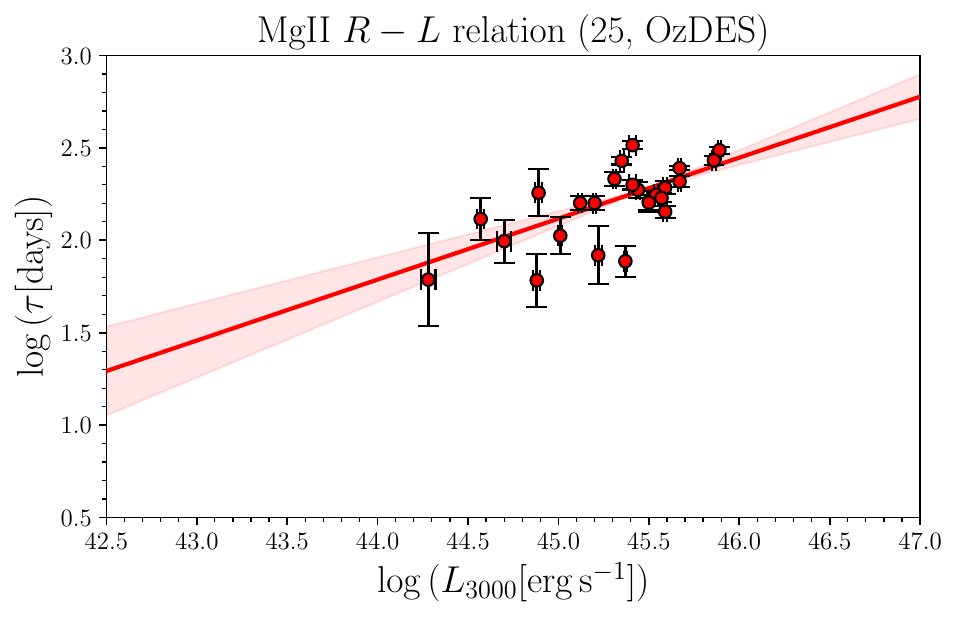}
      \includegraphics[width=0.33\textwidth]{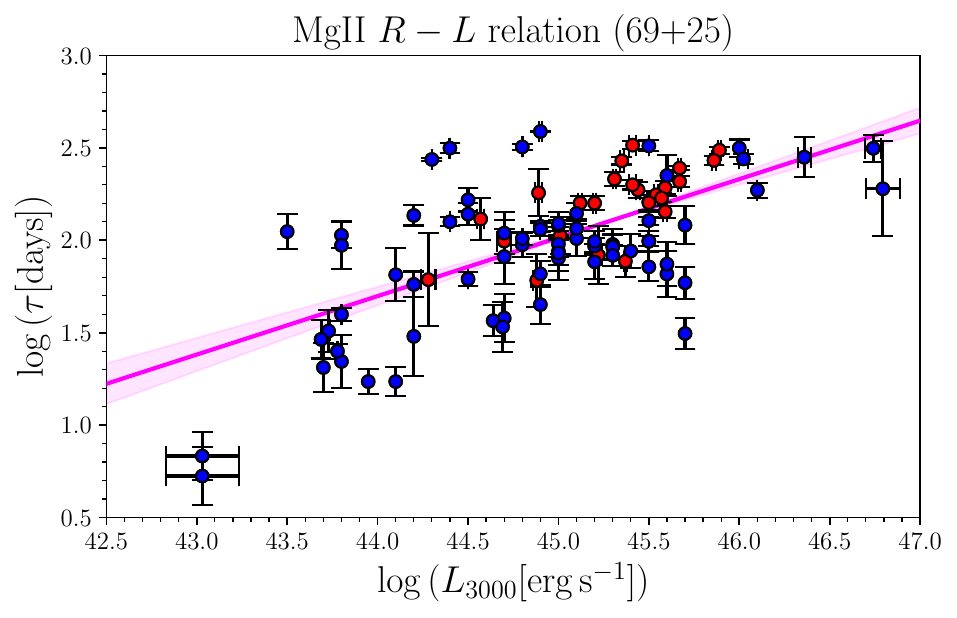}   
  \caption{\mii\ $R-L$ relations in the fixed flat $\Lambda$CDM cosmological model ($\Omega_{ m0}=0.3$, $H_0=70\,{\rm km\,s^{-1}\,Mpc^{-1}}$). {\bf Left panel:} The sample of 69 measurements. The best-fit $R-L$ relation is $\log{\tau}=(0.29^{+0.05}_{-0.05})\log{L_{44}}+1.69^{+0.05}_{-0.06}$ with $\sigma=0.29^{+0.03}_{-0.03}$. {\bf Middle panel:} The sample of 25 OzDES measurements. The best-fit $R-L$ relation is $\log{\tau}=0.33^{+0.08}_{-0.08})\log{L_{44}}+1.79^{+0.12}_{-0.12}$ with $\sigma=0.13^{+0.03}_{-0.02}$. {\bf Right panel:} The combined sample of 94 measurements. The best-fit $R-L$ relation is $\log{\tau}=(0.32^{+0.04}_{-0.04})\log{L_{44}}+1.70^{+0.05}_{-0.05}$ with $\sigma=0.26^{+0.02}_{-0.02}$. In each panel, we indicate the best-fit $R-L$ relation by a solid line. The shaded regions show the 1$\sigma$ confidence intervals. }
    \label{fig_RL_MgII}
\end{figure*}

\begin{table}
    \centering
    \caption{\civ\, sources whose redshifts were corrected for peculiar velocities. We include the original redshift value (old), corrected value (new), the difference, and the relative error.}
    \begin{tabular}{c|c|c|c|c}
    \hline
    \hline
    Source  & $z_{\rm old}$ & $z_{\rm new}$ & $\Delta z$ & $|\Delta z|/z_{\rm old}$  \\
    \hline
    NGC 4395 &  0.001064 &  0.001952 &  0.000888 & 0.835\\
    NGC 3783 &  0.009730 &  0.010787 &  0.001057 & 0.109\\
    NGC 7469 &  0.016320 &  0.015084 & -0.001236 & 0.076\\
    3C 390.3 &  0.056100 &  0.055868 & -0.000232 & 0.004\\
    NGC 4151 &  0.003320 &  0.004143 & 0.000823 & 0.248\\
    NGC 5548 &  0.016760 &  0.017462 &  0.000702 & 0.042 \\
    \hline
    \end{tabular} 
    \label{tab_z_corr}
\end{table}

\begin{itemize}
\item[]{\textit{C \textsl{\textsc{iv}} QSO data.}} These are a subset of QSOs, for which a significant time delay of the broad high-ionization \civ\ line (at 1549\,\AA\, in the rest frame) was measured. The primary \civ\ sample consists of high-quality data for 22 QSOs (more precisely 23 measurements with two time-delay measurements for NGC 4151) listed in Table~\ref{tab:civdata1} \citep[for a detailed description of the original papers and analyses of an earlier version of these data, see][]{Caoetal_2022}. These data cover a broad redshift range although the sample is not uniformly spaced in redshift. Several measurements were done for nearby QSOs ($z < 0.055868$) in the far-UV from space, and the rest are from large redshift QSOs ($z$ between 1.486 and 3.368) monitored in the optical band with ground-based telescopes. Redshifts of the six nearby QSOs were corrected for peculiar velocities. Usually, the redshifts of AGN are specified at best in the heliocentric reference frame, or just computed directly from data, without correcting for Earth's motion. The motion of a specific AGN relative to the cosmic microwave background (CMB) frame is typically of order of a few hundred km s$^{-1}$ and depends on the position of the source in the sky. For cosmological applications, the redshift of a nearby source must be corrected for this motion. To do this we use the NED Velocity Correction Calculator\footnote{\url{https://ned.ipac.caltech.edu/forms/vel_correction.html}} which corrects for Galactic rotation, the peculiar motion of the Galaxy within the Local Group, ``infall'' of the Local Group towards the centre of the Local Supercluster, and motion relative to the CMB reference frame. We have six nearby sources which need this correction (see Table~\ref{tab:civdata1}). For these six sources, we list old redshifts, new, corrected ones, the differences ($\Delta z=z_{\rm new}-z_{\rm old}$), and the relative errors in Table~\ref{tab_z_corr}. For four out of six sources, the correction is positive (the new redshift is larger). Given the other sources of scatter around the best-fit $R-L$ relation (measurement uncertainties of time-delays and flux densities for all the sources), the application of this correction, which is at the level of $\lesssim 10\%$ except for NGC 4395, to a subsample of nearby sources does not lead to the significant decrease in scatter and tightening of cosmological constraints. This sample is supplemented with a sample of 41 large redshift QSOs selected from a larger sample monitored from the ground as part of the SDSS-RM AGN project. These are listed in Table~\ref{tab:civdata2}. The selection was based on data quality, adopting the 10\% limit after \citet{2019ApJ...887...38G}. The delay in this sample is measured with two methods which give slightly different results so we test both measurement sets independently, marking them according to the method (CC stands for Interpolation Cross Correlation method and CR for CREAM software; for details see \citealt{2019ApJ...887...38G}). In addition, during the selection of SDSS QSOs, we required that both CC and CREAM time delays be positive, which led to the exclusion of a few QSOs and a final source count of 41. In general, among \civ\ sources we can distinguish higher-quality data, specifically combined from the 23 measurements of Table~\ref{tab:civdata1} and the 16 class-4 and class-5 SDSS measurements of Table~\ref{tab:civdata2}, hence overall 39 higher-quality (HQ) time-delay measurements (of 38 QSOs). These sources have a rather broad redshift distribution with a median redshift of 2.051, while the 16-\% and the 84-\% percentiles are 1.244 and 2.647. We show the redshift distribution of these HQ QSOs in Fig.~\ref{fig_CIV_redshift} (left panel). Among the 41 SDSS sources, we have 25 QSOs in quality categories 1, 2, and 3, which are considered lower-quality (LQ) sources in terms of time-delay detection. Their redshift distribution is more compact (Fig.~\ref{fig_CIV_redshift}, middle panel) with a median of 1.966, a 16\%-percentile of 1.729 and an 84\%-percentile of 2.332. The combined sample of 64 measurements (of 63 QSOs) has a redshift median of 1.966, a 16\%-percentile of 1.590 and an 84\%-percentile of 2.595. The redshift distribution of the combined \civ\ sample is shown in Fig.~\ref{fig_CIV_redshift} (right panel). The 1350\,\AA\, luminosities of the combined sample fall in the range of $\log{(L_{1350}\,[{\rm erg\,s^{-1}])}}\in (40.439, 47.655)$ with a median of $\log{(L_{1350}\,[{\rm erg\,s^{-1}])}}=45.490$. The two \civ\ subsamples (HQ and LQ), as well as the combined sample, approximately follow a power-law radius-luminosity ($R-L$) relation, which combines the observationally determined rest-frame time-delay of the \civ\ line and monochromatic flux density at 1350\,\AA. In Fig.~\ref{fig_RL_CIV} we show the $R-L$ relations in the fixed flat $\Lambda$CDM model ($\Omega_{m0}=0.3$, $H_0=70\,{\rm km\,s^{-1}\,Mpc^{-1}}$) for the HQ subsample of 39 measurements (left panel), the LQ subsample of 25 QSOs (middle panel), and the combined sample of 64 points (right panel). In this figure, the time delays for the 41 SDSS QSOs were determined based on the CC method. Within the 1$\sigma$ uncertainties these $R-L$ relations are consistent with those for the case when the SDSS RM time delays are instead inferred using the CREAM method, see Fig.~\ref{fig_RL_CIV_cr} for CREAM results.

\item[]{\textit{Mg \textsl{\textsc{ii}} QSO data.}} These are a subset of QSOs that were reverberation-mapped using the low-ionization broad \mii\ line (at 2798\,\AA\, in the rest frame). As listed in Tables \ref{tab:MgII69} and \ref{tab:MgII25}, the \mq\ sample consists of 69 + 25 measurements and spans the redshift range $0.004143 \leq z \leq 1.89$. The 69 measurements consist of 68 from \citet{martinezAldama2020} and the measurement of the source HE0435-4312 from \citet{zajacek2021}. For this sample of 69 measurements (from 68 QSOs with two time-delay measurements for NGC 4151), the median redshift is 0.990, while the 16-\% percentile is 0.542 and the 84-\% percentile is 1.387. The redshift of one low-redshift source (NGC 4151) was corrected for peculiar velocity. In terms of the 3000\,\AA\, luminosity (computed in the flat $\Lambda$CDM model with $\Omega_{ m0}=0.3$ and $H_0=70\,{\rm km\,s^{-1}\,Mpc^{-1}}$), the range is $\log{(L_{3000}\,[{\rm erg\,s^{-1}])}}\in (43.030, 46.794)$, with a median of $\log{(L_{3000}\,[{\rm erg\,s^{-1}}])}=44.900$. The sample of 25 OzDES QSOs is from \citet{Yuetal2023}. Its redshift median is  1.492, while the 16-\% percentile is 1.133 and the 84-\% percentile is 1.667. The 3000\,\AA\, luminosity range is $\log{(L_{3000}\,[{\rm erg\,s^{-1}}])}\in (43.930, 45.556)$ with a median of $\log{(L_{3000}\,[{\rm erg\,s^{-1}}])}=44.979$. The combined sample of 94 measurements has a redshift median of 1.155, with a 16-\% percentile of $0.671$ and an 84-\% percentile of $1.531$. The 3000\,\AA\, luminosity range is $\log{(L_{3000}\,[{\rm erg\,s^{-1}])}}\in (43.030, 46.794)$ with a median of $\log{(L_{3000}\,[{\rm erg\,s^{-1}}])}=44.972$. The redshift distributions for the subsamples of 69 and 25 measurements and the combined sample of 94 measurements are shown in Fig.~\ref{fig_MgII_redshift}. We also test that these \mq\ data obey the QSO $R-L$ relation, where the measured quantities are the \mii\ time-delay $\tau$ and the QSO flux $F_{3000}$ measured at 3000 \(\text{\r{A}}\). The $R-L$ relations in the fixed flat $\Lambda$CDM model ($\Omega_{m0}=0.3$, $H_0=70\,{\rm km\,s^{-1}\,Mpc^{-1}}$) for \mq\ data are shown in Fig.~\ref{fig_RL_MgII} for the subsample of 69 measurements, the subsample of 25 measurements, and the combined 94 measurement sample in the left, middle, and right panels, respectively. \citet{Khadkaetal_2021a} and \citet{Caoetal_2022} analyzed the sample of 69 measurements described above in combination with 9 OzDES QSOs presented in \citet{Yuetal2021}. These 9 sources are also included here as a part of 25 OzDES QSOs \citep{Yuetal2023} but with better-established time delays using longer time series. The time delays for the sample of \mii\ sources were inferred using different methods. The largest SDSS-RM subsample of 57 sources was analyzed using the JAVELIN method \citep{2020ApJ...901...55H}. The time delays for the subsample of 6 SDSS-RM QSOs were found using the ICCF method \citep[centroids;][]{2016ApJ...818...30S}. The low-luminosity source NGC 4151 (two measurements) was investigated using the ICCF as well \citep[centroid values;][]{2006ApJ...647..901M}. The time delay for the high-luminosity source CTS 252 was inferred using the ICCF method \citep[centroid value;][]{2018ApJ...865...56L}. The three luminous QSOs, CTS C30.10, HE0413-4031, and HE0435-4312, have rest-frame time delays that are determined as the means of the best time-delays inferred based on six or seven different methodologies \citep[ICCF, DCF, zDCF, JAVELIN, $\chi^2$, von Neumann, and Bartels, see][]{2019ApJ...880...46C,2020ApJ...896..146Z,zajacek2021,2022arXiv220111062P,2023A&A...678A.189P,2023arXiv231003544Z}. For the subsample of 25 OzDES sources, we adopted the rest-frame \mii\ time delays inferred using the ICCF method \citep{Yuetal2023}. Hence, in summary, 57 time delays were determined using the JAVELIN method, 34 time delays were inferred using the ICCF method, and three time delays were determined using a combination of these and other methods.

\item[]{$H(z)\ +\ BAO\ data$.} We also use 32 $H(z)$ and 12 BAO measurements listed in Tables 1 and 2 of \cite{CaoRatra2022}, spanning the redshift ranges $0.07 \leq z \leq 1.965$ and $0.122 \leq z \leq 2.334$, respectively, to determine cosmological parameter constraints and compare these with those from \civ\ and \mq\ data.

\end{itemize}

\section{Data Analysis Methodology}
\label{sec:analysis}

The QSO radius-luminosity relation ($R-L$ ) for \civ\ (denoted by subscript ``\textsc{c}'') or \mii\ (denoted by subscript ``\textsc{m}'') QSOs can be expressed as 
\begin{equation}
    \label{eq:civ}
    \log{\frac{\tau_{\textsc{c/m}}}{\rm days}}=\beta_{\rm\textsc{c/m}}+\gamma_{\rm\textsc{c/m}} \log{\left(\frac{L_{\rm\textsc{c/m}}}{10^{44}\,{\rm erg\ s^{-1}}}\right)},
\end{equation}
where $\tau_{\rm\textsc{c/m}}$, $\beta_{\rm\textsc{c/m}}$, and $\gamma_{\rm\textsc{c/m}}$ are the \civ/\mii\ time-lag, the intercept parameter, and the slope parameter, respectively, and the monochromatic luminosity $L_{\rm\textsc{c/m}}$ at 1350\,\AA\ for \civ\ and at 3000\,\AA\ for \mii\
\be
\label{eq:L1350}
    L_{\rm\textsc{c/m}}=4\pi D_L^2F_{\rm\textsc{c/m}},
\ee
with measured quasar flux $F_{\rm\textsc{c/m}}$ in units of $\rm erg\ s^{-1}\ cm^{-2}$. The luminosity distance is a function of redshift $z$ and the cosmological parameters,
\begin{equation}
  \label{eq:DL}
\resizebox{0.475\textwidth}{!}{%
    $D_L(z) = 
    \begin{cases}
    \frac{c(1+z)}{H_0\sqrt{\Omega_{\rm k0}}}\sinh\left[\frac{H_0\sqrt{\Omega_{\rm k0}}}{c}D_C(z)\right] & \text{if}\ \Omega_{\rm k0} > 0, \\
    \vspace{1mm}
    (1+z)D_C(z) & \text{if}\ \Omega_{\rm k0} = 0,\\
    \vspace{1mm}
    \frac{c(1+z)}{H_0\sqrt{|\Omega_{\rm k0}|}}\sin\left[\frac{H_0\sqrt{|\Omega_{\rm k0}|}}{c}D_C(z)\right] & \text{if}\ \Omega_{\rm k0} < 0,
    \end{cases}$%
    }
\end{equation}
where the comoving distance is
\begin{equation}
\label{eq:gz}
   D_C(z) = c\int^z_0 \frac{dz'}{H(z')},
\end{equation}
with $c$ being the speed of light.

The natural log of the \civ/\mii\ likelihood function \citep{D'Agostini_2005} is
\be
\label{eq:LH_cm}
    \ln\mathcal{L}_{\rm\textsc{c/m}}= -\frac{1}{2}\Bigg[\chi^2_{\rm\textsc{c/m}}+\sum^{N}_{i=1}\ln\left(2\pi\sigma^2_{\mathrm{tot,\textsc{c/m}},i}\right)\Bigg],
\ee
where
\be
\label{eq:chi2_cm}
    \chi^2_{\rm\textsc{c/m}} = \sum^{N}_{i=1}\bigg[\frac{(\log \tau_{\mathrm{obs,\textsc{c/m}},i} - \beta_{\rm\textsc{c/m}}  -\gamma_{\rm\textsc{c/m}}\log L_{\rm\textsc{c/m},i})^2}{\sigma^2_{\mathrm{tot,\textsc{c/m}},i}}\bigg]
\ee
with total uncertainty
\be
\label{eq:sigma_civ}
\sigma^2_{\mathrm{tot,\textsc{c/m}},i}=\sigma_{\rm int,\,\textsc{c/m}}^2+\sigma_{{\log \tau_{\mathrm{obs,\textsc{c/m}},i}}}^2+\gamma_{\rm\textsc{c/m}}^2\sigma_{{\log F_{\mathrm{\textsc{c/m}},i}}}^2,
\ee
where $\sigma_{\rm int,\,\textsc{c/m}}$ is the \civ/\mii\ intrinsic scatter parameter which also contains the unknown systematic uncertainty, $\sigma_{{\log \tau_{\mathrm{obs,\textsc{c/m}},i}}}$ and $\sigma_{{\log F_{\mathrm{\textsc{c/m}},i}}}$ are the time-delay and the flux density measurement uncertainties for individual \civ/\mii\, QSOs, and $N$ is the number of data points.

The likelihood functions of $H(z)$ and BAO data are described in \cite{CaoRyanRatra2020}. The flat priors of the free cosmological model and QSO $R-L$ relation parameters are listed in Table \ref{tab:priors}. We use the Markov chain Monte Carlo (MCMC) code \textsc{MontePython} \citep{Audrenetal2013,Brinckmann2019} to maximize the likelihood functions and to obtain the unmarginalized best-fitting values and posterior distributions of all free cosmological-model and QSO $R-L$ relation parameters. The \textsc{python} package \textsc{getdist} \citep{Lewis_2019} is used to compute the posterior mean values and uncertainties of the constraints and to plot the figures. The definitions of the Akaike Information Criterion (AIC), the Bayesian Information Criterion (BIC), and the Deviance Information Criterion (DIC) can be found in our previous papers \citep[see, e.g.][]{CaoDainottiRatra2022}. $\Delta \mathrm{IC}$ is computed as the differences between the IC value of the flat \lcdm\ reference model and that of the other five cosmological dark energy models, where a positive (negative) value of $\Delta \mathrm{IC}$ indicates that the model under consideration fits the data set worse (better) than does the flat \lcdm\ reference model. Relative to the model with the minimum IC value, $\Delta \mathrm{IC} \in (0, 2]$ shows weak evidence against the model under consideration, $\Delta \mathrm{IC} \in (2, 6]$ shows positive evidence against the model under consideration, $\Delta \mathrm{IC} \in (6, 10] $ shows strong evidence against the model under consideration, and $\Delta \mathrm{IC}>10$ shows very strong evidence against the model under consideration.

\begin{table}
\centering
\resizebox{\columnwidth}{!}{%
\begin{threeparttable}
\caption{Flat priors of the constrained parameters.}
\label{tab:priors}
\begin{tabular}{lcc}
\toprule
Parameter & & Prior\\
\midrule
 & Cosmological-Model Parameters & \\
\midrule
$H_0$\,\tnote{a} &  & [None, None]\\
\obhs\,\tnote{b} &  & [0, 1]\\
\ochs\,\tnote{c} &  & [0, 1]\\
\ok &  & [-2, 2]\\
$\alpha$ &  & [0, 10]\\
\wx &  & [-5, 0.33]\\
\midrule
 & $R-L$ Relation Parameters & \\
\midrule
$\gamma$ &  & [0, 5]\\
$\beta$ &  & [0, 10]\\
$\sigma_{\rm int}$ &  & [0, 5]\\
\bottomrule
\end{tabular}
\begin{tablenotes}[flushleft]
\item [a] \hunit. In the cases excluding $H(z)$ + BAO data, $H_0$ is set to be 70 \hunit, while in other cases, the prior range is irrelevant (unbounded).
\item [b] In the cases excluding $H(z)$ + BAO data, \obhs\ is set to be 0.0245, i.e. $\Omega_{b}=0.05$.
\item [c] In the cases excluding $H(z)$ + BAO data, $\Om\in[0,1]$ is ensured.
\end{tablenotes}
\end{threeparttable}%
}
\end{table}

\section{Constraints on $R-L$ parameters in the standard flat \lcdm\ cosmological model}
\label{sec:results_fixedLCDM}

We first study the radius-luminosity $(R-L)$ relations --- for a number of combinations of QSO measurements --- in a fixed flat $\Lambda$CDM cosmological model with $\Omega_{m0}=0.3$ and $H_0=70\,{\rm km\,s^{-1}\,Mpc^{-1}}$. This provides insight into the dispersion properties of individual QSO subsamples and the combined sample. It also indicates the magnitude of systematic differences among the subsamples, and even between the two methods used for the time-delay measurements. This is an approximate analysis, as it holds cosmological parameters fixed, only allowing for the variation of $R-L$ relation parameters. Results from the more correct analyses --- that allow both sets of parameters to vary --- are described in Section \ref{sec:results_general}.

\begin{table*}
   \caption{Correlation between the rest-frame time delay and monochromatic luminosity in the fixed flat $\Lambda$CDM model for different \civ\ and \mii\ QSO samples. For each subsample, we list the number of sources $N$, Pearson correlation coefficient $r$, and Spearman correlation coefficient $s$. In parentheses, we list the corresponding $p$-values, i.e. the probability that an uncorrelated data set would yield a correlation coefficient as large as the one measured.}
   \begin{tabular}{c|c|c|c}
   \hline
   \hline
    Sample   & \civ\ (CC)  & \civ\ (CR) & \mii\  \\
   \hline 
   \multirow{3}{*}{Subsample 1}     &   $N=23+16$ CC   &  $N=23+16$ CR  &   $N=69$  \\
                                    &   $r=0.90\,(p=1.40 \times 10^{-14})$   &  $r=0.89\, (p=4.24 \times 10^{-14})$  &  $r=0.62\,(p=1.59 \times 10^{-8})$   \\
                                    &   $s=0.81\,(p=3.08 \times 10^{-10})$   &  $s=0.78\, (p=3.33 \times 10^{-9})$  &  $s=0.46\, (p=5.79 \times 10^{-5})$   \\
   \hline 
   \multirow{3}{*}{Subsample 2}     &   $N=25$ CC   &  $N=25$ CR  &  $N=25$  \\
                                    &   $r=0.28\,(p=0.17)$   & $r=-0.04\, (p=0.87)$ &  $r=0.69\, (p=1.55 \times 10^{-4})$   \\
                                    &   $s=0.31\,(p=0.14)$   & $s=0.19\, (p=0.36)$ &  $s=0.65, (p=4.73 \times 10^{-4})$   \\
   \hline
   \multirow{3}{*}{Combined sample}     &   $N=23+41$ CC   &  $N=23+41$ CR &  $N=69+25$  \\
                                        &   $r=0.83\, (p=3.89 \times 10^{-17})$   & $r=0.82\, (p=5.47 \times 10^{-17})$ &  $r=0.66\,(p=7.38 \times 10^{-13})$   \\
                                        &   $s=0.67\, (p=1.57 \times 10^{-9})$   &  $s=0.68\, (p=8.77 \times 10^{-10})$ &  $s=0.54\,(p=2.02 \times 10^{-8})$   \\
   \hline                                     
   \end{tabular} 
   \label{tab_correlation}
\end{table*}

We first compute the corresponding monochromatic luminosities $L_{1350}$ and $L_{3000}$ for \civ\ and \mii\ QSOs, respectively, in the fixed flat $\Lambda$CDM model with $\Omega_{m0}=0.3$ and $H_0=70\,{\rm km\,s^{-1}\,Mpc^{-1}}$. With the rest-frame time delays of \civ\ and \mii\ lines, $\tau_{\rm C}$ and $\tau_{\rm M}$, we then compute Pearson and Spearman rank-order correlation coefficients of the quantity pairs $\tau_{\rm C}-L_{1350}$ and $\tau_{\rm M}-L_{3000}$ for the larger \mii\ and \civ\ subsamples (Subsample 1), smaller subsamples consisting of lower-quality \civ\ QSOs and new OzDES \mii\ sources (Subsample 2), which were not considered previously in cosmological applications, and the combined sample (Subsample 1 $+$ Subsample 2). These three data sets are described in Section \ref{sec:data} and summarized in Table~\ref{tab_correlation}, where we list their $R-L$ correlation properties. 

For the \civ\ case the high-quality sample of 39 measurements is significantly positively correlated, while the lower-quality sample of 25 sources does not exhibit a significant correlation. This leads to a lower correlation between the time delay and the monochromatic luminosity when the two sets are combined, irrespective of the time-delay method used. For the \mii\ case the sample of 69 measurements is significantly positively correlated, although a bit less than for the HQ \civ\ data set. The smaller sample of 25 OzDES QSOs is also positively correlated, however with lower significance than for the larger \mii\ subsample. After combining both \mii\ sets the correlation between the time delay and the monochromatic luminosity drops slightly but remains significant.

We next use the \textsc{python} package \textsc{emcee} to infer the $R-L$ relation parameters in the fixed flat $\Lambda$CDM model, specifically the slope, intercept, and intrinsic dispersion, see Figs.~\ref{fig_RL_CIV}, \ref{fig_RL_CIV_cr} and \ref{fig_RL_MgII} for the \civ\ (CC), \civ\ (CR), and \mii\ data samples, respectively. For HQ \civ\ data, the chosen time-delay method does not significantly affect the $R-L$ relation parameter values. However, the LQ sample of 25 \civ\ measurements exhibits a significantly flatter relation with a larger intercept for the CREAM case compared to the CC case. After combining both LQ and HQ subsamples, the intercept of the $R-L$ relation of the combined \civ\ sample is slightly increased to the HQ sample for both time-delay methods ($\beta_{\textsc{c}}=1.11$ vs. $\beta_{\textsc{c}}=1.02$). For the \mii\ case the smaller sample of 25 QSOs from the OzDES survey has an even larger intercept to the subsample of 69 \mii\ measurements. After combining both \mii\ samples, the $R-L$ relation has an increased intercept ($\beta_{\textsc{m}}=1.83$ vs.\ $\beta_{\textsc{m}}=1.69$) as well as a significantly larger scatter ($\sigma_{\rm int,\textsc{m}}=0.39$ vs.\ $\sigma_{\rm int,\textsc{m}}=0.29$), relative to the 69 measurements case. 

These results compare well with results in the literature. For \mii\, we \citep{martinezAldama2020} previously obtained a slope $0.30 \pm 0.05$ if no third parameter was included. \citet{2020ApJ...901...55H} give slope values of $0.22^{+0.05}_{-0.06}$  for the entire sample with significant delays, and $0.31 \pm 0.10$ for their golden sample. \citet{Yuetal2023} determined the value $0.39 \pm 0.08 $ from their latest OzDES results. For \civ\ the SDSS RM campaign give a slope of $0.51 \pm
0.05$ \citep{2019ApJ...887...38G}, consistent with the earlier measurements. \citet{OzDES2022} in their most recent analysis obtained a slope for the OzDES \civ\ sample of $0.454 \pm 0.016$. Finally, \citet{2021ApJ...915..129K} constrained the slope of the \civ\ $R-L$ relation to $\sim 0.45 \pm 0.05$.

\section{Simultaneous constraints on $R-L$ and cosmological model parameters}
\label{sec:results_general}

\begin{figure*}
\centering
 \subfloat[]{%
    \includegraphics[width=0.5\textwidth,height=0.5\textwidth]{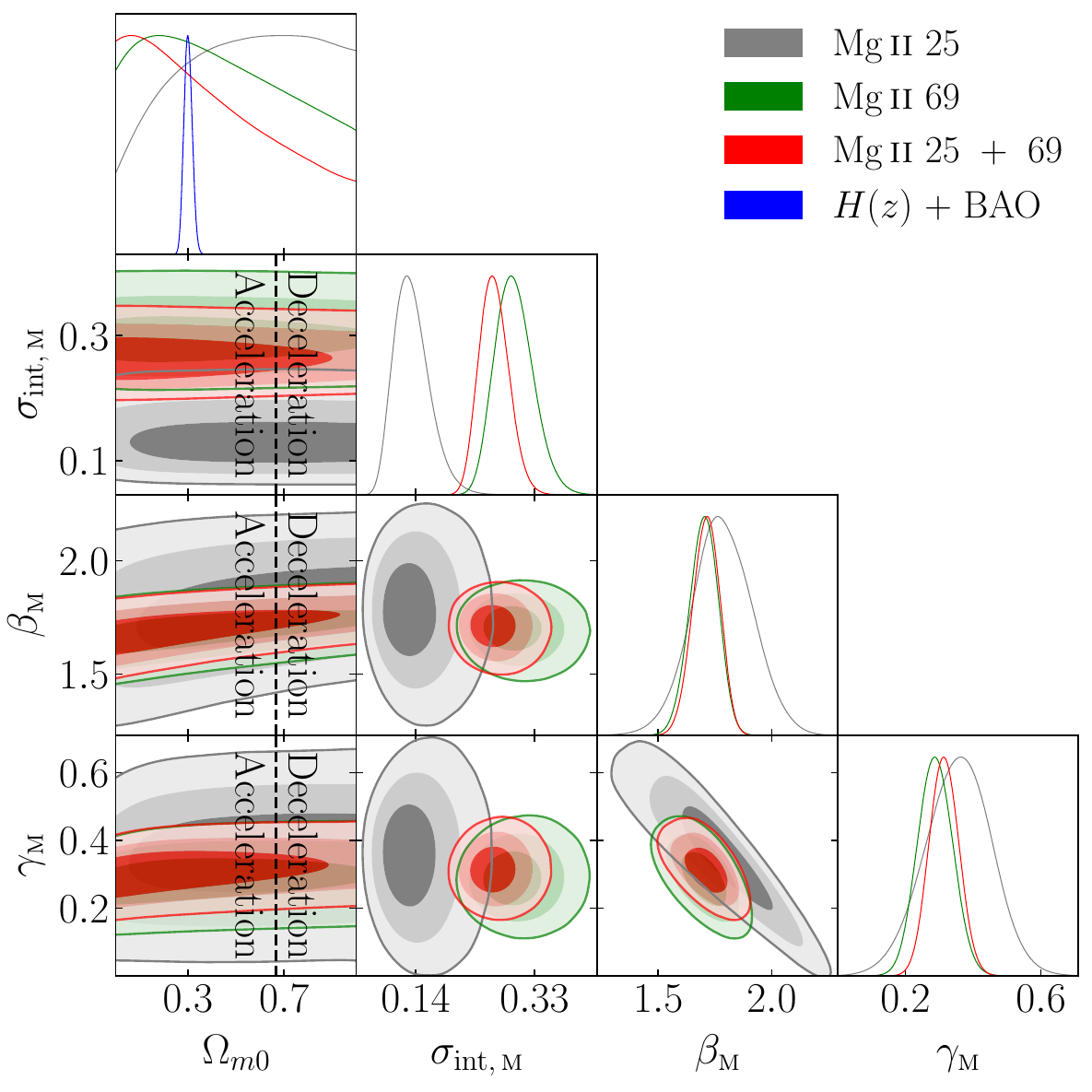}}
 \subfloat[]{%
    \includegraphics[width=0.5\textwidth,height=0.5\textwidth]{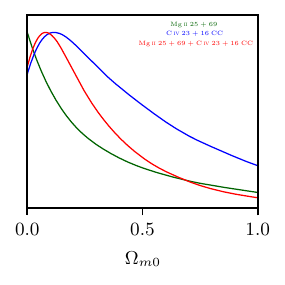}}\\
 \subfloat[]{%
    \includegraphics[width=0.5\textwidth,height=0.5\textwidth]{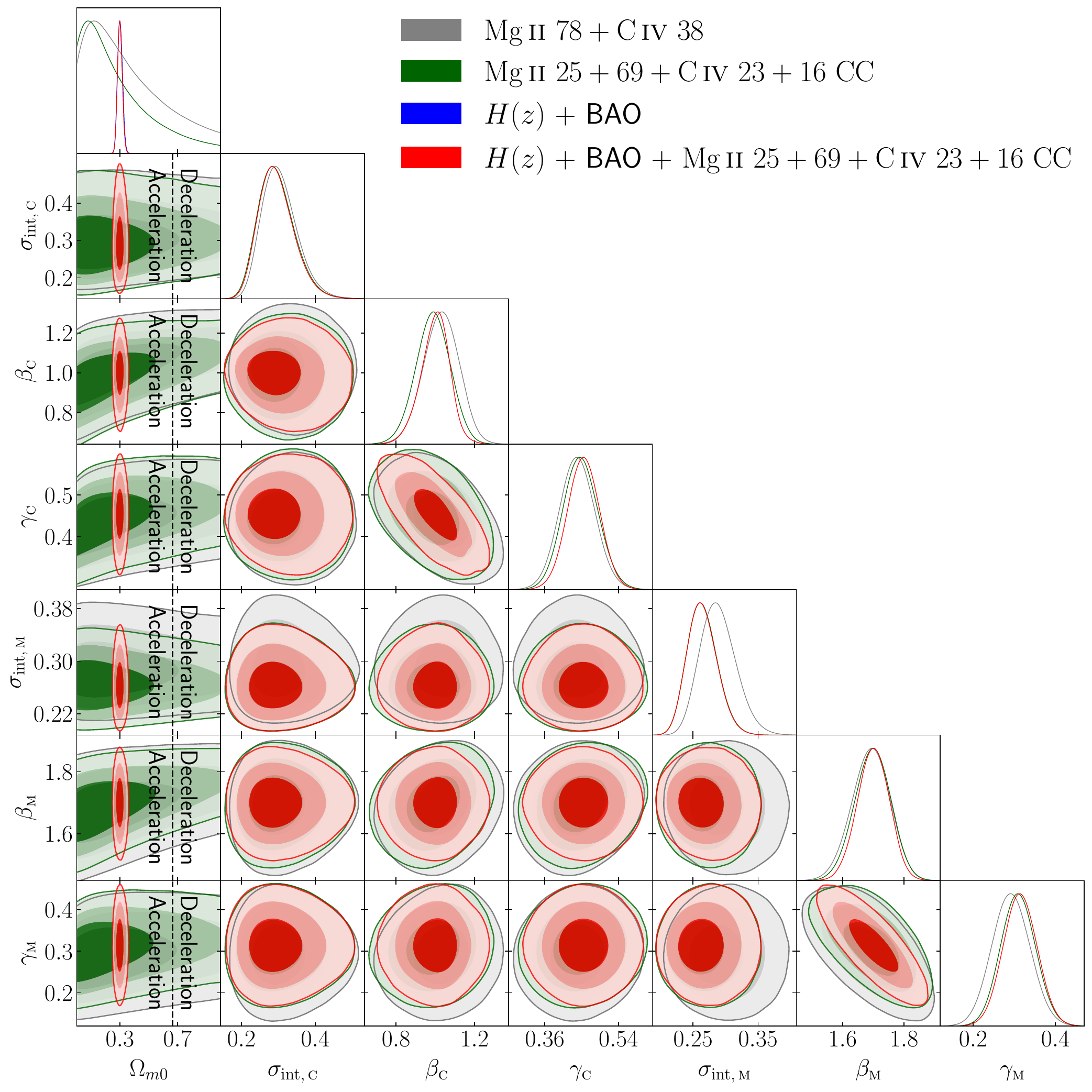}}
 \subfloat[]{%
    \includegraphics[width=0.5\textwidth,height=0.5\textwidth]{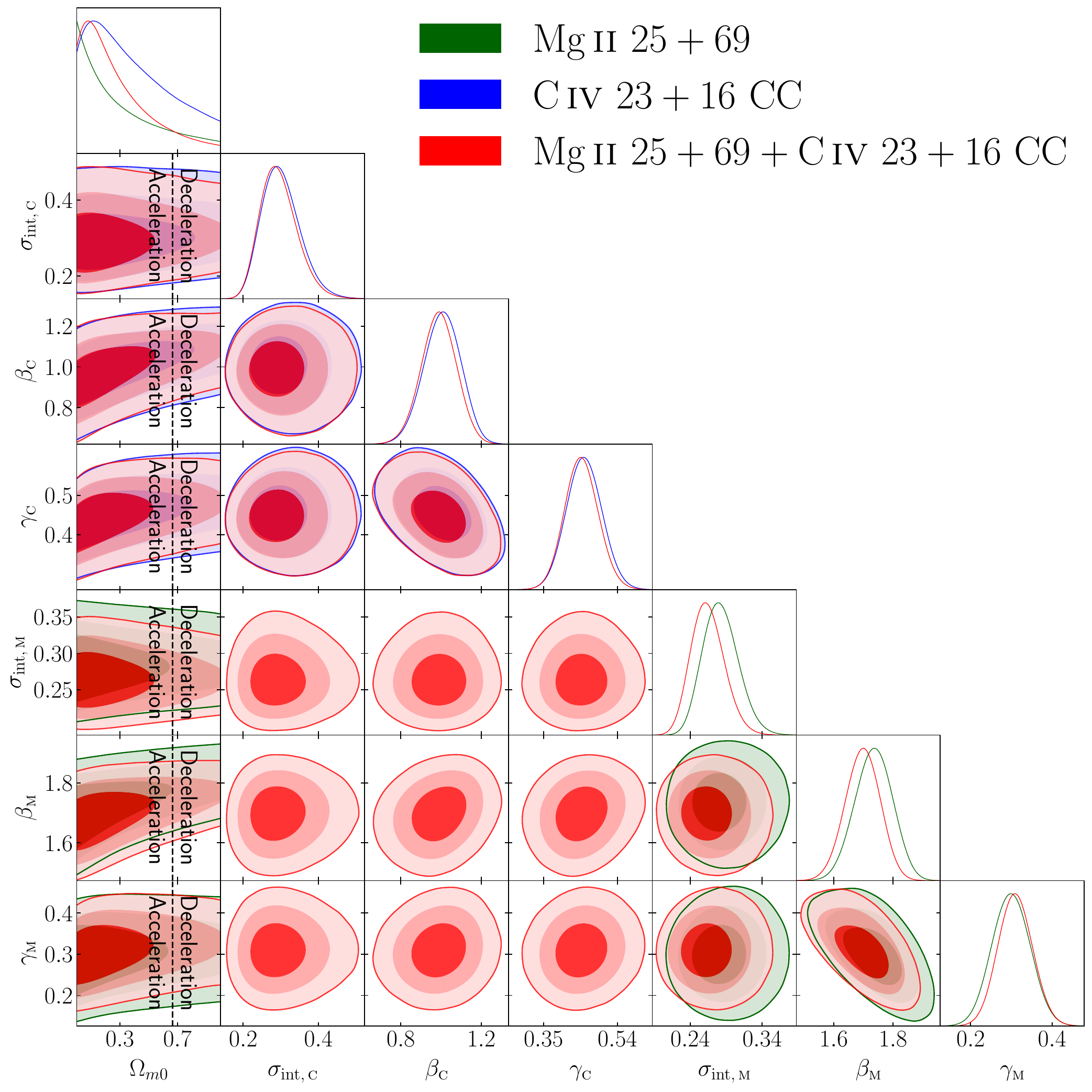}}\\
\caption{One-dimensional likelihood distributions and 1$\sigma$, 2$\sigma$, and 3$\sigma$ two-dimensional likelihood confidence contours for flat \lcdm\ from various combinations of data. The black dashed zero-acceleration lines divide the parameter space into regions associated with currently accelerating (left) and currently decelerating (right) cosmological expansion.}
\label{fig01}
\end{figure*}

\begin{figure*}
\centering
 \subfloat[]{%
    \includegraphics[width=0.5\textwidth,height=0.5\textwidth]{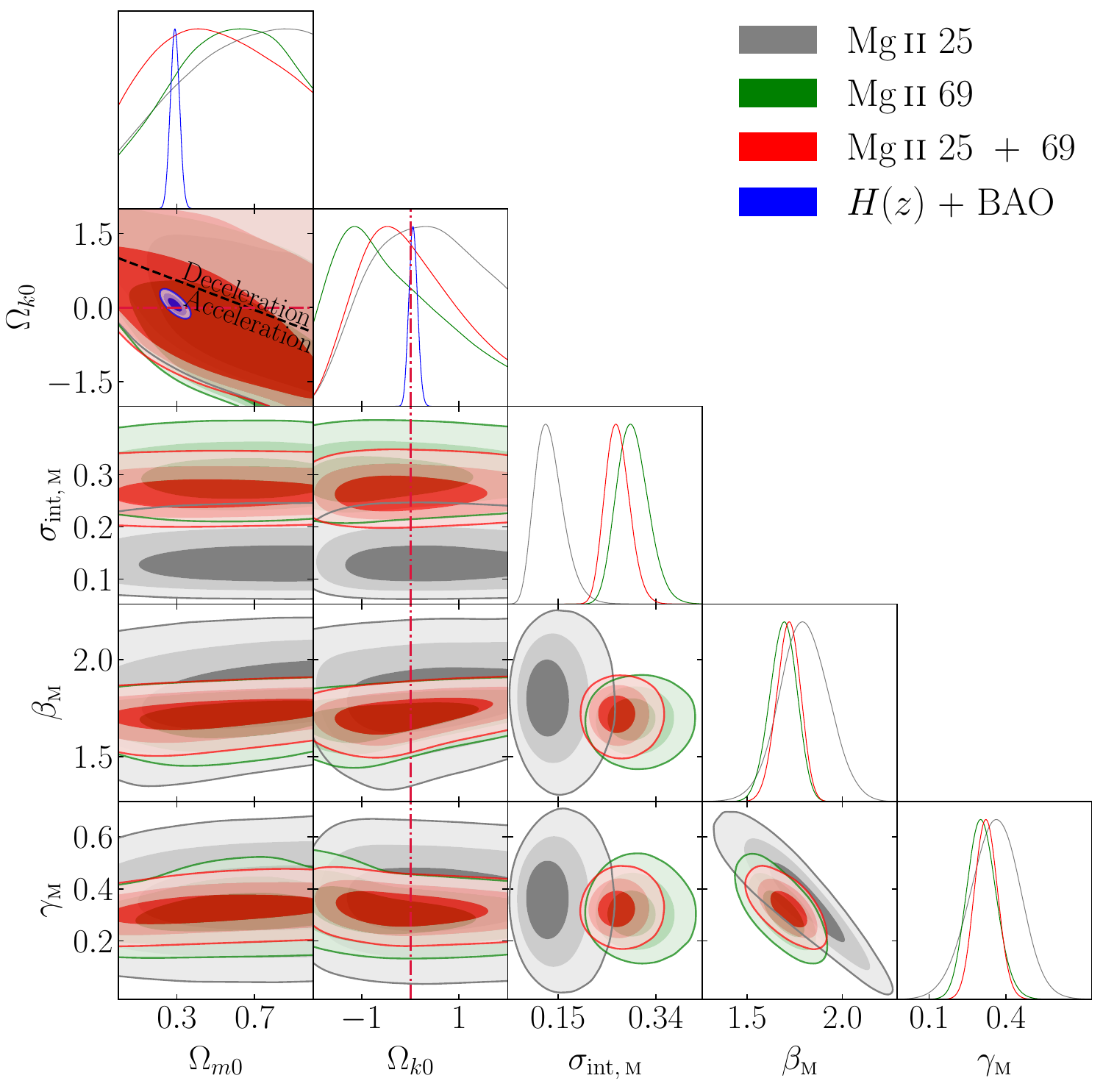}}
 \subfloat[]{%
    \includegraphics[width=0.5\textwidth,height=0.5\textwidth]{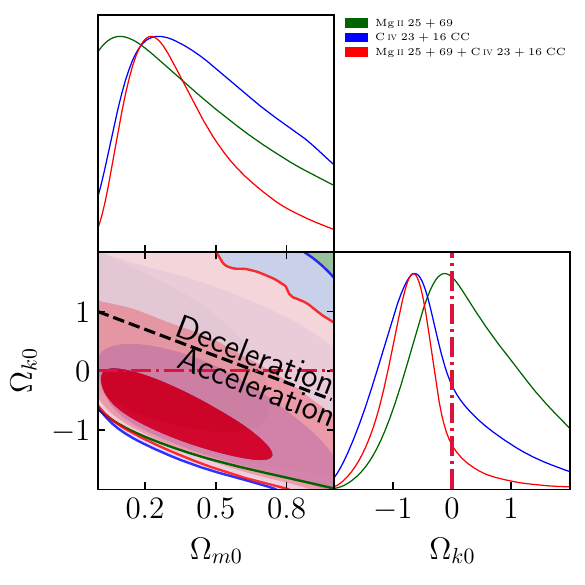}}\\
 \subfloat[]{%
    \includegraphics[width=0.5\textwidth,height=0.5\textwidth]{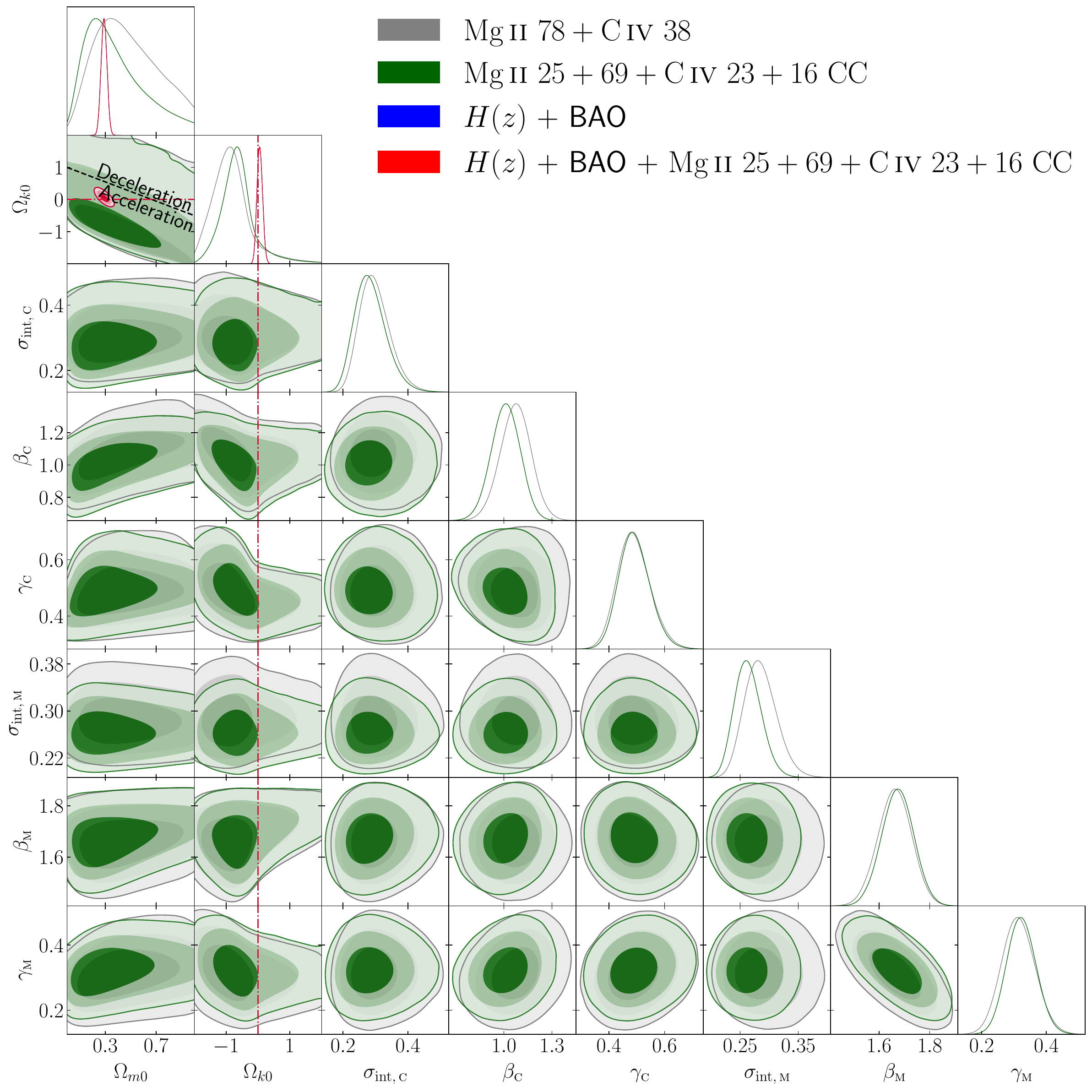}}
 \subfloat[]{%
    \includegraphics[width=0.5\textwidth,height=0.5\textwidth]{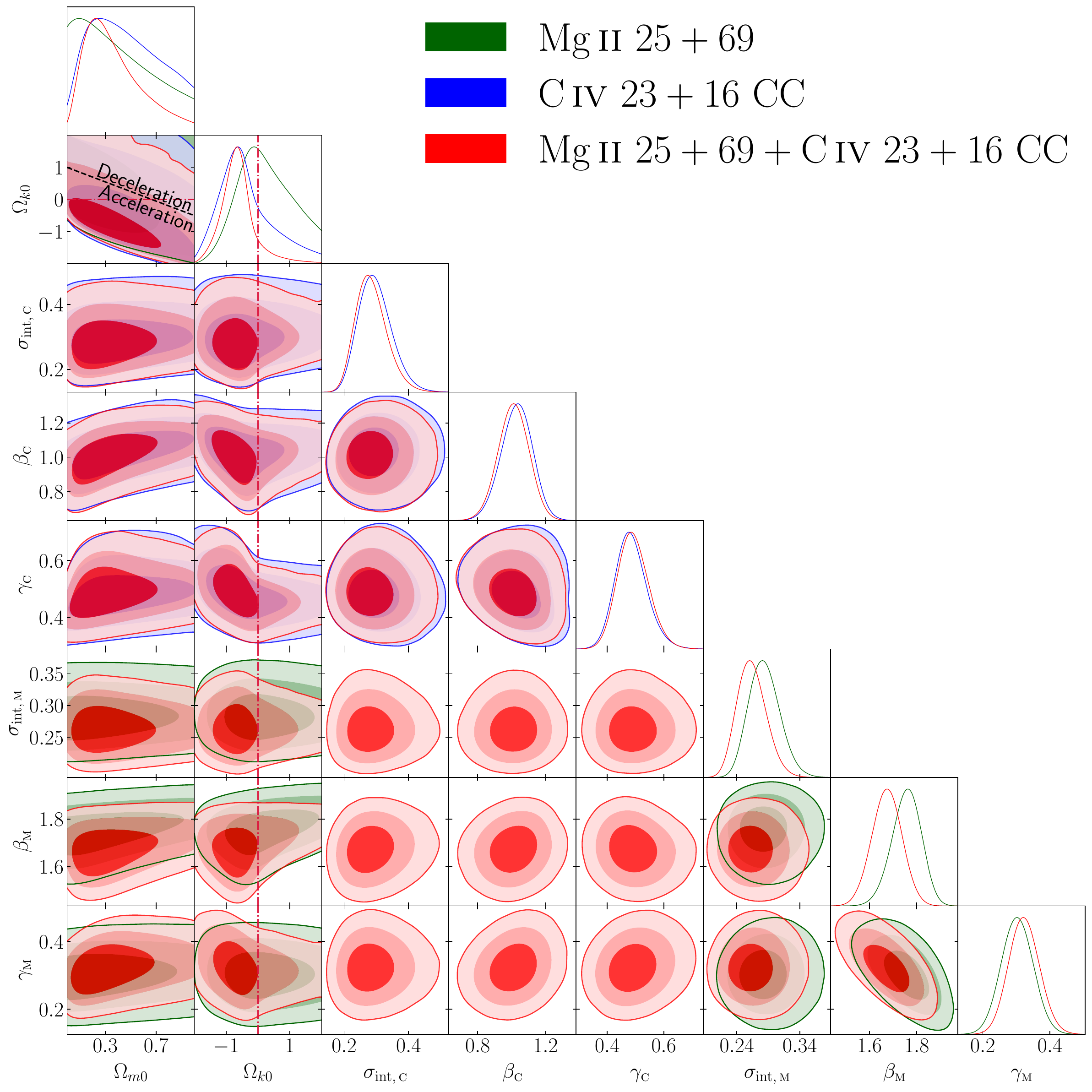}}\\
\caption{Same as Fig.\ \ref{fig01} but for non-flat \lcdm. The black dashed zero-acceleration lines, computed for the third cosmological parameter set to the $H(z)$ + BAO data best-fitting values listed in Table \ref{tab:BFP}, divide the parameter space into regions associated with currently accelerating (below) and currently-decelerating (above) cosmological expansion. The crimson dash-dot lines represent flat hypersurfaces, with closed spatial hypersurfaces either below or to the left.}
\label{fig02}
\end{figure*}

\begin{figure*}
\centering
 \subfloat[]{%
    \includegraphics[width=0.5\textwidth,height=0.5\textwidth]{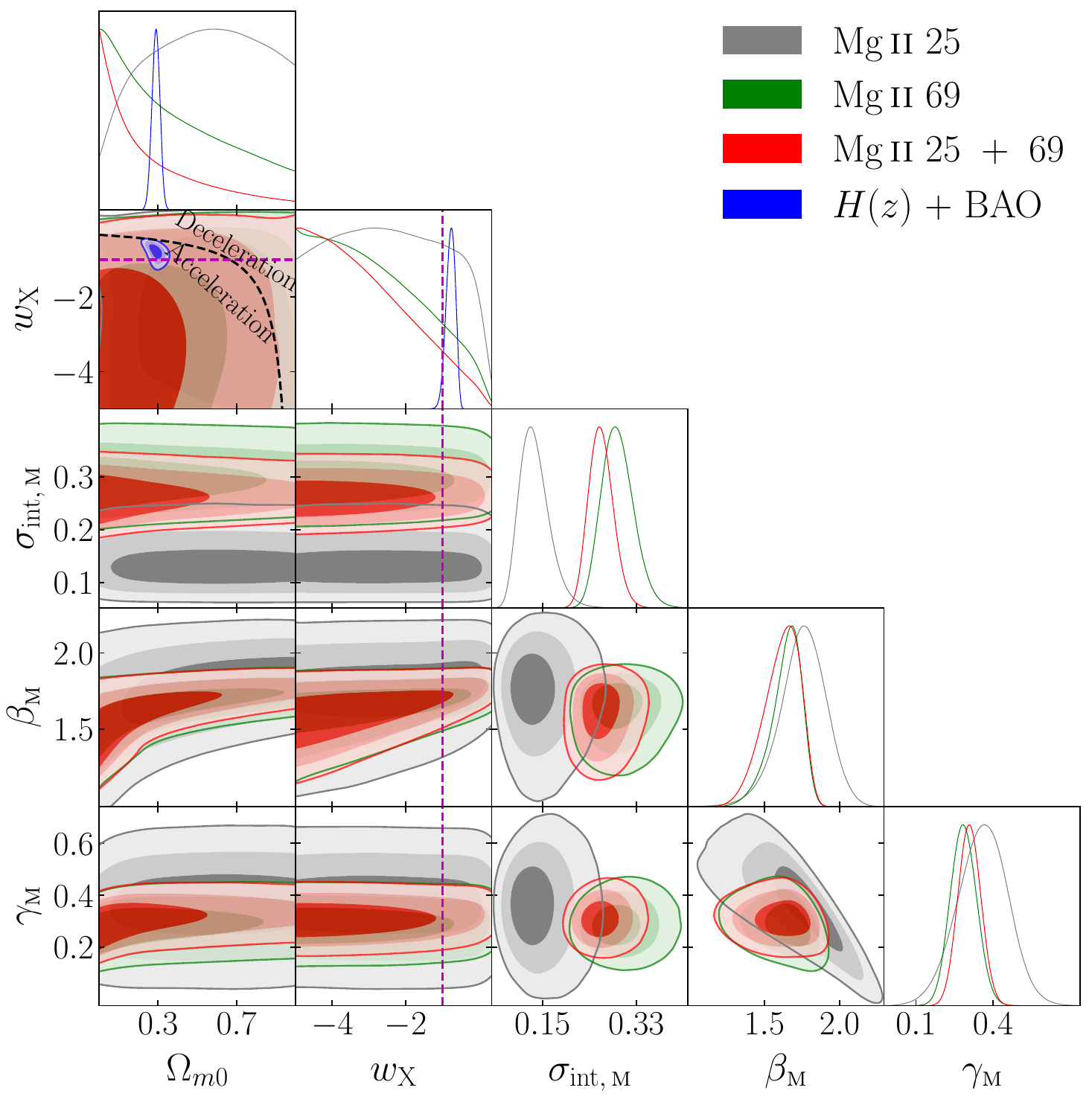}}
 \subfloat[]{%
    \includegraphics[width=0.5\textwidth,height=0.5\textwidth]{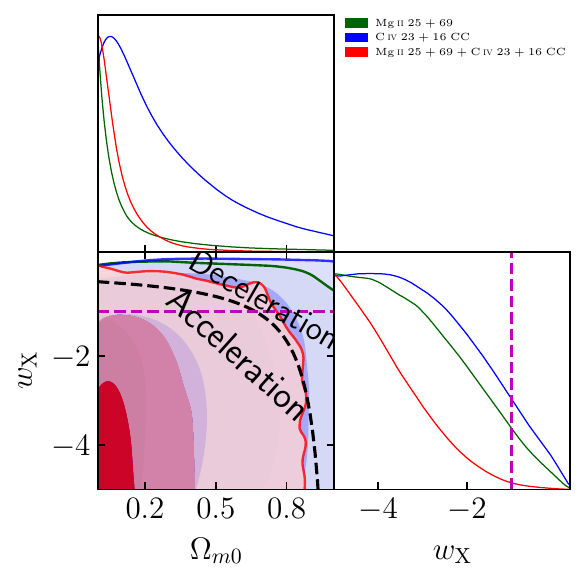}}\\
 \subfloat[]{%
    \includegraphics[width=0.5\textwidth,height=0.5\textwidth]{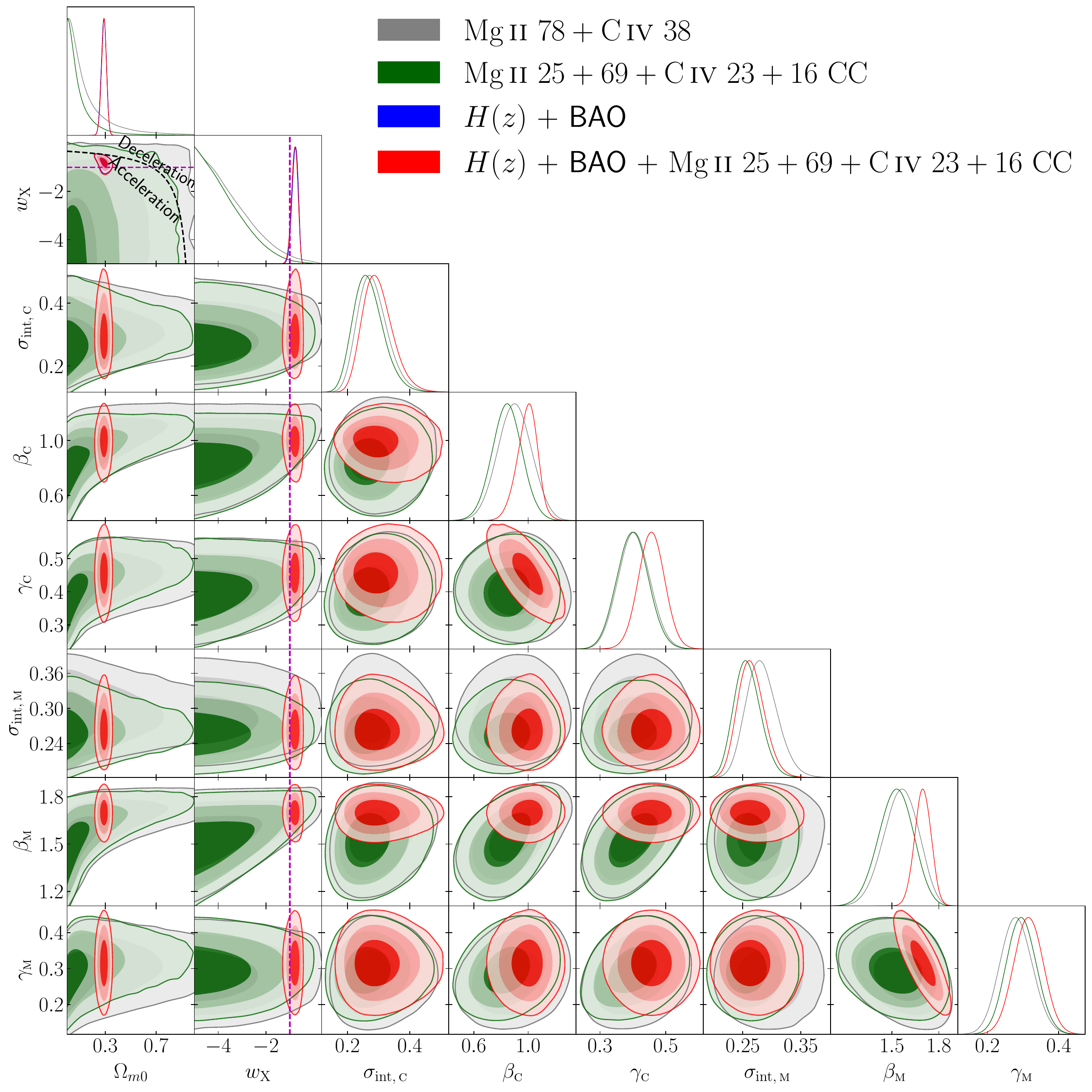}}
 \subfloat[]{%
    \includegraphics[width=0.5\textwidth,height=0.5\textwidth]{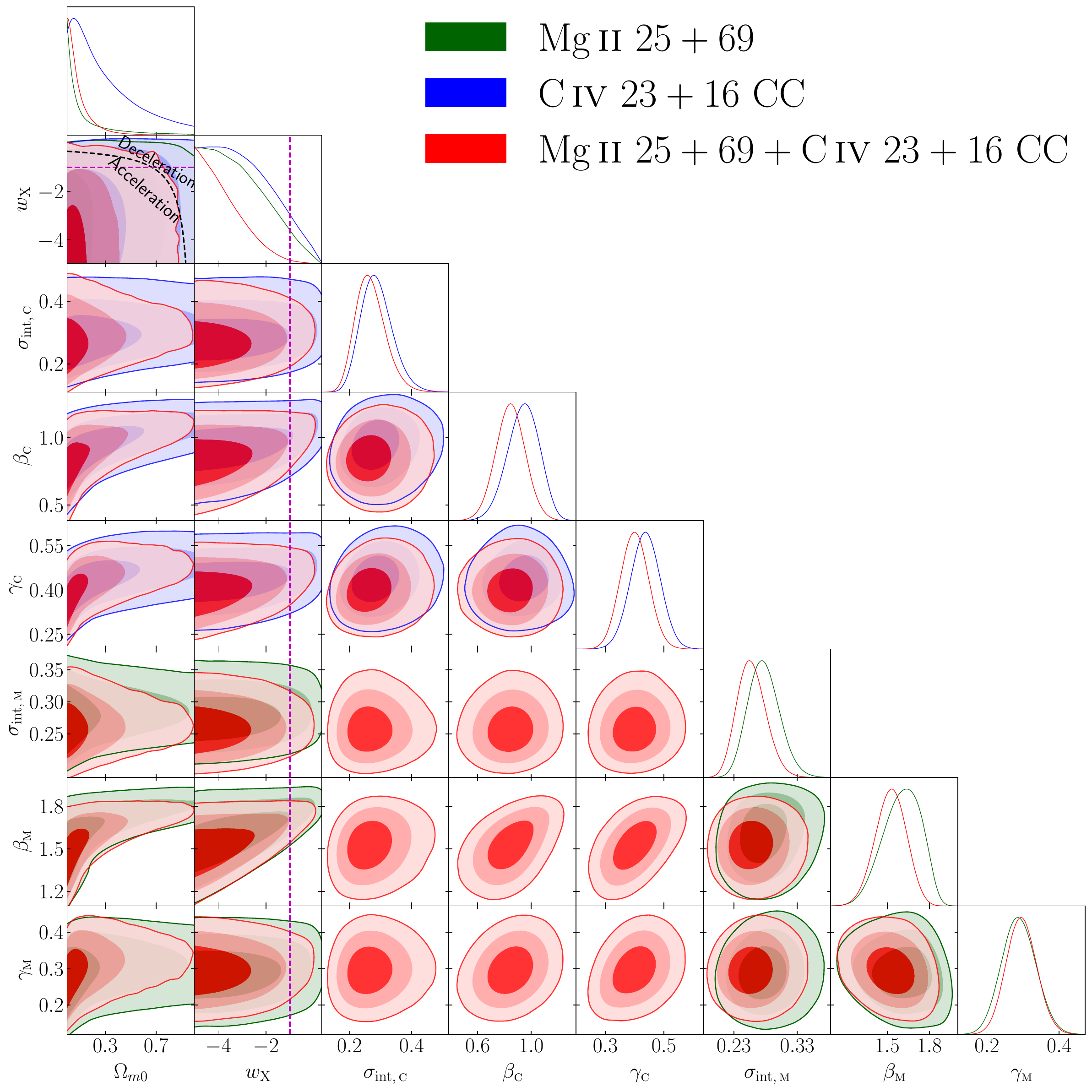}}\\
\caption{One-dimensional likelihood distributions and 1$\sigma$, 2$\sigma$, and 3$\sigma$ two-dimensional likelihood confidence contours for flat XCDM from various combinations of data. The black dashed zero-acceleration lines divide the parameter space into regions associated with currently accelerating (below left) and currently decelerating (above right) cosmological expansion. The magenta lines represent $w_{\rm X}=-1$, i.e.\ flat \lcdm\ model.}
\label{fig03}
\end{figure*}

\begin{figure*}
\centering
 \subfloat[]{%
    \includegraphics[width=0.5\textwidth,height=0.5\textwidth]{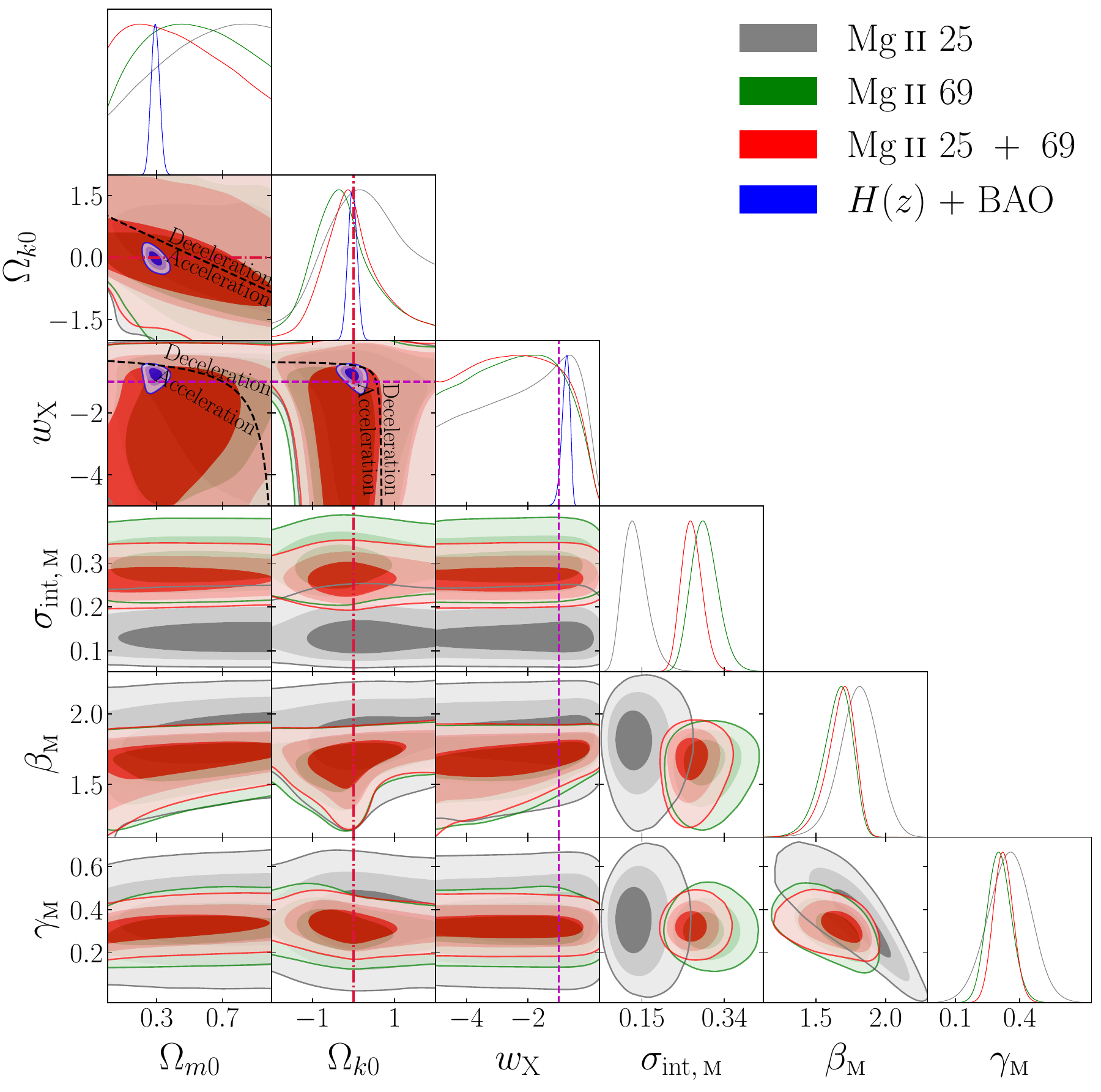}}
 \subfloat[]{%
    \includegraphics[width=0.5\textwidth,height=0.5\textwidth]{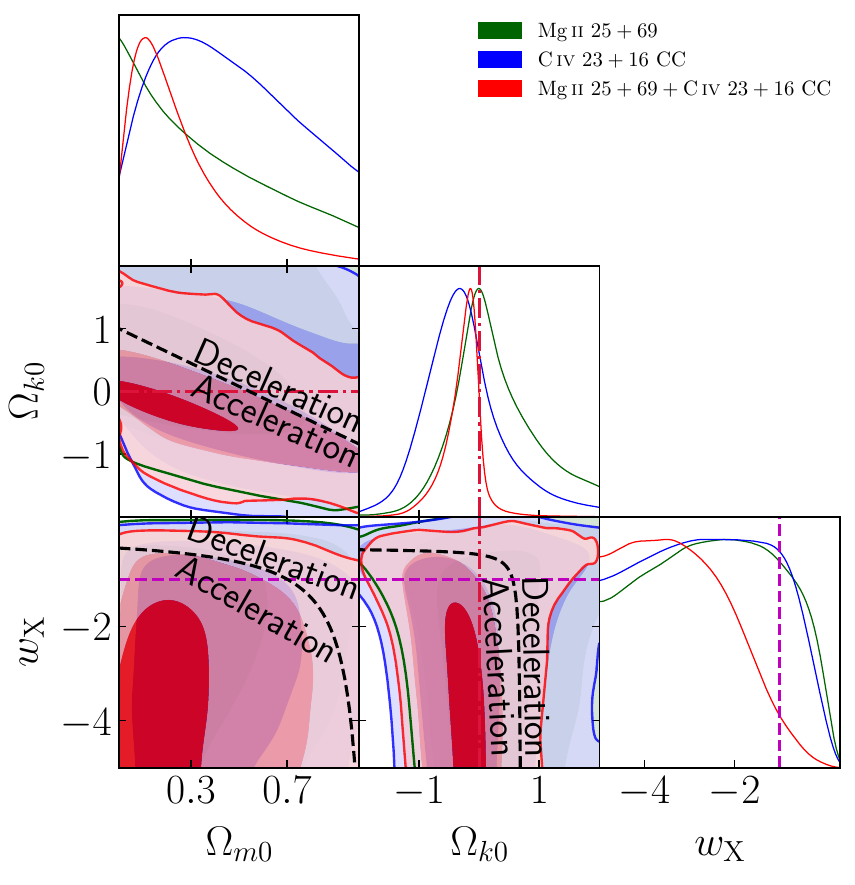}}\\
 \subfloat[]{%
    \includegraphics[width=0.5\textwidth,height=0.5\textwidth]{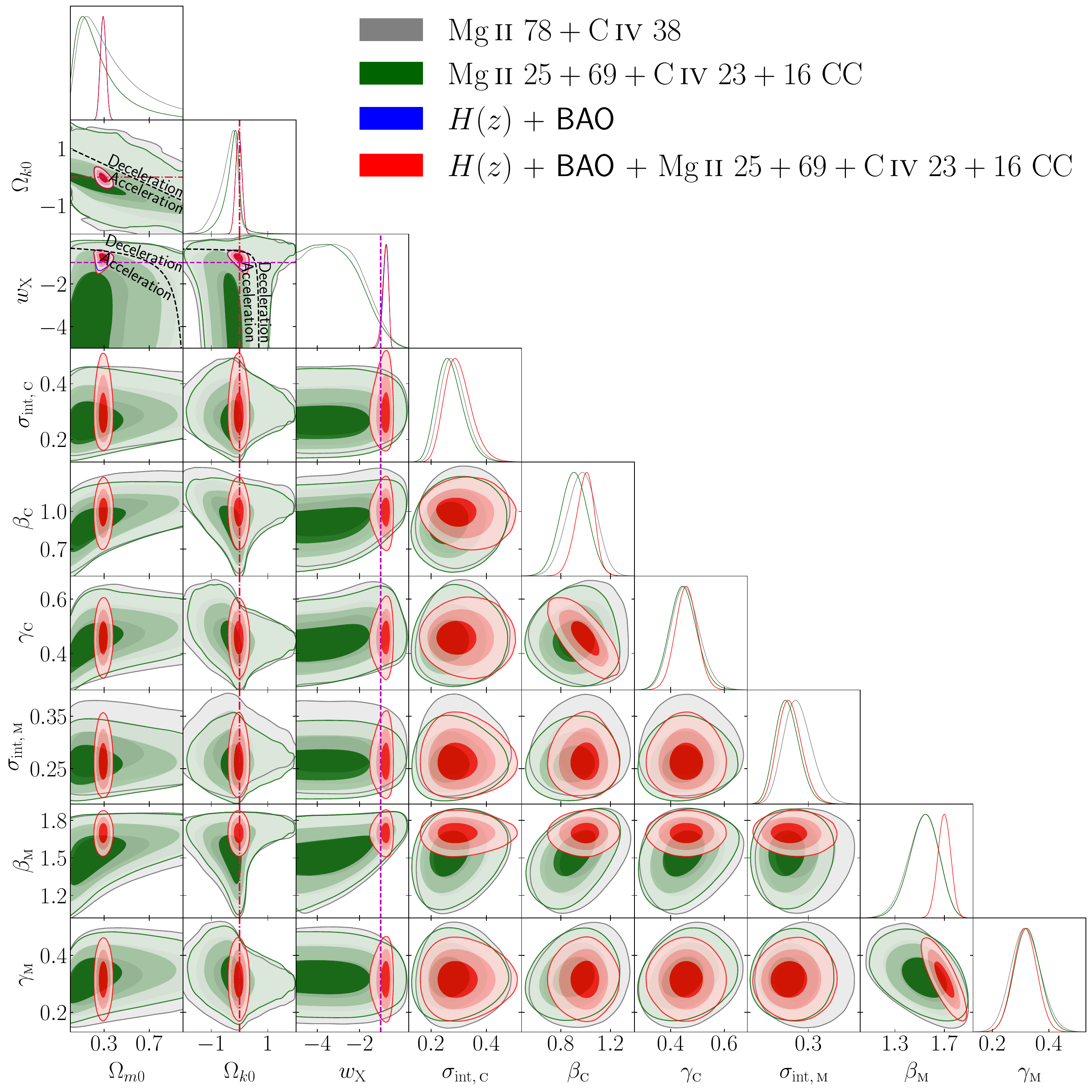}}
 \subfloat[]{%
    \includegraphics[width=0.5\textwidth,height=0.5\textwidth]{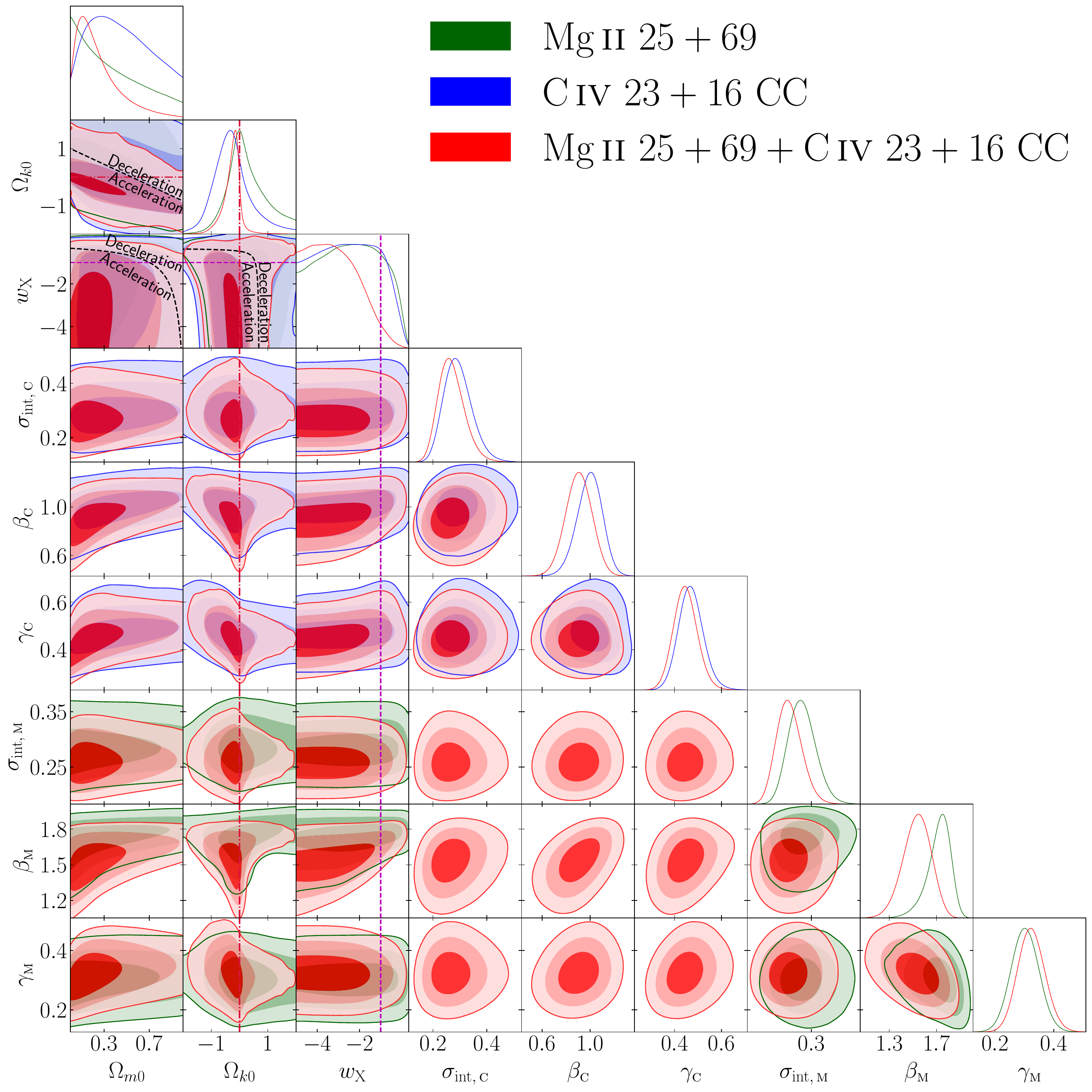}}\\
\caption{Same as Fig.\ \ref{fig03} but for non-flat XCDM. The black dashed zero-acceleration lines, computed for the third cosmological parameter set to the $H(z)$ + BAO data best-fitting values listed in Table \ref{tab:BFP}, divide the parameter space into regions associated with currently accelerating (below or below left) and currently-decelerating (above or above right) cosmological expansion. The crimson dash-dot lines represent flat hypersurfaces, with closed spatial hypersurfaces either below or to the left. The magenta lines represent $w_{\rm X}=-1$, i.e.\ non-flat \lcdm\ model.}
\label{fig04}
\end{figure*}

\begin{figure*}
\centering
 \subfloat[]{%
    \includegraphics[width=0.5\textwidth,height=0.5\textwidth]{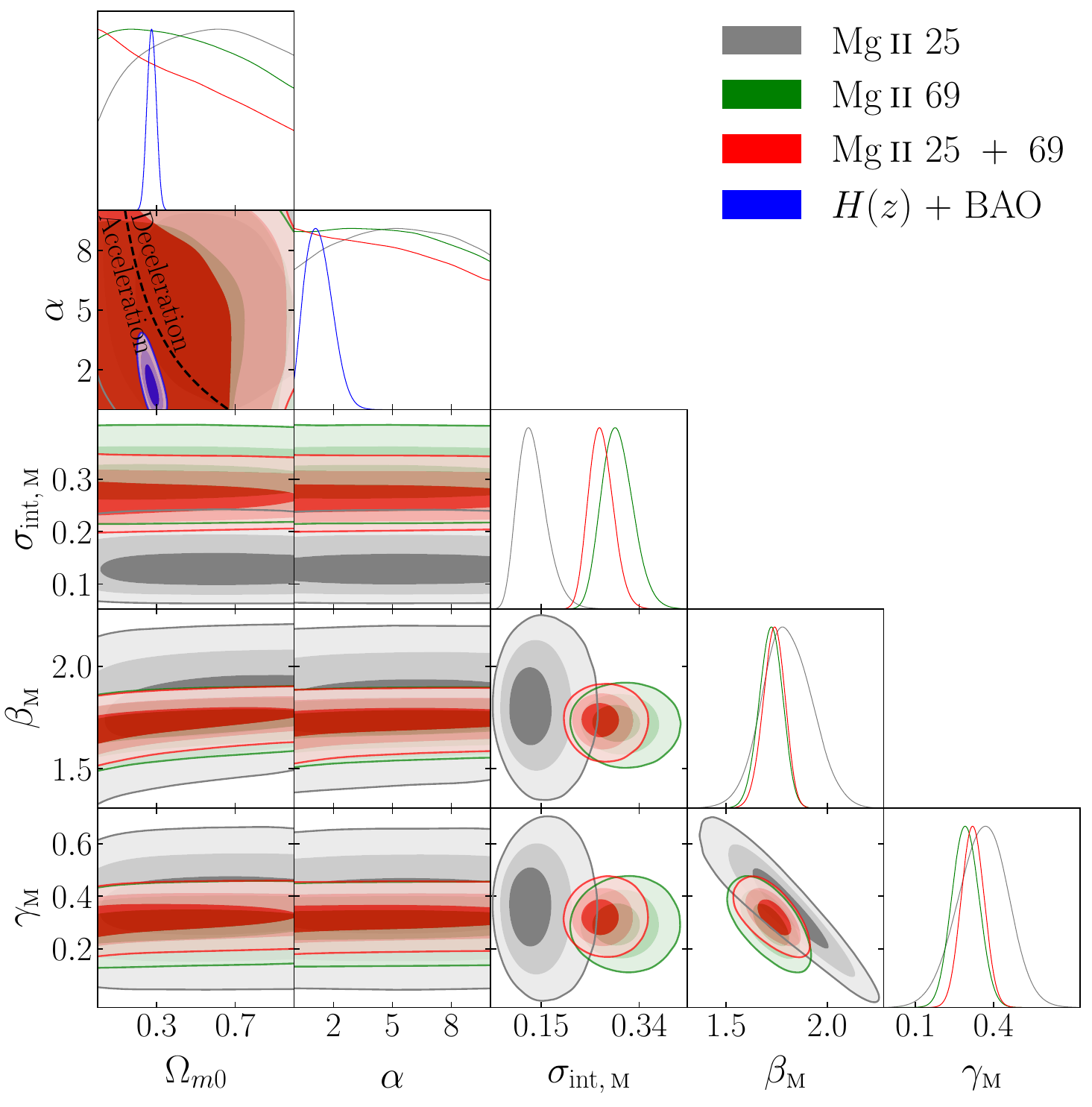}}
 \subfloat[]{%
    \includegraphics[width=0.5\textwidth,height=0.5\textwidth]{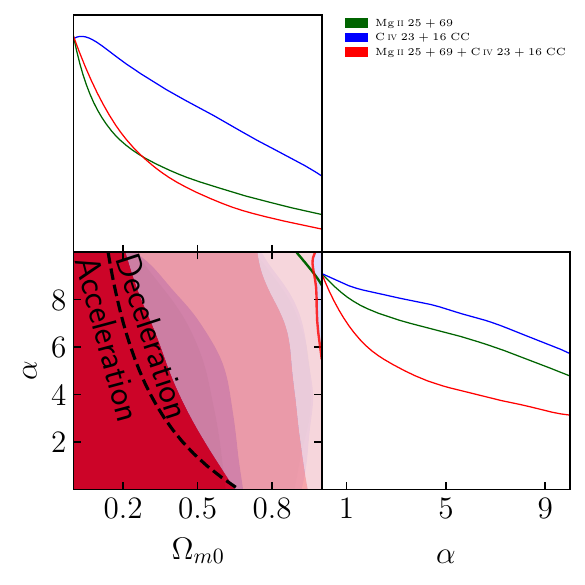}}\\
 \subfloat[]{%
    \includegraphics[width=0.5\textwidth,height=0.5\textwidth]{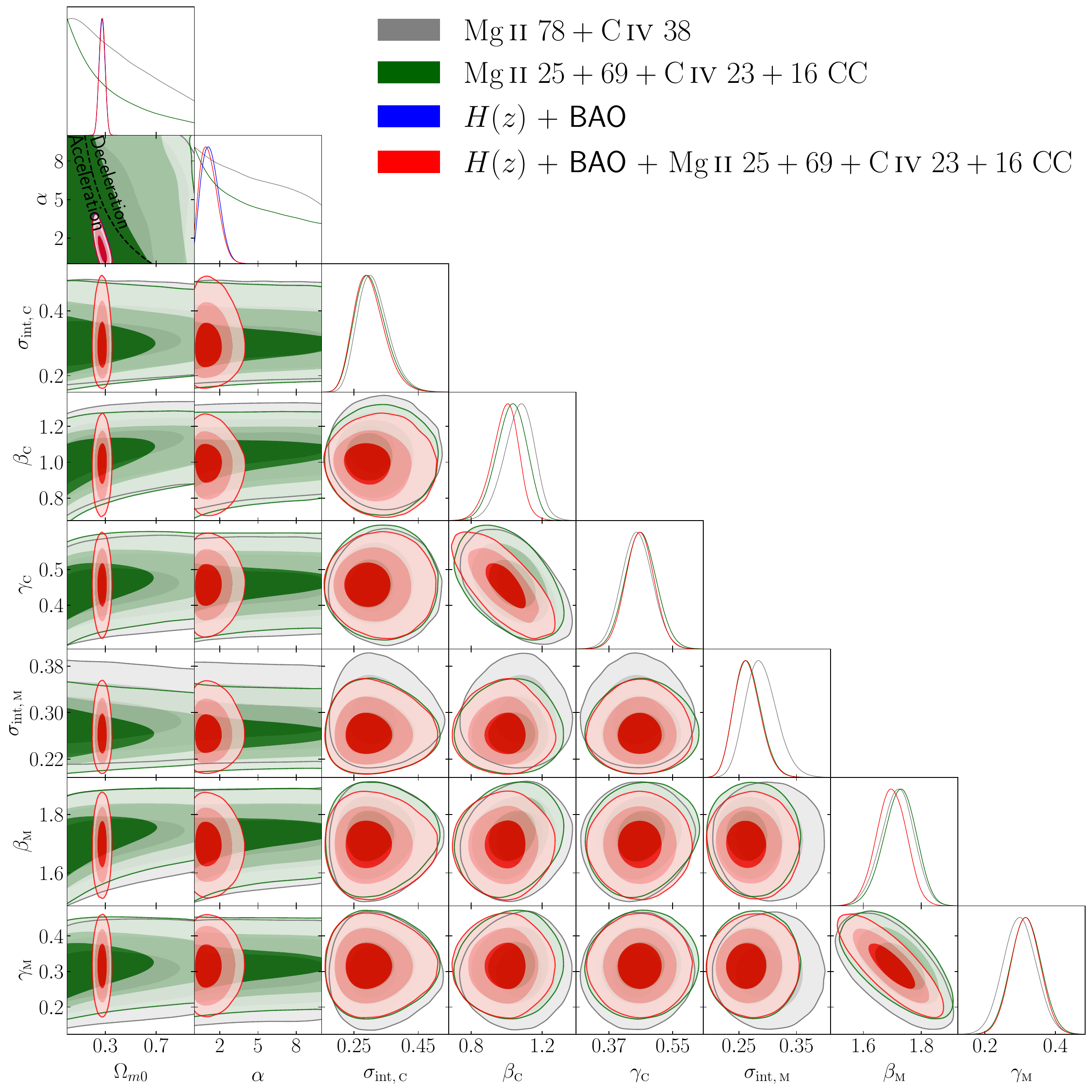}}
 \subfloat[]{%
    \includegraphics[width=0.5\textwidth,height=0.5\textwidth]{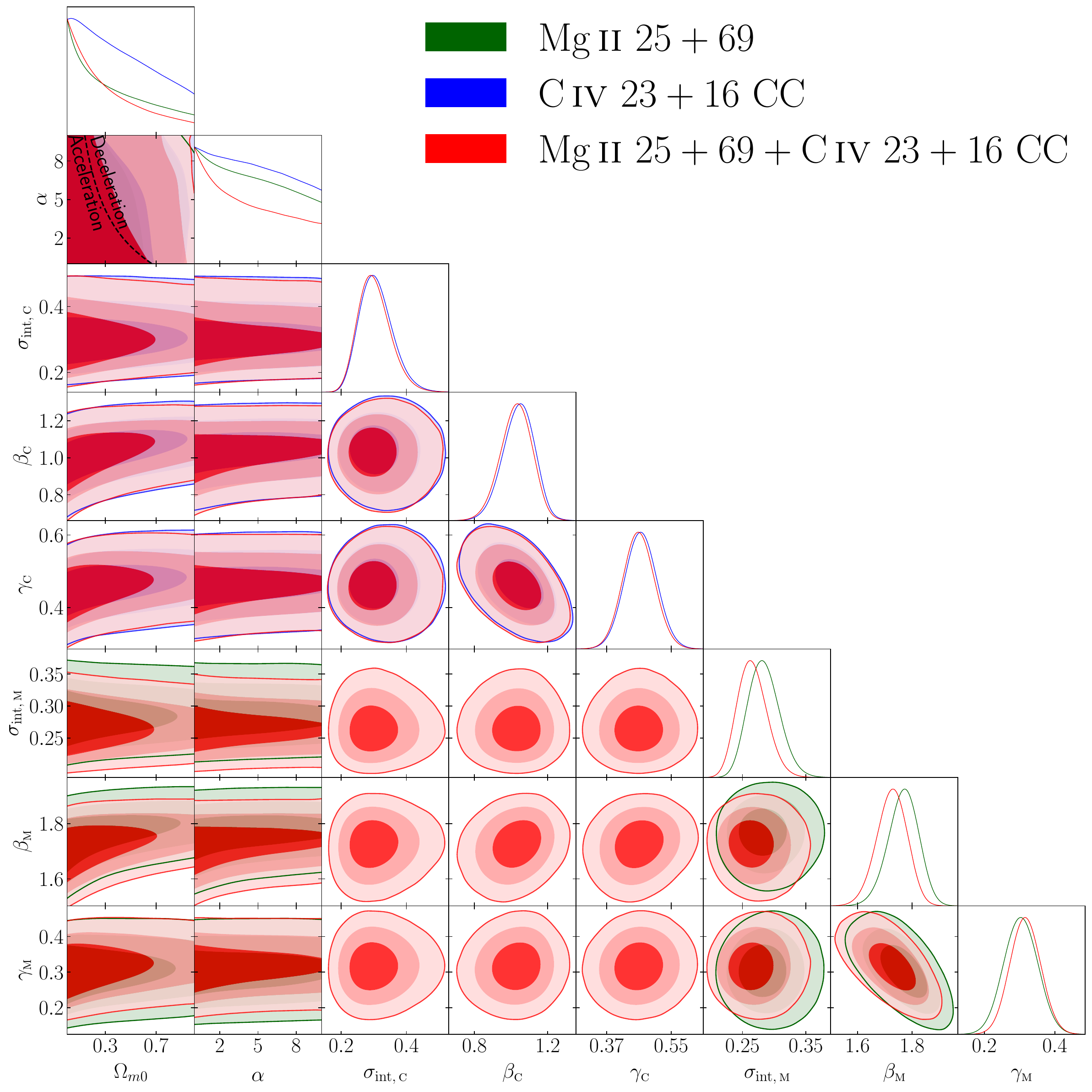}}\\
\caption{One-dimensional likelihood distributions and 1$\sigma$, 2$\sigma$, and 3$\sigma$ two-dimensional likelihood confidence contours for flat \pcdm\ from various combinations of data. The black dashed zero-acceleration lines divide the parameter space into regions associated with currently accelerating (below left) and currently decelerating (above right) cosmological expansion. The $\alpha = 0$ axes correspond to the flat \lcdm\ model.}
\label{fig05}
\end{figure*}

\begin{figure*}
\centering
 \subfloat[]{%
    \includegraphics[width=0.5\textwidth,height=0.5\textwidth]{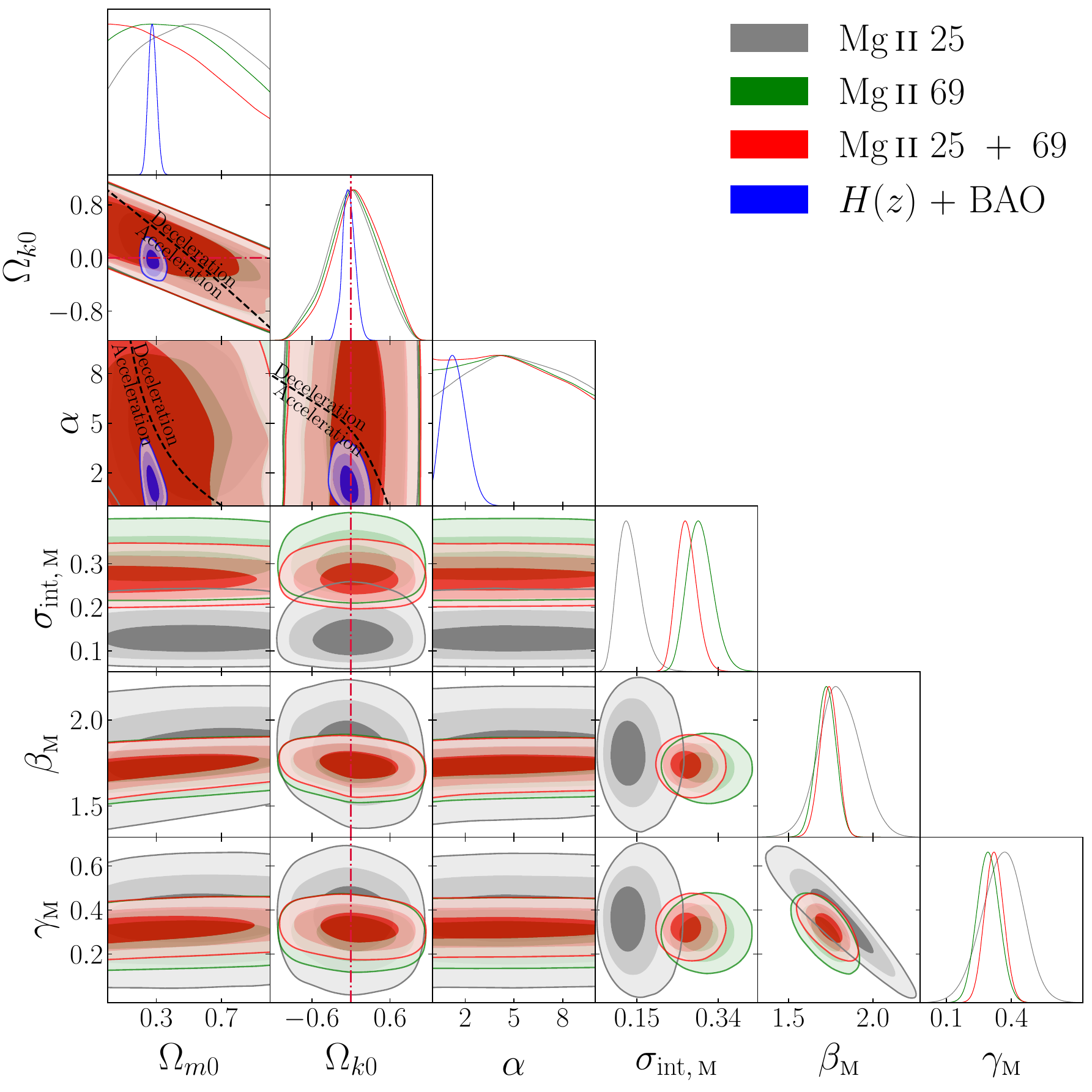}}
 \subfloat[]{%
    \includegraphics[width=0.5\textwidth,height=0.5\textwidth]{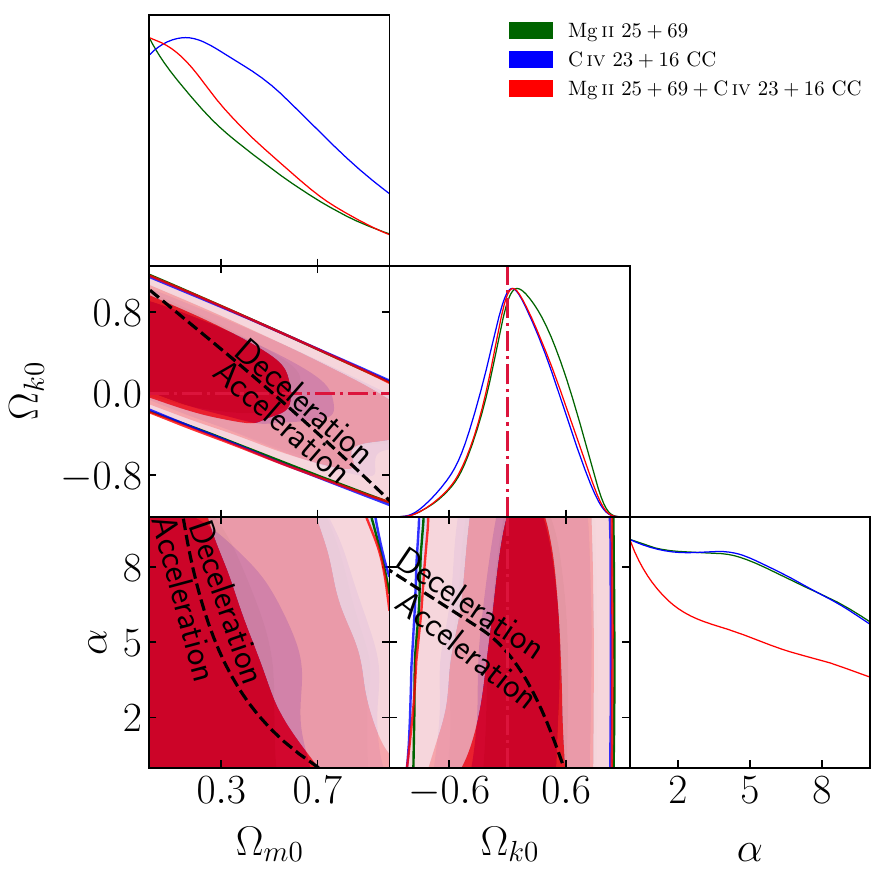}}\\
 \subfloat[]{%
    \includegraphics[width=0.5\textwidth,height=0.5\textwidth]{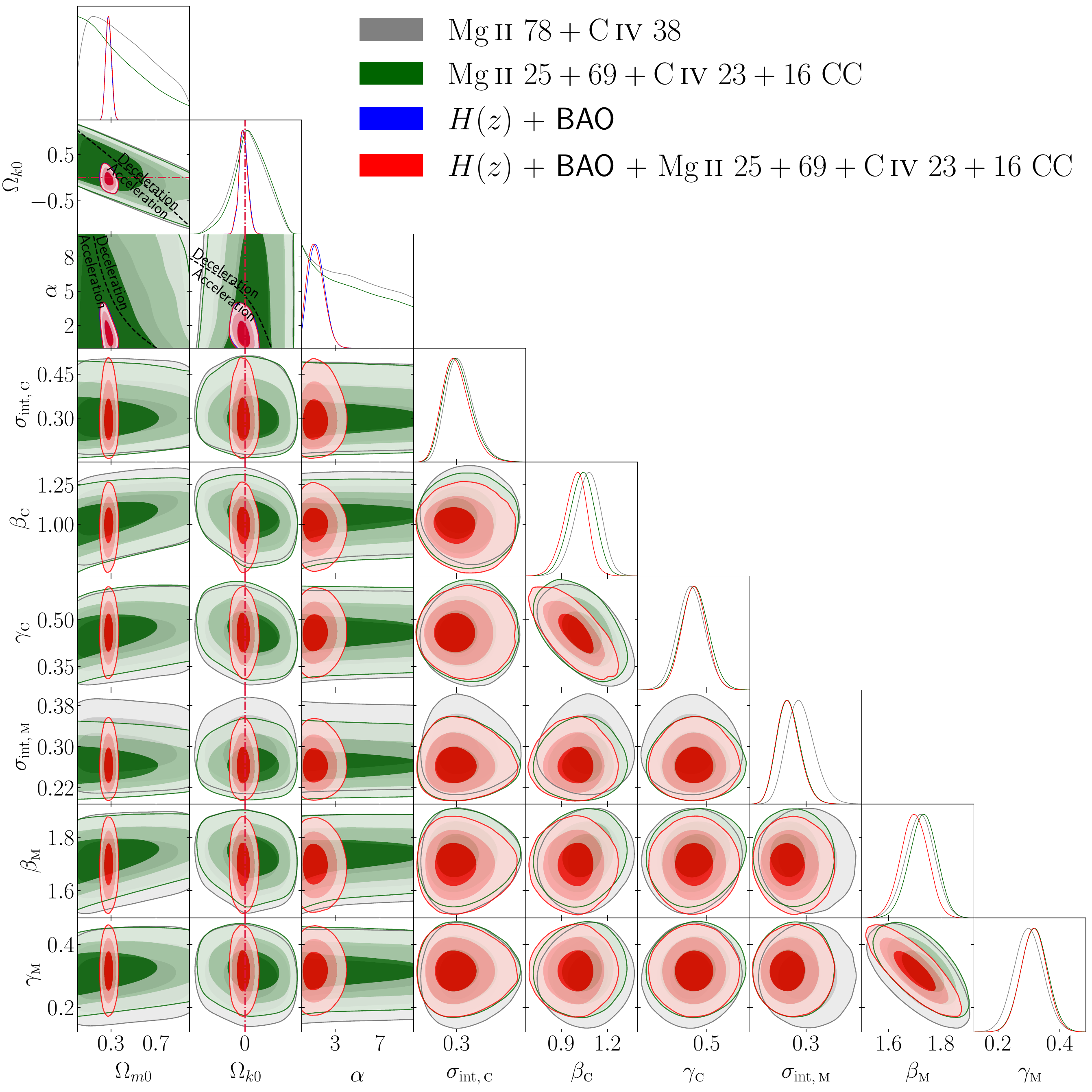}}
 \subfloat[]{%
    \includegraphics[width=0.5\textwidth,height=0.5\textwidth]{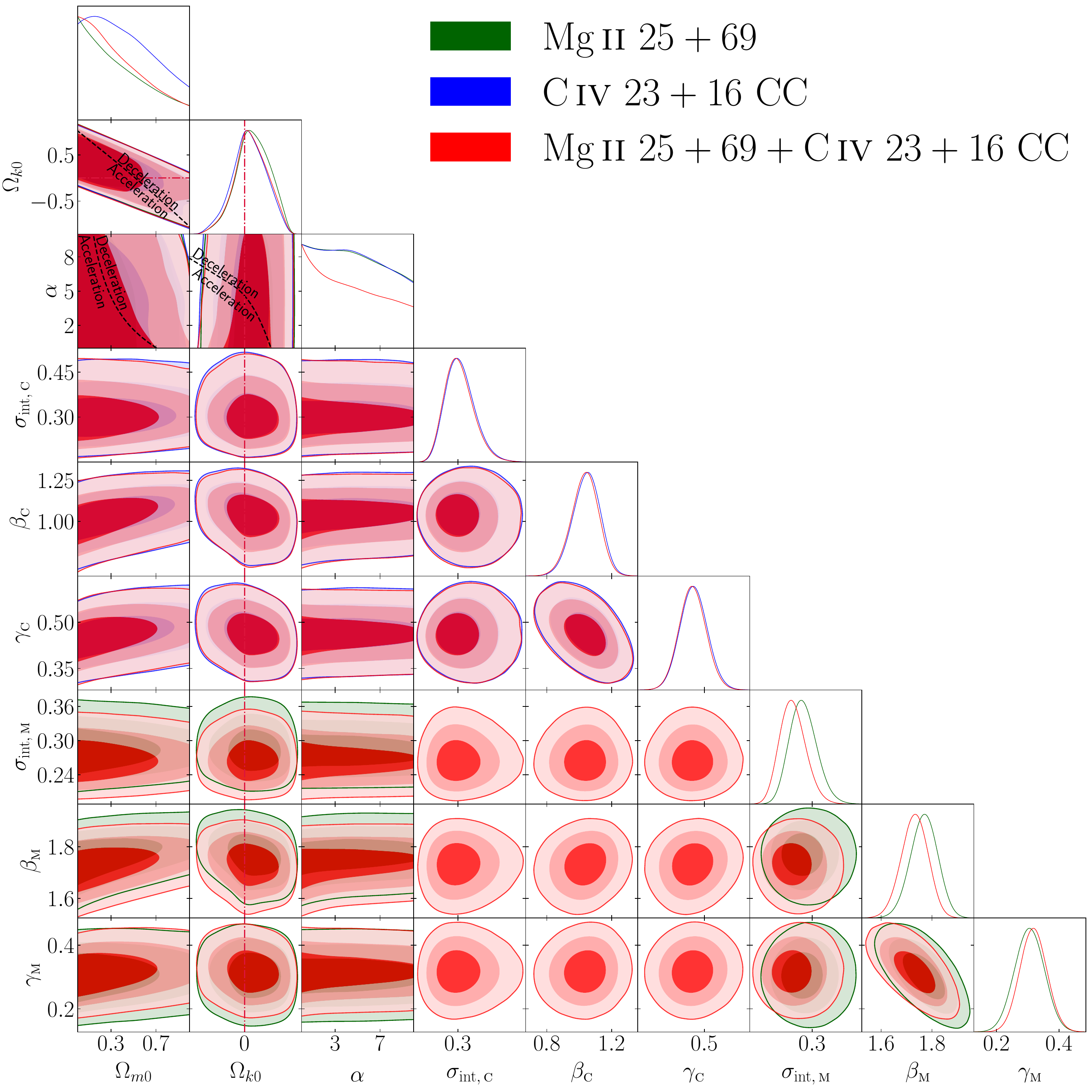}}\\
\caption{Same as Fig.\ \ref{fig05} but for non-flat \pcdm. The black dashed zero-acceleration lines, computed for the third cosmological parameter set to the $H(z)$ + BAO data best-fitting values listed in Table \ref{tab:BFP}, divide the parameter space into regions associated with currently accelerating (below or below left) and currently-decelerating (above or above right) cosmological expansion. The crimson dash-dot lines represent flat hypersurfaces, with closed spatial hypersurfaces either below or to the left. The $\alpha = 0$ axes correspond to the non-flat \lcdm\ model.}
\label{fig06}
\end{figure*}

The posterior one-dimensional probability distributions and two-dimensional confidence regions for the parameters of six cosmological models and the $R-L$ relation are depicted in Figs.\ \ref{fig01}--\ref{fig06}. Note that the asymmetric errors of the \civ\ and \mii\ measurements are accounted for in our analyses. In panel (a) of each figure results of the \mii\ 25, \mii\ 69, \mii\ 25 + 69, and $H(z)$ + BAO data analyses are shown in grey, green, red, and blue, respectively; in panels (b) and (d) of each figure results of the \mii\ 25 + 69, \civ\ 23 + 16 CC, and joint \mii\ 25 + 69 + \civ\ 23 + 16 CC data analyses are shown in green, blue, and red, respectively; in panel (c) of each figure the results of the \mii\ 78 +  \civ\ 38 \citep[results from][]{Caoetal_2022}, \mii\ 25 + 69 + \civ\ 23 + 16 CC, $H(z)$ + BAO, and joint $H(z)$ + BAO + \mii\ 25 + 69 + \civ\ 23 + 16 CC data analyses are shown in grey, green, blue, and red, respectively. Unmarginalized best-fitting parameter values, as well as maximum likelihood values $\mathcal{L}_{\rm max}$, AIC, BIC, DIC, $\Delta \mathrm{AIC}$, $\Delta \mathrm{BIC}$, and $\Delta \mathrm{DIC}$ for all models and data sets are tabulated in Table \ref{tab:BFP}. The marginalized posterior mean parameters and their uncertainties ($\pm 1\sigma$ error bars and 1 or 2$\sigma$ limits) for all models and data sets are presented in Table \ref{tab:1d_BFP}. In Table~\ref{tab:diff} we show the largest differences for $R-L$ correlation parameters and intrinsic scatter parameters between the cosmological models under study. We note that \civ\ 23 + 16 CC data are considered more reliable; thus only their posterior distributions are shown in the figures, although other \civ\ data set analysis results are included in the tables.

In the six cosmological models, most of the data combinations more favour currently accelerating cosmological expansion, except for the cases of non-flat XCDM (\mii\ 25), flat \pcdm\ (\mii\ 25), \om---$\alpha$ and \ok---$\alpha$ panels of non-flat \pcdm\ (\mii\ 25, \mii\ 69, and \mii\ 25 + 69), \ok---$\alpha$ panels of non-flat \pcdm\ (\civ\ 23 + 16 CC and \mii\ 25 + 69 + \civ\ 23 + 16 CC).

\subsection{Constraints from \mq\ data}
\label{subsec:MQ}

As demonstrated in panel (a) of Figs.\ \ref{fig01}--\ref{fig06} and Table \ref{tab:1d_BFP}, the results for both cosmological and $R-L$ relation parameters derived from \mii\ 25 and \mii\ 69 data are in mutual agreement. Consequently, we perform a joint analysis of \mii\ 25 + 69 data. While the \mii\ 25 + 69 data constraints on cosmological parameters are weak, they align well with those derived from $H(z)$ + BAO data.

The majority of \om\ constraints based on individual \mii\ 25 and 69 data sets are notably weak. For instance, \om\ constraints are absent in flat \lcdm\ and flat \pcdm\ for \mii\ 25 and 69 data, respectively. Nonetheless, the joint \mii\ 25 + 69 data analysis yields relatively stronger \om\ constraints, varying from a low of $<0.814$ (2$\sigma$, flat XCDM) to a high of $<0.537$ (1$\sigma$, flat \lcdm).

For the \ok\ parameter, constraints derived from \mii\ 25 and \mii\ 69 data are $0.244^{+1.105}_{-1.076}$ and $-0.386^{+0.591}_{-1.457}$ for non-flat \lcdm, $0.264^{+1.021}_{-0.950}$ and $-0.207^{+0.544}_{-0.919}$ for non-flat XCDM, and $0.018^{+0.390}_{-0.379}$ and $0.056^{+0.394}_{-0.380}$ for non-flat \pcdm, respectively. Although these constraints are mutually consistent and align with flat hypersurfaces within a $1\sigma$ limit, they exhibit divergent preferences for spatial hypersurfaces in non-flat \lcdm\ and non-flat XCDM, while converging on the same preference in non-flat \pcdm. In contrast, joint \mii\ 25 + 69 data yield \ok\ constraints of $-0.057^{+0.776}_{-1.169}$, $-0.009^{+0.513}_{-0.830}$, and $0.099^{+0.391}_{-0.372}$ for non-flat \lcdm, XCDM, and \pcdm, respectively, where spatial hypersurfaces preferences are consistent with \mii\ 69 data, consistently, but mildly closer to flat, favouring closed hypersurfaces for non-flat \lcdm\ and XCDM and open hypersurfaces for non-flat \pcdm.

Constraints on dark energy dynamics, represented by the \wx\ and $\alpha$ parameters, are weak. Specifically, the \mii\ 25, 69, and 25 + 69 data sets provide \wx\ constraints of $-2.386^{+2.276}_{-1.263}$ ($1\sigma$), $<-0.439$ ($2\sigma$), and $<-0.670$ ($2\sigma$) for flat XCDM, respectively, while for non-flat XCDM the respective constraints are $-2.175^{+2.212}_{-0.925}$ ($1\sigma$), $-2.492^{+1.749}_{-1.390}$ ($1\sigma$), and $-2.521^{+1.601}_{-1.619}$ ($1\sigma$). Notably, constraints of $\alpha$ are absent for and non-flat \pcdm\ for all three data combinations.

As indicated in Table \ref{tab:diff}, the maximal discrepancies in the slope ($\gamma$), intercept ($\beta$), and intrinsic scatter ($\sigma_{\rm int}$) parameters of the \mii\ 25 + 69 data $R-L$ relations among different cosmological models do not exceed $0.85\sigma$ (corresponding to $\beta$), however, $\beta$ differences increase the most compared with individual data. This indicates support for a cosmological-model-independent $R-L$ relation for these \mii\ data.

\subsection{Constraints from \cq\ data in conjunction with \mq\ and $H(z)$ + BAO data}
\label{subsec:CQ}

Panels (b), (c), and (d) of Figs.\ \ref{fig01}--\ref{fig06} and Table \ref{tab:1d_BFP} highlight the congruence in results for both cosmological and $R-L$ relation parameters derived from \mii\ 25 + 69, \civ\ 23 + 16 CC, and $H(z)$ + BAO data sets. This consistency prompts joint analyses of \mii\ 25 + 69 + \civ\ 23 + 16 CC and $H(z)$ + BAO + \mii\ 25 + 69 + \civ\ 23 + 16 CC data. Other combinations of \civ\ data, such as \civ\ 23 + 16 CR, \civ\ 23 + 41 CC, and \civ\ 23 + 41 CR, while standardizable individually, showed discrepancies when analyzed jointly with \mii\ 25 + 69 data, as evidenced by the tensions ($\Delta\beta_{\rm M}>2\sigma$) in Table \ref{tab:diff}. Notably \mii\ 25 + 69 + \civ\ 23 + 41 CC data yielded tighter cosmological model parameter constraints compared to \mii\ 25 + 69 + \civ\ 23 + 16 CC data. In this section, we present constraints from the standardizable \mii\ 25 + 69 + \civ\ 23 + 16 CC data and the combined $H(z)$ + BAO + \mii\ 25 + 69 + \civ\ 23 + 16 CC data. Additional data constraints are listed in Table \ref{tab:1d_BFP}. In contrast to the previously studied \mii\ 78 + \civ\ 38 data constraints \citep{Caoetal_2022}, \mii\ 25 + 69 + \civ\ 23 + 16 CC data offer slightly more precise and consistent constraints on both cosmological-model and $R-L$ relation parameters.

\begin{sidewaystable*}
\centering
\resizebox*{\columnwidth}{0.74\columnwidth}{%
\begin{threeparttable}
\caption{Unmarginalized best-fitting parameter values for all models from various combinations of data.}\label{tab:BFP}
\begin{tabular}{lcccccccccccccccccccc}
\toprule
Model & Data set & $\Omega_{b}h^2$ & $\Omega_{c}h^2$ & $\Omega_{m0}$ & $\Omega_{k0}$ & $w_{\mathrm{X}}$/$\alpha$\tnote{a} & $H_0$\tnote{b} & $\sigma_{\mathrm{int,\,\textsc{c}}}$ & $\gamma_{\rm\textsc{c}}$ & $\beta_{\rm\textsc{c}}$ & $\sigma_{\mathrm{int,\,\textsc{m}}}$ & $\gamma_{\rm\textsc{m}}$ & $\beta_{\rm\textsc{m}}$ & $-2\ln\mathcal{L}_{\mathrm{max}}$ & AIC & BIC & DIC & $\Delta \mathrm{AIC}$ & $\Delta \mathrm{BIC}$ & $\Delta \mathrm{DIC}$ \\
\midrule
 & $H(z)$ + BAO & 0.0244 & 0.1181 & 0.301 & -- & -- & 68.98 & -- & -- & -- & -- & -- & -- & 25.64 & 31.64 & 36.99 & 32.32 & 0.00 & 0.00 & 0.00\\
 & \mii\ 25 & -- & 0.4337 & 0.936 & -- & -- & -- & -- & -- & -- & 0.116 & 0.382 & 1.806 & $-28.42$ & $-20.42$ & $-8.17$ & $-19.95$ & 0.00 & 0.00 & 0.00\\
 & \mii\ 69 & -- & 0.0157 & 0.083 & -- & -- & -- & -- & -- & -- & 0.285 & 0.273 & 1.666 & 28.94 & 36.94 & 49.19 & 37.05 & 0.00 & 0.00 & 0.00\\
 & \mii\ 25 + 69 & -- & $-0.0148$ & 0.021 & -- & -- & -- & -- & -- & -- & 0.255 & 0.290 & 1.666 & 21.62 & 29.62 & 41.87 & 30.75 & 0.00 & 0.00 & 0.00\\
 & \civ\ 23 + 16 CC & -- & $-0.0049$ & 0.041 & -- & -- & -- & 0.261 & 0.418 & 0.962 & -- & -- & -- & 22.34 & 30.34 & 36.99 & 34.04 & 0.00 & 0.00 & 0.00\\
 & \mii\ 25 + 69 + \civ\ 23 + 16 CC & -- & $-0.0028$ & 0.046 & -- & -- & -- & 0.250 & 0.419 & 0.963 & 0.255 & 0.287 & 1.674 & 44.11 & 58.11 & 79.55 & 62.22 & 0.00 & 0.00 & 0.00\\
 & $H(z)$ + BAO + \mii\ 25 + 69 + \civ\ 23 + 16 CC & 0.0243 & 0.1161 & 0.299 & -- & -- & 68.73 & 0.274 & 0.446 & 1.018 & 0.259 & 0.320 & 1.697 & 72.37 & 90.37 & 118.96 & 91.32 & 0.00 & 0.00 & 0.00\\
Flat \lcdm & \civ\ 23 + 16 CR & -- & $-0.0009$ & 0.050 & -- & -- & -- & 0.281 & 0.419 & 0.966 & -- & -- & -- & 23.61 & 31.61 & 38.26 & 34.13 & 0.00 & 0.00 & 0.00\\
 & \mii\ 25 + 69 + \civ\ 23 + 16 CR & -- & 0.0046 & 0.061 & -- & -- & -- & 0.288 & 0.419 & 0.972 & 0.257 & 0.294 & 1.674 & 45.32 & 59.32 & 80.76 & 62.56 & 0.00 & 0.00 & 0.00\\
 & \civ\ 25 CC & -- & $-0.0250$ & 0.000 & -- & -- & -- & 0.291 & 0.203 & 1.543 & -- & -- & -- & 21.06 & 29.06 & 33.94 & 28.42 & 0.00 & 0.00 & 0.00\\
 & \civ\ 25 CR & -- & 0.4517 & 0.973 & -- & -- & -- & 0.160 & 0.453 & 1.384 & -- & -- & -- & $-8.16$ & $-0.16$ & 4.72 & 0.63 & 0.00 & 0.00 & 0.00\\
 & \civ\ 23 + 41 CC & -- & $-0.0162$ & 0.018 & -- & -- & -- & 0.312 & 0.411 & 0.950 & -- & -- & -- & 52.97 & 60.97 & 69.61 & 65.41 & 0.00 & 0.00 & 0.00\\
 & \mii\ 25 + 69 + \civ\ 23 + 41 CC & -- & $-0.0151$ & 0.020 & -- & -- & -- & 0.317 & 0.412 & 0.950 & 0.254 & 0.292 & 1.667 & 74.73 & 88.73 & 110.17 & 93.18 & 0.00 & 0.00 & 0.00\\
 & \civ\ 23 + 41 CR & -- & $-0.0142$ & 0.022 & -- & -- & -- & 0.241 & 0.414 & 1.019 & -- & -- & -- & 28.82 & 36.82 & 45.46 & 40.81 & 0.00 & 0.00 & 0.00\\
 & \mii\ 25 + 69 + \civ\ 23 + 41 CR & -- & $-0.0127$ & 0.025 & -- & -- & -- & 0.235 & 0.416 & 1.019 & 0.255 & 0.300 & 1.643 & 50.64 & 64.64 & 86.08 & 68.33 & 0.00 & 0.00 & 0.00\\
\midrule
 & $H(z)$ + BAO & 0.0260 & 0.1098 & 0.292 & 0.048 & -- & 68.35 & -- & -- & -- & -- & -- & -- & 25.30 & 33.30 & 40.43 & 33.87 & 1.66 & 3.44 & 1.54\\
 & \mii\ 25 & -- & 0.4484 & 0.966 & $-1.056$ & -- & -- & -- & -- & -- & 0.114 & 0.418 & 1.747 & $-28.44$ & $-18.44$ & $-3.12$ & $-20.06$ & 1.98 & 5.05 & $-0.11$\\
 & \mii\ 69 & -- & 0.1638 & 0.386 & $-1.128$ & -- & -- & -- & -- & -- & 0.275 & 0.358 & 1.617 & 24.13 & 34.13 & 49.44 & 39.87 & $-2.81$ & 0.25 & 2.82\\
 & \mii\ 25 + 69 & -- & 0.1085 & 0.273 & $-0.772$ & -- & -- & -- & -- & -- & 0.254 & 0.332 & 1.637 & 20.66 & 30.66 & 45.98 & 32.06 & 1.04 & 4.10 & 1.31\\
 & \civ\ 23 + 16 CC & -- & 0.0185 & 0.089 & $-0.374$ & -- & -- & 0.242 & 0.500 & 0.880 & -- & -- & -- & 14.85 & 24.85 & 33.17 & 38.98 & $-5.48$ & $-3.82$ & 4.95\\
 & \mii\ 25 + 69 + \civ\ 23 + 16 CC & -- & 0.0360 & 0.125 & $-0.465$ & -- & -- & 0.239 & 0.499 & 0.928 & 0.254 & 0.293 & 1.659 & 35.81 & 51.81 & 76.31 & 63.81 & $-6.30$ & $-3.24$ & 1.59\\
 & $H(z)$ + BAO + \mii\ 25 + 69 + \civ\ 23 + 16 CC & 0.0256 & 0.1139 & 0.298 & 0.038 & -- & 68.61 & 0.268 & 0.436 & 1.033 & 0.256 & 0.307 & 1.714 & 72.23 & 92.23 & 123.99 & 93.23 & 1.86 & 5.04 & 1.91\\
Non-flat \lcdm & \civ\ 23 + 16 CR & -- & 0.0178 & 0.088 & $-0.378$ & -- & -- & 0.251 & 0.510 & 0.958 & -- & -- & -- & 13.65 & 23.65 & 31.97 & 40.20 & $-7.96$ & $-6.30$ & 6.07\\
 & \mii\ 25 + 69 + \civ\ 23 + 16 CR & -- & 0.0219 & 0.096 & $-0.402$ & -- & -- & 0.251 & 0.516 & 0.949 & 0.251 & 0.295 & 1.640 & 34.62 & 50.62 & 75.12 & 63.42 & $-8.70$ & $-5.64$ & 0.86\\
 & \civ\ 25 CC & -- & $-0.0235$ & 0.003 & $-0.023$ & -- & -- & 0.288 & 0.187 & 1.573 & -- & -- & -- & 21.07 & 31.07 & 37.16 & 28.30 & 2.01 & 3.23 & $-0.13$\\
 & \civ\ 25 CR & -- & 0.3847 & 0.836 & $-1.719$ & -- & -- & 0.148 & 0.579 & 1.345 & -- & -- & -- & $-9.87$ & 0.13 & 6.22 & 1.64 & 0.29 & 1.51 & 1.01\\
 & \civ\ 23 + 41 CC & -- & 0.0003 & 0.052 & $-0.268$ & -- & -- & 0.284 & 0.481 & 0.950 & -- & -- & -- & 45.44 & 55.44 & 66.24 & 72.44 & $-5.53$ & $-3.37$ & 7.03\\
 & \mii\ 25 + 69 + \civ\ 23 + 41 CC & -- & 0.0126 & 0.077 & $-0.348$ & -- & -- & 0.283 & 0.507 & 0.996 & 0.255 & 0.298 & 1.630 & 67.07 & 83.07 & 107.57 & 97.62 & $-5.66$ & $-2.60$ & 4.44\\
 & \civ\ 23 + 41 CR & -- & 0.0105 & 0.073 & $-0.334$ & -- & -- & 0.204 & 0.508 & 0.983 & -- & -- & -- & 13.28 & 23.28 & 34.07 & 41.88 & $-13.54$ & $-11.38$ & 1.07\\
 & \mii\ 25 + 69 + \civ\ 23 + 41 CR & -- & 0.0100 & 0.072 & $-0.332$ & -- & -- & 0.204 & 0.503 & 0.989 & 0.257 & 0.279 & 1.652 & 34.46 & 50.46 & 74.96 & 64.48 & $-14.17$ & $-11.11$ & $-3.65$\\
\midrule
 & $H(z)$ + BAO & 0.0296 & 0.0951 & 0.290 & -- & $-0.754$ & 65.79 & -- & -- & -- & -- & -- & -- & 22.39 & 30.39 & 37.52 & 30.63 & $-1.25$ & 0.53 & $-1.69$\\
 & \mii\ 25 & -- & 0.4231 & 0.915 & -- & $-2.775$ & -- & -- & -- & -- & 0.116 & 0.383 & 1.800 & $-28.42$ & $-18.42$ & $-3.10$ & $-20.00$ & 2.00 & 5.06 & $-0.06$\\
 & \mii\ 69 & -- & $-0.0239$ & 0.003 & -- & $-4.993$ & -- & -- & -- & -- & 0.274 & 0.232 & 1.362 & 22.28 & 32.28 & 47.60 & 40.90 & $-4.65$ & $-1.59$ & 3.85\\
 & \mii\ 25 + 69 & -- & $-0.0241$ & 0.002 & -- & $-4.941$ & -- & -- & -- & -- & 0.246 & 0.244 & 1.349 & 13.34 & 23.34 & 38.65 & 33.93 & $-6.29$ & $-3.22$ & 3.18\\
 & \civ\ 23 + 16 CC & -- & $-0.0186$ & 0.013 & -- & $-4.998$ & -- & 0.228 & 0.358 & 0.764 & -- & -- & -- & 17.46 & 27.46 & 35.77 & 34.78 & $-2.88$ & $-1.22$ & 0.75\\
 & \mii\ 25 + 69 + \civ\ 23 + 16 CC & -- & $-0.0209$ & 0.009 & -- & $-4.989$ & -- & 0.224 & 0.340 & 0.756 & 0.246 & 0.270 & 1.391 & 31.63 & 47.63 & 72.13 & 57.25 & $-10.49$ & $-7.42$ & $-4.97$\\
 & $H(z)$ + BAO + \mii\ 25 + 69 + \civ\ 23 + 16 CC & 0.0292 & 0.0975 & 0.291 & -- & $-0.768$ & 66.19 & 0.269 & 0.453 & 1.030 & 0.255 & 0.318 & 1.705 & 70.07 & 90.07 & 118.97 & 90.72 & $-0.30$ & 0.02 & $-0.60$\\
Flat XCDM & \civ\ 23 + 16 CR & -- & $-0.0214$ & 0.008 & -- & $-4.954$ & -- & 0.245 & 0.341 & 0.751 & -- & -- & -- & 16.30 & 26.30 & 34.62 & 35.19 & $-5.31$ & $-3.64$ & 1.06\\
 & \mii\ 25 + 69 + \civ\ 23 + 16 CR & -- & $-0.0237$ & 0.003 & -- & $-4.919$ & -- & 0.247 & 0.318 & 0.721 & 0.243 & 0.258 & 1.340 & 30.08 & 46.08 & 70.58 & 55.24 & $-13.24$ & $-10.18$ & $-7.32$\\
 & \civ\ 23 + 41 CC & -- & $-0.0251$ & 0.000 & -- & $-4.636$ & -- & 0.255 & 0.280 & 0.713 & -- & -- & -- & 44.53 & 54.53 & 65.33 & 66.07 & $-6.44$ & $-4.28$ & 0.67\\
 & \mii\ 25 + 69 + \civ\ 23 + 41 CC & -- & $-0.0247$ & 0.001 & -- & $-4.983$ & -- & 0.268 & 0.301 & 0.745 & 0.245 & 0.226 & 1.348 & 58.03 & 74.03 & 98.53 & 84.93 & $-14.71$ & $-11.64$ & $-8.25$\\
 & \civ\ 23 + 41 CR & -- & $-0.0241$ & 0.002 & -- & $-4.985$ & -- & 0.198 & 0.320 & 0.747 & -- & -- & -- & 14.89 & 24.89 & 35.68 & 35.34 & $-11.93$ & $-9.77$ & $-5.47$\\
 & \mii\ 25 + 69 + \civ\ 23 + 41 CR & -- & $-0.0243$ & 0.002 & -- & $-4.963$ & -- & 0.195 & 0.311 & 0.752 & 0.240 & 0.238 & 1.348 & 28.38 & 44.38 & 68.89 & 53.13 & $-22.25$ & $-17.19$ & $-15.20$\\
\midrule
 & $H(z)$ + BAO & 0.0289 & 0.0985 & 0.296 & $-0.053$ & $-0.730$ & 65.76 & -- & -- & -- & -- & -- & -- & 22.13 & 32.13 & 41.05 & 32.51 & 0.49 & 4.06 & 0.19\\
 & \mii\ 25 & -- & 0.2962 & 0.656 & $-0.663$ & 0.087 & -- & -- & -- & -- & 0.116 & 0.418 & 1.841 & $-28.47$ & $-16.47$ & 1.91 & $-19.59$ & 3.95 & 10.08 & 0.36\\
 & \mii\ 69 & -- & $-0.0133$ & 0.024 & $-0.046$ & $-3.601$ & -- & -- & -- & -- & 0.267 & 0.281 & 1.337 & 17.46 & 29.46 & 47.84 & 45.64 & $-7.48$ & $-1.35$ & 8.59\\
 & \mii\ 25 + 69 & -- & $-0.0174$ & 0.016 & $-0.021$ & $-4.907$ & -- & -- & -- & -- & 0.244 & 0.278 & 1.330 & 12.33 & 24.33 & 42.71 & 38.75 & $-5.29$ & 0.84 & 8.00\\
 & \civ\ 23 + 16 CC & -- & 0.0240 & 0.100 & $-0.161$ & $-3.860$ & -- & 0.232 & 0.488 & 0.869 & -- & -- & -- & 13.36 & 25.36 & 38.31 & 38.76 & $-4.98$ & 1.32 & 4.72\\
 & \mii\ 25 + 69 + \civ\ 23 + 16 CC & -- & $-0.0203$ & 0.010 & $-0.019$ & $-3.877$ & -- & 0.210 & 0.365 & 0.728 & 0.240 & 0.266 & 1.307 & 26.26 & 44.26 & 71.82 & 61.98 & $-13.86$ & $-7.73$ & $-0.24$\\
 & $H(z)$ + BAO + \mii\ 25 + 69 + \civ\ 23 + 16 CC & 0.0270 & 0.1057 & 0.301 & $-0.060$ & $-0.789$ & 66.57 & 0.270 & 0.464 & 0.991 & 0.253 & 0.325 & 1.691 & 70.04 & 92.04 & 126.98 & 92.53 & 1.67 & 8.02 & 1.21\\
Non-flat XCDM & \civ\ 23 + 16 CR & -- & $-0.0103$ & 0.030 & $-0.063$ & $-2.946$ & -- & 0.239 & 0.447 & 0.844 & -- & -- & -- & 11.16 & 23.16 & 36.12 & 40.38 & $-8.45$ & $-2.15$ & 6.25\\
 & \mii\ 25 + 69 + \civ\ 23 + 16 CR & -- & $-0.0214$ & 0.008 & $-0.015$ & $-4.135$ & -- & 0.241 & 0.378 & 0.760 & 0.235 & 0.272 & 1.279 & 23.75 & 41.75 & 69.32 & 60.60 & $-17.57$ & $-11.44$ & $-1.96$\\
 & \civ\ 23 + 41 CC & -- & $-0.0161$ & 0.019 & $-0.035$ & $-3.461$ & -- & 0.271 & 0.376 & 0.784 & -- & -- & -- & 43.32 & 55.32 & 68.27 & 71.36 & $-5.65$ & $-1.34$ & 5.95\\
 & \mii\ 25 + 69 + \civ\ 23 + 41 CC & -- & $-0.0219$ & 0.007 & $-0.012$ & $-4.305$ & -- & 0.254 & 0.333 & 0.773 & 0.240 & 0.250 & 1.316 & 54.50 & 72.50 & 100.07 & 92.60 & $-16.23$ & $-10.11$ & $-0.58$\\
 & \civ\ 23 + 41 CR & -- & $-0.0166$ & 0.017 & $-0.046$ & $-2.254$ & -- & 0.196 & 0.407 & 0.808 & -- & -- & -- & 9.22 & 21.22 & 34.17 & 41.51 & $-15.60$ & $-11.29$ & 0.70\\
 & \mii\ 25 + 69 + \civ\ 23 + 41 CR & -- & $-0.0221$ & 0.006 & $-0.012$ & $-4.286$ & -- & 0.194 & 0.370 & 0.745 & 0.240 & 0.260 & 1.274 & 21.63 & 39.63 & 67.19 & 55.35 & $-25.01$ & $-18.89$ & $-12.98$\\
\midrule
 & $H(z)$ + BAO & 0.0310 & 0.0900 & 0.280 & -- & 1.010 & 65.89 & -- & -- & -- & -- & -- & -- & 22.31 & 30.31 & 37.45 & 29.90 & $-1.33$ & 0.46 & $-2.42$\\
 & \mii\ 25 & -- & 0.4579 & 0.986 & -- & 1.111 & -- & -- & -- & -- & 0.115 & 0.385 & 1.810 & $-28.42$ & $-18.42$ & $-3.97$ & $-20.43$ & 2.00 & 4.20 & $-0.49$\\
 & \mii\ 69 & -- & 0.0246 & 0.101 & -- & 0.012 & -- & -- & -- & -- & 0.286 & 0.278 & 1.663 & 28.95 & 38.95 & 53.41 & 37.81 & 2.02 & 4.22 & 0.77\\
 & \mii\ 25 + 69 & -- & $-0.0056$ & 0.040 & -- & 0.005 & -- & -- & -- & -- & 0.257 & 0.293 & 1.670 & 21.65 & 31.65 & 46.10 & 32.15 & 2.03 & 4.23 & 1.41\\
 & \civ\ 23 + 16 CC & -- & 0.0088 & 0.069 & -- & 0.007 & -- & 0.265 & 0.420 & 0.978 & -- & -- & -- & 22.43 & 32.43 & 43.23 & 35.99 & 2.10 & 6.24 & 1.95\\
 & \mii\ 25 + 69 + \civ\ 23 + 16 CC & -- & $-0.0169$ & 0.017 & -- & 0.144 & -- & 0.255 & 0.414 & 0.954 & 0.261 & 0.282 & 1.675 & 44.38 & 60.38 & 83.50 & 66.19 & 2.27 & 3.95 & 3.90\\
 & $H(z)$ + BAO + \mii\ 25 + 69 + \civ\ 23 + 16 CC & 0.0311 & 0.0915 & 0.282 & -- & 0.982 & 66.13 & 0.267 & 0.457 & 1.012 & 0.260 & 0.310 & 1.707 & 70.06 & 90.06 & 118.96 & 89.82 & $-0.31$ & 0.01 & $-1.50$\\
Flat \pcdm & \civ\ 23 + 16 CR & -- & $-0.0048$ & 0.042 & -- & 0.005 & -- & 0.280 & 0.410 & 0.977 & -- & -- & -- & 23.65 & 33.65 & 44.44 & 36.47 & 2.04 & 6.18 & 2.34\\
 & \mii\ 25 + 69 + \civ\ 23 + 16 CR & -- & $-0.0022$ & 0.047 & -- & 0.032 & -- & 0.282 & 0.416 & 0.975 & 0.262 & 0.301 & 1.658 & 45.43 & 61.43 & 84.55 & 66.78 & 2.11 & 3.80 & 4.21\\
 & \civ\ 23 + 41 CC & -- & $-0.0170$ & 0.017 & -- & 0.072 & -- & 0.311 & 0.413 & 0.949 & -- & -- & -- & 53.02 & 63.02 & 73.82 & 67.75 & 2.05 & 4.21 & 2.34\\
 & \mii\ 25 + 69 + \civ\ 23 + 41 CC & -- & $-0.0023$ & 0.047 & -- & 0.070 & -- & 0.305 & 0.413 & 0.977 & 0.259 & 0.296 & 1.658 & 74.97 & 90.97 & 114.09 & 97.24 & 2.23 & 3.92 & 4.07\\
 & \civ\ 23 + 41 CR & -- & $-0.0035$ & 0.044 & -- & 0.011 & -- & 0.241 & 0.422 & 1.024 & -- & -- & -- & 28.94 & 38.94 & 49.73 & 45.09 & 2.11 & 4.27 & 4.28\\
 & \mii\ 25 + 69 + \civ\ 23 + 41 CR & -- & $-0.0030$ & 0.045 & -- & 0.004 & -- & 0.234 & 0.425 & 1.023 & 0.258 & 0.298 & 1.660 & 50.83 & 66.83 & 89.95 & 73.61 & 2.19 & 3.88 & 5.29\\
\midrule
 & $H(z)$ + BAO & 0.0306 & 0.0920 & 0.284 & $-0.058$ & 1.200 & 65.91 & -- & -- & -- & -- & -- & -- & 22.05 & 32.05 & 40.97 & 31.30 & 0.41 & 3.98 & $-1.02$\\
 & \mii\ 25 & -- & 0.3956 & 0.859 & $-0.833$ & 5.538 & -- & -- & -- & -- & 0.116 & 0.416 & 1.804 & $-28.42$ & $-16.42$ & 0.92 & $-20.51$ & 4.00 & 9.09 & $-0.57$\\
 & \mii\ 69 & -- & 0.1002 & 0.256 & $-0.254$ & 0.188 & -- & -- & -- & -- & 0.286 & 0.283 & 1.678 & 28.81 & 40.81 & 58.15 & 38.01 & 3.87 & 8.97 & 0.97\\
 & \mii\ 25 + 69 & -- & 0.0396 & 0.132 & $-0.092$ & 0.021 & -- & -- & -- & -- & 0.254 & 0.300 & 1.682 & 21.69 & 33.69 & 51.03 & 32.21 & 4.07 & 9.16 & 1.47\\
 & \civ\ 23 + 16 CC & -- & 0.0003 & 0.052 & $-0.051$ & 0.087 & -- & 0.253 & 0.424 & 0.965 & -- & -- & -- & 22.33 & 34.33 & 47.28 & 36.52 & 3.99 & 10.29 & 2.48\\
 & \mii\ 25 + 69 + \civ\ 23 + 16 CC & -- & 0.0126 & 0.077 & $-0.029$ & 0.100 & -- & 0.266 & 0.418 & 0.976 & 0.256 & 0.306 & 1.660 & 44.31 & 62.31 & 88.32 & 67.22 & 4.20 & 8.77 & 5.00\\
 & $H(z)$ + BAO + \mii\ 25 + 69 + \civ\ 23 + 16 CC & 0.0307 & 0.0898 & 0.277 & $-0.046$ & 1.091 & 66.16 & 0.271 & 0.441 & 1.026 & 0.253 & 0.317 & 1.692 & 69.79 & 91.79 & 123.59 & 90.98 & 1.42 & 4.63 & $-0.33$\\
Non-flat \pcdm & \civ\ 23 + 16 CR & -- & 0.0357 & 0.124 & $-0.124$ & 0.126 & -- & 0.282 & 0.424 & 0.992 & -- & -- & -- & 23.45 & 35.45 & 48.40 & 37.14 & 3.84 & 10.14 & 3.01\\
 & \mii\ 25 + 69 + \civ\ 23 + 16 CR & -- & 0.0392 & 0.131 & $-0.127$ & 0.029 & -- & 0.286 & 0.433 & 0.969 & 0.247 & 0.301 & 1.689 & 45.37 & 63.37 & 89.38 & 67.80 & 4.05 & 8.63 & 5.24\\
 & \civ\ 23 + 41 CC & -- & 0.0261 & 0.105 & $-0.095$ & 0.063 & -- & 0.313 & 0.418 & 1.000 & -- & -- & -- & 53.22 & 65.22 & 78.17 & 67.74 & 4.25 & 8.57 & 2.34\\
 & \mii\ 25 + 69 + \civ\ 23 + 41 CC & -- & 0.0027 & 0.057 & $-0.056$ & 0.403 & -- & 0.309 & 0.411 & 0.997 & 0.256 & 0.302 & 1.676 & 75.13 & 93.13 & 119.15 & 98.23 & 4.40 & 8.97 & 5.05\\
 & \civ\ 23 + 41 CR & -- & 0.0034 & 0.058 & $-0.056$ & 0.064 & -- & 0.243 & 0.414 & 1.051 & -- & -- & -- & 28.78 & 40.78 & 53.74 & 46.02 & 3.96 & 8.28 & 5.21\\
 & \mii\ 25 + 69 + \civ\ 23 + 41 CR & -- & $-0.0040$ & 0.043 & $-0.035$ & 0.065 & -- & 0.237 & 0.406 & 1.048 & 0.254 & 0.274 & 1.689 & 50.77 & 68.77 & 94.78 & 75.92 & 4.13 & 8.70 & 7.60\\
\bottomrule
\end{tabular}
\begin{tablenotes}[flushleft]
\item [a] \wx\ corresponds to flat/non-flat XCDM and $\alpha$ corresponds to flat/non-flat \pcdm.
\item [b] \hunit. $\Omega_b$ and $H_0$ are set to be 0.05 and 70 \hunit, respectively.
\end{tablenotes}
\end{threeparttable}%
}
\end{sidewaystable*}

\begin{sidewaystable*}
\centering
\resizebox*{\columnwidth}{0.74\columnwidth}{%
\begin{threeparttable}
\caption{One-dimensional marginalized posterior mean values and uncertainties ($\pm 1\sigma$ error bars or $2\sigma$ limits) of the parameters for all models from various combinations of data.}\label{tab:1d_BFP}
\begin{tabular}{lccccccccccccc}
\toprule
Model & Data set & $\Omega_{b}h^2$ & $\Omega_{c}h^2$ & $\Omega_{m0}$ & $\Omega_{k0}$ & $w_{\mathrm{X}}$/$\alpha$\tnote{a} & $H_0$\tnote{b} & $\sigma_{\mathrm{int,\,\textsc{c}}}$ & $\gamma_{\rm\textsc{c}}$ & $\beta_{\rm\textsc{c}}$ & $\sigma_{\mathrm{int,\,\textsc{m}}}$ & $\gamma_{\rm\textsc{m}}$ & $\beta_{\rm\textsc{m}}$ \\
\midrule
 & $H(z)$ + BAO & $0.0247\pm0.0030$ & $0.1186^{+0.0076}_{-0.0083}$ & $0.301^{+0.016}_{-0.018}$ & -- & -- & $69.14\pm1.85$ & -- & -- & -- & -- & -- & -- \\
 & \mii\ 25 & -- & -- & -- & -- & -- & -- & -- & -- & -- & $0.135^{+0.021}_{-0.033}$ & $0.355^{+0.100}_{-0.091}$ & $1.782^{+0.131}_{-0.128}$ \\
 & \mii\ 69 & -- & -- & $<0.612$\tnote{c} & -- & -- & -- & -- & -- & -- & $0.298^{+0.026}_{-0.033}$ & $0.289^{+0.050}_{-0.051}$ & $1.700^{+0.066}_{-0.059}$ \\
 & \mii\ 25 + 69 & -- & -- & $<0.537$\tnote{c} & -- & -- & -- & -- & -- & -- & $0.266^{+0.021}_{-0.026}$ & $0.314\pm0.043$ & $1.711^{+0.062}_{-0.056}$ \\
 & \civ\ 23 + 16 CC & -- & -- & $<0.500$\tnote{c} & -- & -- & -- & $0.301^{+0.040}_{-0.057}$ & $0.455\pm0.045$ & $1.006\pm0.090$ & -- & -- & -- \\
 & \mii\ 25 + 69 + \civ\ 23 + 16 CC & -- & -- & $<0.779$ & -- & -- & -- & $0.296^{+0.038}_{-0.055}$ & $0.448\pm0.042$ & $0.989\pm0.086$ & $0.265^{+0.020}_{-0.025}$ & $0.310\pm0.041$ & $1.697\pm0.058$ \\
 & $H(z)$ + BAO + \mii\ 25 + 69 + \civ\ 23 + 16 CC & $0.0247\pm0.0028$ & $0.1182^{+0.0073}_{-0.0081}$ & $0.300^{+0.015}_{-0.017}$ & -- & -- & $69.17\pm1.78$ & $0.296^{+0.037}_{-0.054}$ & $0.455\pm0.038$ & $0.999^{+0.078}_{-0.063}$ & $0.265^{+0.020}_{-0.025}$ & $0.314\pm0.040$ & $1.700\pm0.051$ \\
Flat \lcdm & \civ\ 23 + 16 CR & -- & -- & $<0.476$\tnote{c} & -- & -- & -- & $0.317^{+0.039}_{-0.055}$ & $0.445\pm0.046$ & $1.032\pm0.094$ & -- & -- & -- \\
 & \mii\ 25 + 69 + \civ\ 23 + 16 CR & -- & -- & $<0.770$ & -- & -- & -- & $0.313^{+0.037}_{-0.054}$ & $0.439\pm0.044$ & $1.016\pm0.092$ & $0.265^{+0.020}_{-0.025}$ & $0.309\pm0.041$ & $1.696\pm0.058$ \\
 & \civ\ 25 CC & -- & -- & -- & -- & -- & -- & $0.338^{+0.052}_{-0.084}$ & $0.288^{+0.113}_{-0.230}$ & $1.525^{+0.321}_{-0.184}$ & -- & -- & -- \\
 & \civ\ 25 CR & -- & -- & $>0.380$\tnote{c} & -- & -- & -- & $0.190^{+0.027}_{-0.044}$ & $0.421\pm0.143$ & $1.320^{+0.190}_{-0.189}$ & -- & -- & -- \\
 & \civ\ 23 + 41 CC & -- & -- & $<0.860$ & -- & -- & -- & $0.335^{+0.035}_{-0.045}$ & $0.448\pm0.049$ & $1.066\pm0.098$ & -- & -- & -- \\
 & \mii\ 25 + 69 + \civ\ 23 + 41 CC & -- & -- & $<0.721$ & -- & -- & -- & $0.331^{+0.033}_{-0.044}$ & $0.442\pm0.046$ & $1.045\pm0.096$ & $0.264^{+0.020}_{-0.025}$ & $0.308\pm0.041$ & $1.692\pm0.058$ \\
 & \civ\ 23 + 41 CR & -- & -- & $<0.748$ & -- & -- & -- & $0.263^{+0.029}_{-0.037}$ & $0.445\pm0.041$ & $1.070\pm0.081$ & -- & -- & -- \\
 & \mii\ 25 + 69 + \civ\ 23 + 41 CR & -- & -- & $<0.583$ & -- & -- & -- & $0.260^{+0.027}_{-0.036}$ & $0.440\pm0.039$ & $1.055\pm0.079$ & $0.264^{+0.020}_{-0.025}$ & $0.306\pm0.041$ & $1.686\pm0.057$ \\
\midrule
 & $H(z)$ + BAO & $0.0266^{+0.0039}_{-0.0045}$ & $0.1088\pm0.0166$ & $0.291\pm0.023$ & $0.059^{+0.081}_{-0.091}$ & -- & $68.37\pm2.10$ & -- & -- & -- & -- & -- & -- \\
 & \mii\ 25 & -- & -- & $>0.431$\tnote{c} & $0.244^{+1.105}_{-1.076}$ & -- & -- & -- & -- & -- & $0.134^{+0.020}_{-0.032}$ & $0.357^{+0.100}_{-0.091}$ & $1.799^{+0.127}_{-0.124}$ \\
 & \mii\ 69 & -- & -- & $0.552^{+0.348}_{-0.221}$ & $-0.386^{+0.591}_{-1.457}$ & -- & -- & -- & -- & -- & $0.296^{+0.026}_{-0.033}$ & $0.307^{+0.052}_{-0.060}$ & $1.690^{+0.073}_{-0.066}$ \\
 & \mii\ 25 + 69 & -- & -- & $0.495^{+0.280}_{-0.345}$ & $-0.057^{+0.776}_{-1.169}$ & -- & -- & -- & -- & -- & $0.266^{+0.020}_{-0.025}$ & $0.323\pm0.044$ & $1.717^{+0.065}_{-0.058}$ \\
 & \civ\ 23 + 16 CC & -- & -- & $0.462^{+0.191}_{-0.373}$ & $-0.374^{+0.480}_{-1.037}$ & -- & -- & $0.297^{+0.042}_{-0.057}$ & $0.486^{+0.047}_{-0.060}$ & $1.030^{+0.095}_{-0.086}$ & -- & -- & -- \\
 & \mii\ 25 + 69 + \civ\ 23 + 16 CC & -- & -- & $0.375^{+0.116}_{-0.293}$ & $-0.611^{+0.341}_{-0.534}$ & -- & -- & $0.285^{+0.038}_{-0.054}$ & $0.493^{+0.048}_{-0.059}$ & $1.011^{+0.089}_{-0.088}$ & $0.264^{+0.020}_{-0.025}$ & $0.323^{+0.045}_{-0.044}$ & $1.674\pm0.062$ \\
 & $H(z)$ + BAO + \mii\ 25 + 69 + \civ\ 23 + 16 CC & $0.0263^{+0.0037}_{-0.0045}$ & $0.1103\pm0.0164$ & $0.292^{+0.022}_{-0.023}$ & $0.047^{+0.079}_{-0.089}$ & -- & $68.56\pm2.07$ & $0.297^{+0.037}_{-0.054}$ & $0.454\pm0.039$ & $0.995^{+0.078}_{-0.064}$ & $0.265^{+0.019}_{-0.025}$ & $0.314\pm0.040$ & $1.698\pm0.051$ \\
Non-flat \lcdm & \civ\ 23 + 16 CR & -- & -- & $0.436^{+0.169}_{-0.360}$ & $-0.484^{+0.390}_{-0.902}$ & -- & -- & $0.310^{+0.040}_{-0.056}$ & $0.484^{+0.051}_{-0.066}$ & $1.054\pm0.098$ & -- & -- & -- \\
 & \mii\ 25 + 69 + \civ\ 23 + 16 CR & -- & -- & $0.344^{+0.097}_{-0.267}$ & $-0.647^{+0.330}_{-0.431}$ & -- & -- & $0.299^{+0.036}_{-0.053}$ & $0.492^{+0.053}_{-0.063}$ & $1.038\pm0.093$ & $0.263^{+0.020}_{-0.025}$ & $0.323^{+0.042}_{-0.046}$ & $1.667\pm0.061$ \\%
 & \civ\ 25 CC & -- & -- & $>0.391$\tnote{c} & $0.291^{+1.234}_{-0.910}$ & -- & -- & $0.337^{+0.051}_{-0.084}$ & $0.279^{+0.104}_{-0.230}$ & $1.564^{+0.295}_{-0.168}$ & -- & -- & -- \\
 & \civ\ 25 CR & -- & -- & $>0.458$\tnote{c} & $0.029^{+0.986}_{-1.313}$ & -- & -- & $0.188^{+0.028}_{-0.045}$ & $0.435\pm0.148$ & $1.343^{+0.182}_{-0.180}$ & -- & -- & -- \\
 & \civ\ 23 + 41 CC & -- & -- & $0.405^{+0.147}_{-0.375}$ & $-0.134^{+0.561}_{-0.946}$ & -- & -- & $0.334^{+0.035}_{-0.045}$ & $0.468^{+0.048}_{-0.053}$ & $1.088^{+0.095}_{-0.096}$ & -- & -- & -- \\
 & \mii\ 25 + 69 + \civ\ 23 + 41 CC & -- & -- & $0.331^{+0.096}_{-0.294}$ & $-0.389^{+0.326}_{-0.635}$ & -- & -- & $0.328^{+0.034}_{-0.044}$ & $0.471^{+0.048}_{-0.054}$ & $1.070\pm0.096$ & $0.264^{+0.020}_{-0.025}$ & $0.317\pm0.043$ & $1.679\pm0.063$ \\
 & \civ\ 23 + 41 CR & -- & -- & $0.255^{+0.041}_{-0.209}$ & $-0.503^{+0.253}_{-0.276}$ & -- & -- & $0.244^{+0.031}_{-0.039}$ & $0.492^{+0.047}_{-0.051}$ & $1.057\pm0.089$ & -- & -- & -- \\
 & \mii\ 25 + 69 + \civ\ 23 + 41 CR & -- & -- & $0.206^{+0.029}_{-0.151}$ & $-0.530^{+0.245}_{-0.152}$ & -- & -- & $0.238^{+0.028}_{-0.037}$ & $0.497\pm0.047$ & $1.045\pm0.082$ & $0.263^{+0.020}_{-0.025}$ & $0.310^{+0.040}_{-0.044}$ & $1.650\pm0.059$ \\
\midrule
 & $H(z)$ + BAO & $0.0295^{+0.0042}_{-0.0050}$ & $0.0969^{+0.0178}_{-0.0152}$ & $0.289\pm0.020$ & -- & $-0.784^{+0.140}_{-0.107}$ & $66.22^{+2.31}_{-2.54}$ & -- & -- & -- & -- & -- & -- \\
 & \mii\ 25 & -- & -- & $0.544^{+0.386}_{-0.219}$ & -- & $-2.386^{+2.276}_{-1.263}$ & -- & -- & -- & -- & $0.135^{+0.020}_{-0.032}$ & $0.356^{+0.101}_{-0.093}$ & $1.765^{+0.155}_{-0.129}$ \\
 & \mii\ 69 & -- & -- & $<0.462$\tnote{c} & -- & $<-0.439$ & -- & -- & -- & -- & $0.295^{+0.026}_{-0.033}$ & $0.287\pm0.050$ & $1.633^{+0.126}_{-0.078}$ \\
 & \mii\ 25 + 69 & -- & -- & $<0.814$ & -- & $<-0.670$ & -- & -- & -- & -- & $0.263^{+0.020}_{-0.025}$ & $0.308\pm0.044$ & $1.610^{+0.143}_{-0.090}$ \\
 & \civ\ 23 + 16 CC & -- & -- & $<0.816$ & -- & $<-0.662$ & -- & $0.289^{+0.041}_{-0.056}$ & $0.437\pm0.049$ & $0.941^{+0.125}_{-0.108}$ & -- & -- & -- \\
 & \mii\ 25 + 69 + \civ\ 23 + 16 CC & -- & -- & $<0.348$ & -- & $<-1.917$ & -- & $0.269^{+0.038}_{-0.054}$ & $0.400\pm0.047$ & $0.845\pm0.108$ & $0.258^{+0.020}_{-0.025}$ & $0.296\pm0.041$ & $1.525^{+0.113}_{-0.104}$ \\
 & $H(z)$ + BAO + \mii\ 25 + 69 + \civ\ 23 + 16 CC & $0.0291^{+0.0040}_{-0.0051}$ & $0.0987^{+0.0181}_{-0.0145}$ & $0.290^{+0.020}_{-0.019}$ & -- & $-0.804^{+0.144}_{-0.113}$ & $66.53^{+2.29}_{-2.55}$ & $0.298^{+0.037}_{-0.054}$ & $0.457\pm0.039$ & $0.993^{+0.080}_{-0.064}$ & $0.265^{+0.020}_{-0.025}$ & $0.315\pm0.041$ & $1.698\pm0.051$ \\
Flat XCDM & \civ\ 23 + 16 CR & -- & -- & $<0.770$ & -- & $<-0.868$ & -- & $0.301^{+0.039}_{-0.055}$ & $0.421\pm0.051$ & $0.952^{+0.128}_{-0.116}$ & -- & -- & -- \\
 & \mii\ 25 + 69 + \civ\ 23 + 16 CR & -- & -- & $<0.271$ & -- & $<-2.191$ & -- & $0.282^{+0.035}_{-0.051}$ & $0.384^{+0.042}_{-0.048}$ & $0.854\pm0.112$ & $0.258^{+0.020}_{-0.025}$ & $0.290\pm0.041$ & $1.499\pm0.108$ \\
 & \civ\ 23 + 41 CC & -- & -- & $<0.710$ & -- & $<-0.999$ & -- & $0.318^{+0.037}_{-0.046}$ & $0.415\pm0.057$ & $0.966^{+0.143}_{-0.125}$ & -- & -- & -- \\
 & \mii\ 25 + 69 + \civ\ 23 + 41 CC & -- & -- & $<0.208$ & -- & $<-2.361$ & -- & $0.301^{+0.032}_{-0.043}$ & $0.375^{+0.042}_{-0.052}$ & $0.870\pm0.117$ & $0.257^{+0.020}_{-0.025}$ & $0.283\pm0.041$ & $1.474^{+0.109}_{-0.108}$ \\
 & \civ\ 23 + 41 CR & -- & -- & $<0.279$ & -- & $<-1.935$ & -- & $0.232^{+0.028}_{-0.037}$ & $0.390^{+0.041}_{-0.048}$ & $0.901\pm0.108$ & -- & -- & -- \\
 & \mii\ 25 + 69 + \civ\ 23 + 41 CR & -- & -- & $<0.104$ & -- & $<-2.876$ & -- & $0.222^{+0.025}_{-0.034}$ & $0.366^{+0.033}_{-0.041}$ & $0.843^{+0.094}_{-0.093}$ & $0.256^{+0.020}_{-0.025}$ & $0.277\pm0.039$ & $1.444\pm0.098$ \\
\midrule
 & $H(z)$ + BAO & $0.0294^{+0.0047}_{-0.0050}$ & $0.0980^{+0.0186}_{-0.0187}$ & $0.292\pm0.025$ & $-0.027\pm0.109$ & $-0.770^{+0.149}_{-0.098}$ & $66.13^{+2.35}_{-2.36}$ & -- & -- & -- & -- & -- & -- \\
 & \mii\ 25 & -- & -- & $>0.418$\tnote{c} & $0.264^{+1.021}_{-0.950}$ & $-2.175^{+2.212}_{-0.925}$ & -- & -- & -- & -- & $0.136^{+0.020}_{-0.033}$ & $0.352^{+0.097}_{-0.098}$ & $1.807^{+0.142}_{-0.125}$ \\
 & \mii\ 69 & -- & -- & $0.509^{+0.314}_{-0.309}$ & $-0.207^{+0.544}_{-0.919}$ & $-2.492^{+1.749}_{-1.390}$ & -- & -- & -- & -- & $0.295^{+0.026}_{-0.033}$ & $0.306^{+0.052}_{-0.059}$ & $1.648^{+0.125}_{-0.079}$ \\
 & \mii\ 25 + 69 & -- & -- & $0.451^{+0.174}_{-0.446}$ & $-0.009^{+0.513}_{-0.830}$ & $-2.521^{+1.601}_{-1.619}$ & -- & -- & -- & -- & $0.265^{+0.021}_{-0.026}$ & $0.324\pm0.045$ & $1.672^{+0.120}_{-0.071}$ \\
 & \civ\ 23 + 16 CC & -- & -- & $0.455^{+0.189}_{-0.383}$ & $-0.247^{+0.428}_{-0.730}$ & $-2.573^{+1.781}_{-1.340}$ & -- & $0.294^{+0.042}_{-0.058}$ & $0.471^{+0.047}_{-0.054}$ & $0.995^{+0.110}_{-0.094}$ & -- & -- & -- \\
 & \mii\ 25 + 69 + \civ\ 23 + 16 CC & -- & -- & $0.258^{+0.052}_{-0.241}$ & $-0.274^{+0.273}_{-0.184}$ & $<-1.217$ & -- & $0.269^{+0.038}_{-0.054}$ & $0.447^{+0.046}_{-0.052}$ & $0.906\pm0.107$ & $0.260^{+0.020}_{-0.025}$ & $0.324^{+0.043}_{-0.047}$ & $1.533^{+0.126}_{-0.106}$ \\
 & $H(z)$ + BAO + \mii\ 25 + 69 + \civ\ 23 + 16 CC & $0.0290^{+0.0044}_{-0.0053}$ & $0.0998^{+0.0185}_{-0.0187}$ & $0.292\pm0.025$ & $-0.029\pm0.108$ & $-0.792^{+0.168}_{-0.103}$ & $66.46^{+2.28}_{-2.51}$ & $0.299^{+0.037}_{-0.054}$ & $0.458\pm0.039$ & $0.994^{+0.081}_{-0.065}$ & $0.265^{+0.020}_{-0.025}$ & $0.316\pm0.041$ & $1.699\pm0.051$ \\
Non-flat XCDM & \civ\ 23 + 16 CR & -- & -- & $0.414^{+0.153}_{-0.377}$ & $-0.267^{+0.379}_{-0.648}$ & $-2.647^{+1.537}_{-1.560}$ & -- & $0.305^{+0.041}_{-0.056}$ & $0.462^{+0.048}_{-0.058}$ & $1.012^{+0.117}_{-0.104}$ & -- & -- & -- \\
 & \mii\ 25 + 69 + \civ\ 23 + 16 CR & -- & -- & $0.212^{+0.034}_{-0.201}$ & $-0.112\pm0.150$ & $-3.267^{+0.594}_{-1.658}$ & -- & $0.281^{+0.036}_{-0.052}$ & $0.435^{+0.046}_{-0.055}$ & $0.919\pm0.112$ & $0.259^{+0.020}_{-0.025}$ & $0.320^{+0.043}_{-0.047}$ & $1.508^{+0.126}_{-0.111}$ \\
 & \civ\ 23 + 41 CC & -- & -- & $<0.498$\tnote{c} & $-0.103^{+0.368}_{-0.642}$ & $-2.709^{+1.080}_{-1.940}$ & -- & $0.329^{+0.035}_{-0.045}$ & $0.453\pm0.052$ & $1.045^{+0.120}_{-0.103}$ & -- & -- & -- \\
 & \mii\ 25 + 69 + \civ\ 23 + 41 CC & -- & -- & $0.179^{+0.021}_{-0.175}$ & $-0.168^{+0.186}_{-0.124}$ & $<-1.557$ & -- & $0.309^{+0.033}_{-0.044}$ & $0.423^{+0.047}_{-0.055}$ & $0.942\pm0.118$ & $0.258^{+0.020}_{-0.025}$ & $0.314^{+0.042}_{-0.047}$ & $1.502^{+0.127}_{-0.115}$ \\
 & \civ\ 23 + 41 CR & -- & -- & $0.220^{+0.035}_{-0.216}$ & $-0.239^{+0.255}_{-0.167}$ & $<-1.035$ & -- & $0.236^{+0.031}_{-0.039}$ & $0.444^{+0.047}_{-0.053}$ & $0.974\pm0.114$ & -- & -- & -- \\
 & \mii\ 25 + 69 + \civ\ 23 + 41 CR & -- & -- & $0.096^{+0.010}_{-0.089}$ & $-0.128^{+0.119}_{-0.034}$ & $-3.637^{+0.401}_{-1.363}$ & -- & $0.219^{+0.025}_{-0.034}$ & $0.415^{+0.040}_{-0.048}$ & $0.891\pm0.097$ & $0.256^{+0.020}_{-0.024}$ & $0.306^{+0.042}_{-0.046}$ & $1.430\pm0.114$ \\
\midrule
 & $H(z)$ + BAO & $0.0320^{+0.0054}_{-0.0041}$ & $0.0855^{+0.0175}_{-0.0174}$ & $0.275\pm0.023$ & -- & $1.267^{+0.536}_{-0.807}$ & $65.47^{+2.22}_{-2.21}$ & -- & -- & -- & -- & -- & -- \\
 & \mii\ 25 & -- & -- & $0.523^{+0.388}_{-0.248}$ & -- & -- & -- & -- & -- & -- & $0.134^{+0.020}_{-0.032}$ & $0.359^{+0.100}_{-0.090}$ & $1.803^{+0.117}_{-0.124}$ \\
 & \mii\ 69 & -- & -- & -- & -- & -- & -- & -- & -- & -- & $0.299^{+0.026}_{-0.033}$ & $0.293\pm0.051$ & $1.721^{+0.059}_{-0.055}$ \\
 & \mii\ 25 + 69 & -- & -- & $<0.594$\tnote{c} & -- & -- & -- & -- & -- & -- & $0.266^{+0.020}_{-0.026}$ & $0.320\pm0.043$ & $1.736^{+0.056}_{-0.050}$ \\
 & \civ\ 23 + 16 CC & -- & -- & $<0.563$\tnote{c} & -- & -- & -- & $0.308^{+0.040}_{-0.057}$ & $0.466\pm0.044$ & $1.040^{+0.091}_{-0.077}$ & -- & -- & -- \\
 & \mii\ 25 + 69 + \civ\ 23 + 16 CC & -- & -- & $<0.871$ & -- & $<5.813$\tnote{c} & -- & $0.303^{+0.039}_{-0.055}$ & $0.460\pm0.043$ & $1.027^{+0.090}_{-0.078}$ & $0.266^{+0.020}_{-0.025}$ & $0.317\pm0.042$ & $1.725^{+0.058}_{-0.051}$ \\
 & $H(z)$ + BAO + \mii\ 25 + 69 + \civ\ 23 + 16 CC & $0.0317^{+0.0050}_{-0.0048}$ & $0.0871^{+0.0189}_{-0.0170}$ & $0.276\pm0.023$ & -- & $1.174^{+0.482}_{-0.847}$ & $65.75\pm2.20$ & $0.299^{+0.037}_{-0.054}$ & $0.457\pm0.039$ & $0.991^{+0.081}_{-0.065}$ & $0.265^{+0.020}_{-0.025}$ & $0.316\pm0.041$ & $1.697\pm0.051$ \\
Flat \pcdm & \civ\ 23 + 16 CR & -- & -- & $<0.554$\tnote{c} & -- & $<6.304$\tnote{c} & -- & $0.324^{+0.039}_{-0.056}$ & $0.457\pm0.047$ & $1.065\pm0.090$ & -- & -- & -- \\
 & \mii\ 25 + 69 + \civ\ 23 + 16 CR & -- & -- & $<0.864$ & -- & $<5.719$\tnote{c} & -- & $0.320^{+0.038}_{-0.055}$ & $0.451\pm0.045$ & $1.051^{+0.091}_{-0.090}$ & $0.266^{+0.020}_{-0.025}$ & $0.316\pm0.042$ & $1.724^{+0.058}_{-0.051}$ \\
 & \civ\ 23 + 41 CC & -- & -- & $<0.507$\tnote{c} & -- & $<6.155$\tnote{c} & -- & $0.340^{+0.034}_{-0.045}$ & $0.460\pm0.047$ & $1.104\pm0.087$ & -- & -- & -- \\
 & \mii\ 25 + 69 + \civ\ 23 + 41 CC & -- & -- & $<0.837$ & -- & $<5.527$\tnote{c} & -- & $0.337^{+0.033}_{-0.044}$ & $0.455\pm0.046$ & $1.086^{+0.090}_{-0.091}$ & $0.265^{+0.020}_{-0.025}$ & $0.315\pm0.042$ & $1.721^{+0.059}_{-0.052}$ \\
 & \civ\ 23 + 41 CR & -- & -- & $<0.849$ & -- & $<5.625$\tnote{c} & -- & $0.270^{+0.028}_{-0.037}$ & $0.458\pm0.041$ & $1.111\pm0.078$ & -- & -- & -- \\
 & \mii\ 25 + 69 + \civ\ 23 + 41 CR & -- & -- & $<0.754$ & -- & $<4.852$\tnote{c} & -- & $0.266^{+0.028}_{-0.037}$ & $0.453\pm0.040$ & $1.095\pm0.078$ & $0.265^{+0.020}_{-0.025}$ & $0.313\pm0.042$ & $1.713^{+0.059}_{-0.053}$ \\
\midrule
 & $H(z)$ + BAO & $0.0320^{+0.0057}_{-0.0038}$ & $0.0865^{+0.0172}_{-0.0198}$ & $0.277^{+0.023}_{-0.026}$ & $-0.034^{+0.087}_{-0.098}$ & $1.360^{+0.584}_{-0.819}$ & $65.53\pm2.19$ & -- & -- & -- & -- & -- & -- \\
 & \mii\ 25 & -- & -- & $0.508^{+0.314}_{-0.316}$ & $0.018^{+0.390}_{-0.379}$ & -- & -- & -- & -- & -- & $0.134^{+0.020}_{-0.032}$ & $0.358^{+0.099}_{-0.089}$ & $1.805^{+0.116}_{-0.123}$ \\
 & \mii\ 69 & -- & -- & $0.463^{+0.163}_{-0.456}$ & $0.056^{+0.394}_{-0.380}$ & -- & -- & -- & -- & -- & $0.299^{+0.025}_{-0.033}$ & $0.292\pm0.051$ & $1.723\pm0.056$ \\
 & \mii\ 25 + 69 & -- & -- & $<0.567$\tnote{c} & $0.099^{+0.391}_{-0.372}$ & -- & -- & -- & -- & -- & $0.267^{+0.020}_{-0.025}$ & $0.320\pm0.043$ & $1.738^{+0.054}_{-0.049}$ \\
 & \civ\ 23 + 16 CC & -- & -- & $<0.550$\tnote{c} & $0.100^{+0.380}_{-0.361}$ & -- & -- & $0.309^{+0.039}_{-0.056}$ & $0.465\pm0.044$ & $1.044^{+0.088}_{-0.077}$ & -- & -- & -- \\
 & \mii\ 25 + 69 + \civ\ 23 + 16 CC & -- & -- & $<0.859$ & $0.142^{+0.374}_{-0.345}$ & $<5.980$\tnote{c} & -- & $0.305^{+0.038}_{-0.055}$ & $0.461\pm0.043$ & $1.034^{+0.087}_{-0.076}$ & $0.266^{+0.020}_{-0.025}$ & $0.318\pm0.042$ & $1.731^{+0.054}_{-0.049}$ \\
 & $H(z)$ + BAO + \mii\ 25 + 69 + \civ\ 23 + 16 CC & $0.0316^{+0.0059}_{-0.0043}$ & $0.0887^{+0.0182}_{-0.0204}$ & $0.278^{+0.024}_{-0.027}$ & $-0.041^{+0.088}_{-0.095}$ & $1.288^{+0.548}_{-0.872}$ & $65.84\pm2.21$ & $0.299^{+0.037}_{-0.054}$ & $0.459\pm0.039$ & $0.992^{+0.081}_{-0.064}$ & $0.265^{+0.020}_{-0.025}$ & $0.316\pm0.041$ & $1.698\pm0.051$ \\
Non-flat \pcdm & \civ\ 23 + 16 CR & -- & -- & $<0.546$\tnote{c} & $0.103^{+0.386}_{-0.365}$ & -- & -- & $0.324^{+0.038}_{-0.056}$ & $0.456\pm0.047$ & $1.069\pm0.089$ & -- & -- & -- \\
 & \mii\ 25 + 69 + \civ\ 23 + 16 CR & -- & -- & $<0.858$ & $0.143^{+0.375}_{-0.342}$ & $<5.947$\tnote{c} & -- & $0.322^{+0.037}_{-0.054}$ & $0.453\pm0.045$ & $1.058^{+0.088}_{-0.089}$ & $0.266^{+0.020}_{-0.025}$ & $0.317\pm0.042$ & $1.731\pm0.052$ \\
 & \civ\ 23 + 41 CC & -- & -- & $<0.484$\tnote{c} & $0.147^{+0.381}_{-0.357}$ & $<6.208$\tnote{c} & -- & $0.340^{+0.034}_{-0.044}$ & $0.460\pm0.047$ & $1.106\pm0.086$ & -- & -- & -- \\
 & \mii\ 25 + 69 + \civ\ 23 + 41 CC & -- & -- & $<0.829$ & $0.181^{+0.372}_{-0.333}$ & $<5.915$\tnote{c} & -- & $0.339^{+0.033}_{-0.044}$ & $0.457\pm0.046$ & $1.095\pm0.086$ & $0.266^{+0.020}_{-0.025}$ & $0.317^{+0.041}_{-0.042}$ & $1.728^{+0.054}_{-0.049}$ \\
 & \civ\ 23 + 41 CR & -- & -- & $<0.847$ & $0.157^{+0.370}_{-0.338}$ & $<5.954$\tnote{c} & -- & $0.272^{+0.028}_{-0.037}$ & $0.460\pm0.040$ & $1.121\pm0.073$ & -- & -- & -- \\
 & \mii\ 25 + 69 + \civ\ 23 + 41 CR & -- & -- & $<0.764$ & $0.187^{+0.338}_{-0.332}$ & $<5.520$\tnote{c} & -- & $0.268^{+0.027}_{-0.037}$ & $0.456\pm0.039$ & $1.108\pm0.073$ & $0.266^{+0.020}_{-0.025}$ & $0.315\pm0.042$ & $1.724\pm0.052$ \\%
\bottomrule
\end{tabular}
\begin{tablenotes}[flushleft]
\item [a] \wx\ corresponds to flat/non-flat XCDM and $\alpha$ corresponds to flat/non-flat \pcdm.
\item [b] \hunit. $\Omega_b$ and $H_0$ are set to be 0.05 and 70 \hunit, respectively.
\item [c] This is the 1$\sigma$ limit. The 2$\sigma$ limit is set by the prior and not shown here.
\end{tablenotes}
\end{threeparttable}%
}
\end{sidewaystable*}

The analysis of \mii\ 25 + 69 + \civ\ 23 + 16 CC data produces \om\ constraints spanning from a low of $<0.348$ (2$\sigma$, flat XCDM) to a high of $<0.871$ (2$\sigma$, flat \pcdm). In the joint analysis of $H(z)$ + BAO + \mii\ 25 + 69 + \civ\ 23 + 16 CC data, the \om\ constraints slightly deviate from those derived solely from $H(z)$ + BAO data, ranging from a low of $0.276\pm0.023$ (flat \pcdm) to a high of $0.300^{+0.015}_{-0.017}$ (flat \lcdm).

For the \ok\ parameter, constraints derived from \mii\ 25 + 69 + \civ\ 23 + 16 CC data and $H(z)$ + BAO + \mii\ 25 + 69 + \civ\ 23 + 16 CC data are $-0.611^{+0.341}_{-0.534}$ and $0.047^{+0.079}_{-0.089}$ for non-flat \lcdm, $-0.274^{+0.273}_{-0.184}$ and $-0.029\pm0.108$ for non-flat XCDM, and $0.142^{+0.374}_{-0.345}$ and $-0.041^{+0.088}_{-0.095}$ for non-flat \pcdm, respectively. The \ok\ constraints derived from $H(z)$ + BAO + \mii\ 25 + 69 + \civ\ 23 + 16 CC data only deviate from $H(z)$ + BAO data by less than $0.1\sigma$, but with same preferences of spatial hypersurfaces, i.e., open for non-flat \lcdm, and closed for non-flat XCDM and non-flat \pcdm, which differ from \mii\ 25 + 69 + \civ\ 23 + 16 CC data preferences for non-flat \lcdm\ and non-flat \pcdm.

Constraints on dark energy dynamics, represented by the \wx\ and $\alpha$ parameters, remain weak for \mii\ 25 + 69 + \civ\ 23 + 16 CC data, so the \wx\ and $\alpha$ constraints derived from $H(z)$ + BAO + \mii\ 25 + 69 + \civ\ 23 + 16 CC data showed marginal ($<0.12\sigma$) differences from those derived solely from $H(z)$ + BAO data. Specifically, \mii\ 25 + 69 + \civ\ 23 + 16 CC data and $H(z)$ + BAO + \mii\ 25 + 69 + \civ\ 23 + 16 CC data constraints on \wx\ are $<-1.917$ ($2\sigma$) and $<-1.217$ ($2\sigma$), and $-0.804^{+0.144}_{-0.113}$ and $-0.792^{+0.168}_{-0.103}$ for flat and non-flat XCDM, respectively; and constraints on $\alpha$ are $<5.813$ ($1\sigma$) and $<5.980$ ($1\sigma$), and $1.174^{+0.482}_{-0.847}$ and $1.288^{+0.548}_{-0.872}$ for flat and non-flat \pcdm, respectively.

According to AIC/BIC values in Table \ref{tab:BFP}, \mii\ 25 + 69 + \civ\ 23 + 16 CC data exhibit a preference for non-flat/flat XCDM. For non-flat \lcdm, the opposing evidence is strong or positive, and for flat \lcdm, it is very strong or strong. Additionally, there is very strong evidence against both flat and non-flat \pcdm. When evaluated with DIC, \mii\ 25 + 69 + \civ\ 23 + 16 CC data show a marked preference for flat XCDM, and there is positive evidence against models such as flat \lcdm\ and non-flat XCDM, and strong evidence against other models.

On the other hand, based on AIC and DIC evaluations, $H(z)$ + BAO + \mii\ 25 + 69 + \civ\ 23 + 16 CC data lean most towards flat \pcdm. There is positive evidence against models like non-flat \lcdm\ and non-flat XCDM (weak for AIC), and weak evidence against other models. However, when assessed using BIC, flat \lcdm\ emerges as the most favoured model, with strong evidence against non-flat XCDM, positive evidence against non-flat \lcdm\ and non-flat \pcdm, and very weak evidence against flat \pcdm\ and flat XCDM.

\begin{table*}
\centering
\resizebox{2\columnwidth}{!}{%
\begin{threeparttable}
\caption{The largest differences between considered cosmological models (flat and non-flat $\Lambda$CDM, XCDM, and $\phi$CDM) from various combinations of data with $1\sigma$ being the quadrature sum of the two corresponding $1\sigma$ error bars.}\label{tab:diff}
\begin{tabular}{lcccccc}
\toprule
 Data set & $\Delta\sigma_{\mathrm{int,\,\textsc{c}}}$ & $\Delta\gamma_{\rm\textsc{c}}$ & $\Delta\beta_{\rm\textsc{c}}$ & $\Delta\sigma_{\mathrm{int,\,\textsc{m}}}$ & $\Delta\gamma_{\rm\textsc{m}}$ & $\Delta\beta_{\rm\textsc{m}}$ \\
\midrule
\mq\ 25 & -- & -- & -- & $0.05\sigma$ & $0.05\sigma$ & $0.14\sigma$\\
\mq\ 69 & -- & -- & -- & $0.10\sigma$ & $0.26\sigma$ & $0.65\sigma$\\
\mq\ 25 + 69 & -- & -- & -- & $0.12\sigma$ & $0.25\sigma$ & $0.85\sigma$\\
\cq\ 23 + 16 CC & $0.29\sigma$ & $0.63\sigma$ & $0.70\sigma$ & -- & -- & --\\
\mq\ 25 + 69 + \cq\ 23 + 16 CC & $0.53\sigma$ & $1.23\sigma$ & $1.43\sigma$ & $0.25\sigma$ & $0.45\sigma$ & $1.67\sigma$\\
\cq\ 23 + 16 CR & $0.34\sigma$ & $0.76\sigma$ & $0.75\sigma$ & -- & -- & --\\
\mq\ 25 + 69 + \cq\ 23 + 16 CR & $0.63\sigma$ & $1.43\sigma$ & $1.43\sigma$ & $0.25\sigma$ & $0.54\sigma$ & $1.94\sigma$\\
\cq\ 25 CC\tnote{a} & $0.01\sigma$ & $0.04\sigma$ & $0.11\sigma$ & -- & -- & --\\
\cq\ 25 CR\tnote{a} & $0.04\sigma$ & $0.07\sigma$ & $0.09\sigma$ & -- & -- & --\\
\cq\ 23 + 41 CC & $0.38\sigma$ & $0.68\sigma$ & $0.84\sigma$ & -- & -- & --\\
\mq\ 25 + 69 + \cq\ 23 + 41 CC &  $0.70\sigma$ & $1.40\sigma$ & $1.55\sigma$ & $0.28\sigma$ & $0.58\sigma$ & $2.13\sigma$\\
\cq\ 23 + 41 CR & $0.86\sigma$ & $1.56\sigma$ & $1.69\sigma$ & -- & -- & --\\
\mq\ 25 + 69 + \cq\ 23 + 41 CR & $1.10\sigma$ & $2.28\sigma$ & $2.23\sigma$ & $0.31\sigma$ & $0.66\sigma$ & $2.35\sigma$\\
\mq\ 78 + \cq\ 38 & $0.45\sigma$ & $1.08\sigma$ & $1.12\sigma$ & $0.26\sigma$ & $0.54\sigma$ & $1.35\sigma$\\
\bottomrule
\end{tabular}
\begin{tablenotes}[flushleft]
\item [a] Only for flat/non-flat \lcdm.

\end{tablenotes}
\label{tab_RL_diff}
\end{threeparttable}%
}
\end{table*}

\section{Discussion}
\label{sec:discussion}

In terms of the $R-L$ correlation, the sample of 78 \mii\ QSO measurements was found to be standardizable and the resulting weak cosmological constraints were consistent with those from better-established cosmological probes \citep{Khadkaetal_2021a,Caoetal_2022}. Similarly, the sample of 38 RM \civ\ QSOs was found to be standardizable and the derived cosmological constraints were in agreement with those from better-established cosmological probes \citep{Caoetal_2022}. Here we include additional sources previously not considered for cosmological applications using the $R-L$ relation, specifically the sample of 25 \mii\ QSOs from the OzDES survey \citep{Yuetal2023} and the sample of lower-quality 25 \civ\ sources from the SDSS RM program \citep{2019ApJ...887...38G}.

We find that these additional subsamples of RM QSOs are standardizable and combining them with the original, larger samples also results in standardizable larger \mii\ and \civ\ data sets, as seen in Table~\ref{tab_RL_diff}. Notably, however, the largest difference between $R-L$ relation parameters (for different cosmological models) increases, especially for the intercept values. However, these differences remain within the 2$\sigma$ range, which is acceptable for such differences. For \civ\ sources, there is also a mild difference in the scatter of $R-L$ parameters among different cosmological models depending on the time-delay method (CC or CR), and this difference is larger for lower-quality \civ\ sources.

\begin{figure}
    \centering
    \includegraphics[width=\columnwidth]{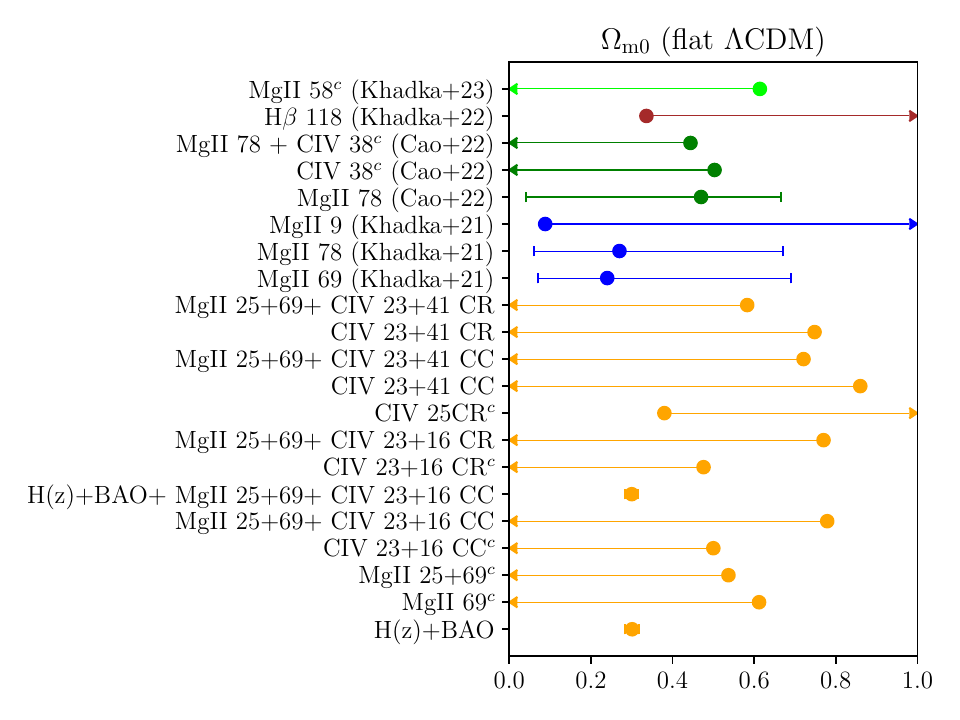}
    \caption{1$\sigma$ confidence intervals and 1$\sigma$ (with the superscript $c$) or 2$\sigma$ limits in the flat $\Lambda$CDM model for the samples in the current study (orange points) as well as in the previous papers (blue, green, brown, and lime points) involving \mii, \civ, and H$\beta$ QSOs \citep{Khadkaetal_2021a,Hbeta_Khadka2022,Caoetal_2022,Khadkaetal2023}.}
    \label{fig_Om0_flatLCDM}
\end{figure}

Combining the \mii\ and \civ\ QSOs results in tightening the cosmological constraints (see Figs.~\ref{fig01}-\ref{fig06}), but also in increasing the difference between $R-L$ relation parameters among different cosmological models (see Table~\ref{tab_RL_diff}). For the most extensive sample of RM \mii\ + \civ\ QSOs (94 + 64 measurements), the most significant difference among $R-L$ relation parameters surpasses the $2\sigma$ threshold, predominantly for the $\beta_{\rm M}$ intercept value and the CREAM time-delay method applied to \civ\ QSOs from the SDSS RM program. This largest sample formally yields the tightest constraint on the \om\ parameter (for the non-flat $\Lambda$CDM model, CC method, $\Omega_{m0}=0.331^{+0.096}_{-0.294}$, 1$\sigma$ limit), which calls for caution when combining different QSO data sets. It is imperative to simultaneously constrain both cosmological and $R-L$ relation parameters and subsequently verify whether the $R-L$ relation parameters remain consistent between different cosmological models. As a result, there is also no straightforward correlation between the sample size and the constraints on $\Omega_{\rm m0}$, see Fig.~\ref{fig_Om0_flatLCDM}, where we compare 1$\sigma$ confidence intervals and 1$\sigma$ or 2$\sigma$ limits in the flat $\Lambda$CDM model for various samples studied in this paper as well as in previous papers.

Of all sample combinations, \mii\ 25 + 69 + \civ\ 23 + 16 CC data yield the tightest cosmological constraints while still maintaining a considerable level of standardizability. Only $\beta_{\rm M}$ shows a departure of 1.67$\sigma$ in Table~\ref{tab_RL_diff}.  The same combination of QSO sources but with SDSS \civ\ time delays inferred using the CREAM method yields $R-L$ relation parameter differences just below the 2$\sigma$ limit, specifically the $\beta_{\rm M}$ parameter values which have a maximum difference between models of $\Delta{\beta_{\rm M}} = 1.94\sigma$. This shows that different time-delay methods mildly impact the $R-L$ relation and cosmological constraints. However, for data from QSOs with robustly detected time delays, the impact is generally not significant. For lower-quality sources, different time-delay methodologies result in larger variations, as expected and as noted above.   

\begin{figure*}
    \centering
    \includegraphics[width=\textwidth]{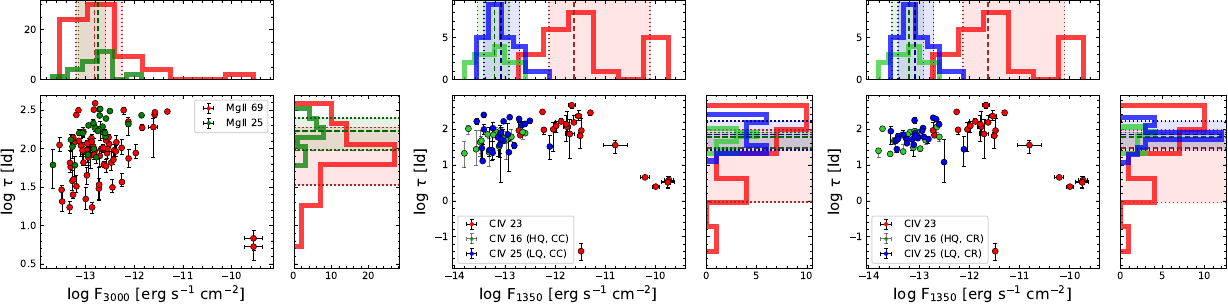}
    \caption{Distributions of the rest-frame time delay versus the monochromatic continuum flux for \mii\, and \civ\, QSOs. \textbf{Left panel:} The joint distribution between the rest-frame time delays (in light days) versus the monochromatic continuum flux at 3000 \AA\ (in erg s$^{-1}$ cm$^{-2}$). The individual measurements are listed in Table \ref{tab:MgII69} (shown in red for the \mii\ sample with 69 measurements) and Table \ref{tab:MgII25} (shown in green for the \mii\ sample with 25 sources). \textbf{Middle panel:} The joint distribution between the rest-frame time delays is estimated using the CC method versus the monochromatic continuum flux at 1350 \AA. The individual measurements are listed in Table \ref{tab:civdata1} (shown in red for the \civ\ sample with 23 measurements) and Table \ref{tab:civdata2} (shown in green for the \civ\ sample with 16 sources classified as high-quality (HQ, quality class 4 and 5), and the rest shown in blue are of low-quality (LQ, quality class 1, 2, and 3)). \textbf{Right panel:} Similar to the previous panel but for the rest-frame time delays estimated using the CREAM method. The corresponding marginal distributions are shown with their respective medians (dashed lines), 16$^{\rm th}$, and 84$^{\rm th}$ percentiles (dotted lines) marked. The histograms are binned based on Sturge's rule.}
    \label{fig:tau_v_flux_MgII_CIV}
\end{figure*}

\begin{figure*}
    \centering
    \includegraphics[width=\textwidth]{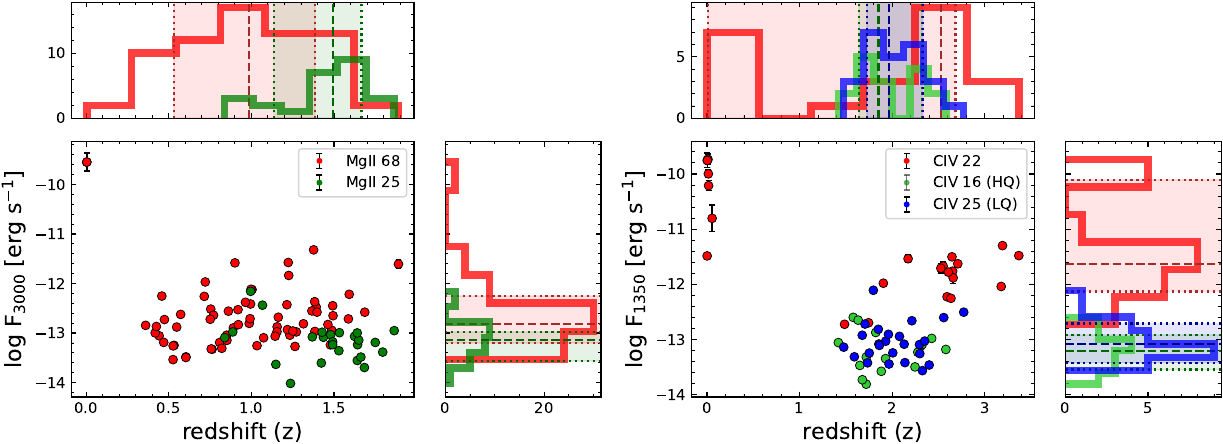}
    \caption{Distributions of the monochromatic continuum flux versus the source redshift for \mii\, and \civ\, QSOs. \textbf{Left panel:} The joint distribution between the monochromatic continuum flux at 3000 \AA\ (in erg s$^{-1}$ cm$^{-2}$) versus redshift. The individual measurements are listed in Table \ref{tab:MgII69} (shown in red for the \mii\ sample with 68 sources --- NGC 4151 has two measurements that are identical for the flux and redshift) and Table \ref{tab:MgII25} (shown in green for the \mii\ sample with 25 sources). \textbf{Right panel:} The joint distribution between the monochromatic continuum flux at 1350 \AA\ versus redshift. The individual measurements are listed in Table \ref{tab:civdata1} (shown in red for the \civ\ sample with 22 sources --- NGC 4151 has two measurements that are identical for the flux and redshift) and Table \ref{tab:civdata2} (shown in green for the \civ\ sample with 16 sources classified as high-quality (HQ, quality class 4 and 5), and the rest shown in blue are of low-quality (LQ, quality class 1, 2, and 3)).}
    \label{fig:flux_v_z_MgII_CIV}
\end{figure*}

From Table~\ref{tab_RL_diff} it is evident that adding more QSO samples from varied surveys, or with differing time-delay detection quality, increases the difference among $R-L$ relation parameters. For the \civ\, sample both the slope and the intercept are similarly affected, while in the case of \mii\ the effect on the intercept is larger in terms of $\sigma$. This must be related to some systematic differences between the samples. In Fig.~\ref{fig:tau_v_flux_MgII_CIV} we plot the distribution in the $\tau - F_{3000}$ plane. There the samples overlap, but some effective shifts (e.g. an overall longer time delay for the \mii\ 25 sample) are seen. 

In Fig.~\ref{fig_RL_MgII}, the intercept for 25 OzDES \mii\ QSOs and the sample of 69 \mii\ measurements differ by 0.1 but a combination of the two samples roughly recovers the original value of the intercept (the difference is within 0.01). A larger effect is observed for the lower-quality SDSS \civ\ sources that elevate the $R-L$ relation intercept for the combined sample by 0.09 in the log scale (refer to Figs.~\ref{fig_RL_CIV} and \ref{fig_RL_CIV_cr}). These trends are inferred for the fixed flat $\Lambda$CDM though they are also noticeable for the $R-L$ relations inferred for cosmological models with free parameters (see Table~\ref{tab:1d_BFP}). This phenomenon could be attributed to a combination of the time-delay systematic shift and the flux-density uncertainty. In Figs.~\ref{fig:tau_v_flux_MgII_CIV} and \ref{fig:flux_v_z_MgII_CIV}, which show time delay versus flux density and flux density versus redshift distributions, respectively, for \mii\ and \civ\ measurements, we see that the SDSS \civ\ measurements have systematically smaller flux densities compared to the 23 \civ\ measurements from other monitoring programs. To a smaller extent, systematically smaller flux densities also characterize the 25 \mii\ OzDES sources. This could potentially lead to a systematic shift in monochromatic luminosity -- meaning a larger $R-L$ relation intercept could arise from a slightly larger time delay for a given luminosity or the luminosity could be slightly smaller for a given time delay. Both effects can contribute to elevating the source above the best-fit $R-L$ relation. Systematic corrections to these subsamples as well as future RM QSO samples, are crucial and warrant a more detailed analysis in future studies of \mii\ and \civ\ $R-L$ relations and their applications.

Overall, an enlargement of the sample without a careful check of the quality of newly included measurements and their methodological consistency with the rest of the sample does not bring fully satisfactory results. If we compare the cosmological constraints from the largest AGN sample alone shown in \citet{Caoetal_2022} with the two largest AGN samples (\mii\ QSO 25 + 69 + \civ\ QSO 23 + 41 CC and \mii\ QSO 25 + 69 + \civ\ QSO 23 + 41 CR) given in Table~\ref{tab:1d_BFP} we see that for the flat $\Lambda$CDM model the cosmological constraints on $\Omega_{m0}$ are weaker now, with upper limits of 0.721 and 0.583, respectively, while previously we had an upper limit of 0.444. However, for non-flat $\Lambda$CDM there is an improvement in constraints on $\Omega_{m0}$, which now are $0.331^{+0.096}_{-0.294}$ and $0.206^{+0.029}_{-0.151}$, respectively, while previously we had $\Omega_{m0}=0.473^{+0.187}_{-0.311}$. Among the remaining cosmological models, two show an improvement, and in two the new results are less constraining. The improvements may also be questionable since the new upper limits are surprisingly small. This, combined with the problem of degradation in $R-L$ relation parameter independence from the cosmological model discussed in Table~\ref{tab_RL_diff}, shows that sample enlargement, despite data scarceness, does not necessarily lead to tighter and reliable cosmological constraints. In summary, the heterogeneity of the sample can be driven by multiple factors, including time-delay methods and systematic shifts of subsamples in different surveys, but we are unable to test them separately and disentangle their effect consistently due to current RM data quality and quantity.
   
\section{Conclusion}
\label{sec:conclusion}

Quasars have recently been more widely studied as possible alternate probes in cosmology \citep{2023Ap&SS.368....8C, 2023FrASS..1030103P}. The primary motivation behind their application is their broad redshift range, bridging the gap between low-redshift local measurements and high-redshift cosmic microwave background measurements. Such alternate probes might eventually help explain why different measurements of some cosmological parameters (in particular, the Hubble constant $H_0$ and the matter fluctuation amplitude $S_8=\sigma_8(\Omega_{m0}/0.3)^{1/2}$) differ \citep[see e.g.][]{PerivolaropoulosSkara2021, Morescoetal2022, Abdallaetal2022, HuWang2023}. A large enough QSO sample with precise enough time delay and flux density measurements could play a significant part in this process.

We have analyzed the largest sample of possibly standardizable \mii\ + \civ\ QSOs in cosmology to date. Previously for the smaller sample of 78 + 38 \mii\ + \civ\ sources, we showed that they are standardizable and their cosmological constraints, albeit weak, are consistent with those from better-established $H(z)$ + BAO data. Here we update the previous sample by (i) including the peculiar-velocity corrections for all low-redshift sources, and (ii) extending the sample size by including an additional 25 \mii\ sources from the OzDES survey and 25 lower-quality \civ\ QSOs from the SDSS RM program. 

Individual \mii\ and \civ\ subsamples as well as the combined sample follow their corresponding radius-luminosity relations. As before, the slope in this relation is much flatter for \mii\ than for \civ. However, the cosmological constraints did not improve satisfactorily with the inclusion of the new sources. In particular, the constraints for $\Omega_{m0}$ from AGN alone are tighter for three cosmological models and less tight for the other three (including flat $\Lambda$CDM) in comparison with the results for a smaller sample analysed by \citet{Caoetal_2022}. We tested the standardizability of the enlarged sample, and we see that this aspect now also looks problematic.

This suggests that adding new sources from various surveys could disrupt the standardizability of the overall sample due to distinct systematic issues like shifted time delays, altered monochromatic luminosities, and to a lesser degree, the effects of the time-delay method. It remains crucial to concurrently constrain both cosmological and $R-L$ relation parameters and to verify the consistency of $R-L$ relation parameters across different cosmological models. This could serve as a serious warning for the future. Dedicated monitoring campaigns slowly bring more line delay measurements of different quality. We also expect a massive flow of delay measurements from the Vera Rubin Observatory. Its Legacy Survey of Space and Time (LSST) will bring hundreds of thousands of line delay estimates but many of them will have biases due to sampling \citep[e.g.][]{czerny2023, Panda_etal_2023}. A sophisticated selection of sources will be needed for tight and reliable cosmological constraints.

\section*{Acknowledgements}
This research was supported in part by the Polish Funding Agency National Science Centre project 2017/26/A/ST9/00756 (Maestro 9), by GAČR Junior Star grant GM24-10599M, and by the Conselho Nacional de Desenvolvimento Científico e Tecnológico Fellowships 164753/2020-6 and 313497/2022-2. BC and MZ acknowledge the OPUS-LAP/GAČR-LA bilateral project (2021/43/I/ST9/01352/OPUS 22 and GF23-04053L). This project received funding from the European Research Council under the European Union’s Horizon 2020 research and innovation programme (grant agreement No. [951549]). Part of the computation for this project was performed on the Beocat Research Cluster at Kansas State University.

\section*{Data availability}
The data underlying this article are tabulated in the Appendixes.







\begin{appendix}
\label{sec:appendix}

\section{\cq\ data}
\label{sec:civdata}
\FloatBarrier
\onecolumn
\addtolength{\tabcolsep}{0pt}
\LTcapwidth=\linewidth
\setlength{\tabcolsep}{2mm}{
\begin{longtable}{lccccc}
\caption{QSO sample with high-quality detections of \civ\ time delay. These 23 measurements are from 22 sources, with two measurements in two different epochs for NGC4151. From left to right the columns list: object name, redshift, flux density at 1350\,\AA, monochromatic luminosity at 1350\,\AA\, for the flat $\Lambda$CDM model ($H_0=70\,{\rm km\,s^{-1}\,Mpc^{-1}}$, $\Omega_{ m0}=0.3$), rest-frame \civ\ time lag (in days), and the original reference. Source redshifts corrected for peculiar velocities are indicated with an asterisk. }
\label{tab:civdata1}
\\
\toprule
Object &  $z$ &  $\log \left(F_{1350}/{\rm erg}\,{\rm s^{-1}}{\rm cm^{-2}}\right)$  &  $\log \left(L_{1350}/{\rm erg}\,{\rm s^{-1}}\right)$  &  $\tau$ (days)  & Reference\\
\midrule
\endfirsthead

\endhead
\endfoot
NGC 4395 &   0.001952*    & $-11.4848 \pm 0.0272$  &    $40.4388    \pm 0.0272$  &    $0.040^{+0.024}_{-0.018}$   & \citet{2005ApJ...632..799P,2006ApJ...641..638P} \\
NGC 3783 &  0.010787*   &  $-9.7341 \pm 0.0918 $ &   $43.6802 \pm    0.0918  $&     $3.80^{+1.0}_{-0.9}$  & \citet{2005ApJ...632..799P,2006ApJ...641..638P}\\
NGC 7469 &  0.015084* &    $ -9.9973 \pm 0.0712$  &  $43.7111 \pm     0.0712 $ &    $2.5^{+0.3}_{-0.2}$    & \citet{2005ApJ...632..799P,2006ApJ...641..638P}\\
3C 390.3 &  0.055868*  &   $-10.8036 \pm 0.2386  $ &   $44.0681 \pm     0.2386$&    $35.7^{+11.4}_{-14.6}$  & \citet{2005ApJ...632..799P,2006ApJ...641..638P}\\
NGC 4151 &   0.004143*  &    $-9.7544 \pm 0.1329$  &  $42.8243 \pm    0.1329 $ &     $3.43^{+1.42}_{-1.24}$   & \citet{2006ApJ...647..901M}\\
NGC 4151 &   0.004143* &    $-9.7544 \pm 0.1329$  &  $42.8243 \pm    0.1329 $ &     $3.27^{+0.83}_{-0.91}$   & \citet{2006ApJ...647..901M}\\
NGC 5548  & 0.017462* &  $-10.2111 \pm 0.0894$  &  $43.6260 \pm     0.0894$&     $4.53^{+0.35}_{-0.34}$     &  \citet{2015ApJ...806..128D}\\
CTS 286 &  2.551 &  $-11.6705 \pm 0.0719$  &  $47.0477 \pm     0.0719$&   $459^{+71}_{-92}$   & \citet{2018ApJ...865...56L} \\
CTS 406 &  3.178 &   $-12.0382 \pm 0.0402$ & $46.9101 \pm 0.0402$ &   $98^{+55}_{-74}$     & \citet{2018ApJ...865...56L}  \\
CTS 564 &  2.653 &   $-11.7615 \pm 0.0664$ & $46.9978 \pm 0.0664$ & $115^{+184}_{-29}$   & \citet{2018ApJ...865...56L} \\
CTS 650 &  2.659 &  $-11.8815 \pm 0.1068$  &  $46.8802 \pm 0.1068$ &  $162^{+33}_{-10}$   & \citet{2018ApJ...865...56L}  \\
CTS 953 &  2.526 &   $-11.7082 \pm 0.0868$ &   $46.9996 \pm 0.0868$ &   $73^{+115}_{-58}$  & \citet{2018ApJ...865...56L}  \\
CTS 1061 &  3.368 &   $-11.4788 \pm 0.0405$ &   $47.5299 \pm 0.0405$ &    $91^{+111}_{-24}$     & \citet{2018ApJ...865...56L}  \\
J 214355 &  2.607 &   $-11.7786 \pm    0.0485$ &   $46.9624 \pm    0.0485$ &  $136^{+100}_{-90}$    &  \citet{2018ApJ...865...56L} \\
J 221516 &  2.709 &   $-11.6263 \pm    0.0569$ &   $47.1550 \pm    0.0569$&   $153^{+91}_{-12}$     &  \citet{2018ApJ...865...56L} \\
DES J0228-04 & 1.905 &   $-11.9791 \pm    0.0405$ &   $46.4298 \pm    0.0405$ &   $123^{+43}_{-42}$   &  \citet{2019MNRAS.487.3650H} \\
DES J0033-42 &  2.593 &   $-12.2248 \pm  0.0201$ &   $46.5105  \pm   0.0201$ &    $95^{+16}_{-23}$     &  \citet{2019MNRAS.487.3650H}  \\
RMID 363 &  2.635 &   $-12.2525   \pm    0.0206$ & $46.4997 \pm     0.0206$ &  $300.4^{+17.1}_{-4.7}$   & \citet{2019ApJ...883L..14S}  \\
RMID 372 &  1.745 &   $-12.6952  \pm     0.0198$ &    $45.6201 \pm     0.0198$ &    $67.0^{+20.4}_{-7.4}$    & \citet{2019ApJ...883L..14S}  \\
RMID 651 &  1.486 &  $-12.7234   \pm    0.0198$ &   $45.4200  \pm    0.0198$ &   $91.7^{+56.3}_{-22.7}$    & \citet{2019ApJ...883L..14S}  \\
S5 0836+71 &  2.172 &  $-11.5354   \pm    0.0680$ &    $47.0128 \pm     0.0680$ &   $230^{+91}_{-59}$   & \citet{2021ApJ...915..129K}  \\
SBS 1116+603  &  2.646 &   $-11.5013   \pm    0.0485$ &   $47.2553 \pm     0.0485$ &    $65^{+17}_{-37}$   &  \citet{2021ApJ...915..129K}   \\
SBS 1425+606  & 3.192 &   $-11.2978  \pm 0.0356$  &    $47.6551  \pm    0.0356$ &   $285^{+30}_{-53}$    & \citet{2021ApJ...915..129K}    \\
\bottomrule
\end{longtable}}

\addtolength{\tabcolsep}{0pt}
\LTcapwidth=\linewidth
\setlength{\tabcolsep}{4mm}{
\begin{longtable}{lccccccc}
\caption{Sample of 41 SDSS-RM QSOs with 10\%  false-positive detection rate of the \civ\ time delay adopted from \citet{2019ApJ...887...38G}. From left to right the columns list: object name, redshift, flux density at 1350\,\AA, monochromatic luminosity at 1350\,\AA\, for the flat $\Lambda$CDM model ($H_0=70\,{\rm km\,s^{-1}\,Mpc^{-1}}$, $\Omega_{m0}=0.3$), rest-frame \civ\ time delay inferred using the cross-correlation function (in days), rest-frame \civ\ time delay inferred using CREAM (in days), and the quality class (from 1 to 5, i.e. from lowest quality to highest quality detection). Sources with quality ranks 4 and 5 are included in the higher-quality (HQ) sample of 39 sources, while those with ranks 3 and smaller are in the lower-quality (LQ) sample of 25 sources (CC- and CREAM-inferred positive time delays).  }
\label{tab:civdata2}\\
\toprule
Object &  $z$ &  $\log \left(F_{1350}/{\rm erg}\,{\rm s^{-1}}{\rm cm^{-2}}\right)$  &  $\log \left(L_{1350}/{\rm erg}\,{\rm s^{-1}}\right)$    & $\tau_{\rm CC}$ (days) & $\tau_{\rm CR}$ (days) & Quality\\
\midrule
\endfirsthead

\endhead
\endfoot
  32  &    1.720 &   $-13.808 \pm  0.021$  &     $44.492 \pm     0.021$  &  $21.1^{+22.7}_{-12.8}$ & $24.8^{+0.8}_{-8.4}$ &  5 \\
 36  &    2.213  &  $-12.659 \pm  0.001$ &     $45.909 \pm     0.001$  &  $129.6^{+32.4}_{-50.6}$ &  $187.1^{+9.6}_{-9.9}$  & 1 \\
 52   &   2.311  &  $-13.115 \pm      0.002$ &    $45.499 \pm 0.002$  &  $32.6^{+6.9}_{-6.6}$ &  $30.3^{+6.6}_{-4.3}$ &   4  \\
  57  &    1.930 &    $-13.030 \pm 0.003$   &  $45.393 \pm 0.003$   & $47.0^{+51.2}_{-5.1}$ & $63.9^{+5.7}_{-6.6}$ &   1 \\
  58   &     2.299  &  $-13.255\pm 0.002$ &    $45.353 \pm 0.002$   &  $25.7^{+18.9}_{-12.7}$ &  $53.8^{+13.3}_{-16.1}$ &  1   \\
  130  &     1.960  &  $-12.905 \pm 0.001$  &    $45.534 \pm 0.001$   & $213.4^{+20.2}_{-18.8}$ & $60.4^{+3.7}_{-9.9}$ &  2\\
   144  &    2.295  &  $-13.091 \pm  0.001$  &  $45.516 \pm 0.001$   & $78.0^{+47.4}_{-57.4}$ &   $174.1^{+29.2}_{-35.1}$ &   2 \\
    145   &   2.138 &   $-13.418\pm 0.004$  &   $45.113 \pm 0.004$   &  $97.8^{+34.9}_{-25.3}$ &  $63.7^{+9.1}_{-9.1}$ &   3\\
    158   &   1.477 &  $-13.138 \pm  0.004$ & $44.999 \pm 0.004$   &  $58.6^{+33.7}_{-41.2}$ &  $51.3^{+18.6}_{-26.8}$ &  3 \\
  181  &  1.678  &  $-13.728 \pm  0.015$  &   $44.545 \pm 0.015$  &  $102.1^{+26.8}_{-30.5}$ & $101.8^{+5.0}_{-7.4}$ &  4\\
 201  &   1.797 &  $-12.107\pm  0.001$  &  $46.240 \pm 0.001$  & $32.5^{+35.6}_{-46.9}$ &  $27.3^{+35.4}_{-36.4}$ &   3  \\
 245  &    1.677  &  $-12.922 \pm 0.004$  &  $45.351 \pm 0.004$  &  $22.5^{+24.2}_{-29.2}$ & $106.3^{+14.7}_{-21.7}$ &  2 \\
  249  &    1.721  &  $-13.317 \pm 0.010$ &  $44.984 \pm 0.010$  &  $22.8^{+31.3}_{-13.5}$ & $23.6^{+12.6}_{-2.0}$ &   4  \\
  256   &   2.247  &  $-13.495 \pm  0.003$ & $45.089 \pm 0.003$  & $43.1^{+49.0}_{-26.1}$ & $46.7^{+10.7}_{-10.7}$ &   5\\
  269  &    2.400  &  $-13.461 \pm  0.003$ & $45.193 \pm 0.003$ & $29.4^{+10.0}_{-14.1}$ & $47.1^{+4.5}_{-3.7}$ &   1 \\
  275  &    1.580  &  $-12.598\pm   0.001$ & $45.611 \pm 0.001$  &  $76.7^{+10.0}_{-9.5}$ & $60.7^{+1.9}_{-16.7}$ &   5\\
 295   &     2.351 &   $-13.027 \pm     0.001$ &     $45.605 \pm  0.001$   & $164.0^{+21.6}_{-18.7}$ & $55.6^{+2.7}_{-6.5}$ &  3 \\
  298  &    1.633 &   $-12.648 \pm  0.001$  &   $45.596 \pm  0.001$   & $82.3^{+64.5}_{-30.7}$ & $113.9^{+10.4}_{-36.1}$ &  4 \\
 312   &   1.929  &  $-13.345 \pm  0.004$ &  $45.077 \pm 0.004$  & $70.9^{+9.6}_{-7.6}$ &  $67.1^{+14.8}_{-9.9}$ &  4  \\
332   &   2.580  &  $-13.179 \pm  0.002$ &  $45.551 \pm 0.002$   & $83.8^{+23.3}_{-19.4}$ & $81.8^{+3.4}_{-9.9}$ &  4\\ 
 346  &    1.592 &    $-13.312 \pm  0.003$  &  $44.905 \pm 0.003$  & $25.9^{+87.2}_{-43.2}$ &  $69.9^{+21.8}_{-11.7}$ &   3\\
  362   &   1.857 &   $-13.081 \pm  0.003$  &  $45.301 \pm  0.003$  & $79.8^{+12.9}_{-10.8}$ & $76.5^{+5.9}_{-11.9}$ &   2\\
  386  &    1.862 &   $-13.106 \pm  0.002$  & $45.279 \pm   0.002$ & $36.0^{+11.5}_{-19.4}$ & $36.5^{+14.3}_{-19.3}$ &   2\\
 387  &    2.427 &    $-12.979  \pm    0.001$ &    $45.687 \pm     0.001$   &  $48.4^{+34.7}_{-34.5}$ & $28.5^{+3.5}_{-4.6}$ &   4\\
  389  &    1.851 &   $-12.814\pm 0.002$ &     $45.564 \pm 0.002$   &  $34.8^{+7.6}_{-6.9}$ & $52.5^{+11.2}_{-12.8}$ &   2 \\
401   &   1.823 &   $-12.872 \pm 0.003$ &   $45.490 \pm 0.003$   & $60.6^{+36.7}_{-14.7}$ & $48.9^{+12.6}_{-10.6}$ &   4 \\
411  &    1.734  &    $-13.422 \pm  0.007$  & $44.887 \pm 0.007$  & $248.0^{+19.5}_{-40.6}$ &  $52.9^{+12.7}_{-7.1}$ &  2 \\
418   &   1.419 &   $-13.054 \pm 0.003$  &  $45.040 \pm 0.003$  & $58.6^{+51.6}_{-13.6}$ & $84.0^{+11.9}_{-18.0}$ &   4  \\
 470   &   1.883 &  $-13.575 \pm 0.006$  &  $44.821 \pm 0.006$   & $27.4^{+63.5}_{-17.6}$ &  $20.3^{+1.8}_{-2.5}$ &   4  \\
 485 &   2.557 &   $-12.602 \pm  0.001$  &  $46.119 \pm 0.001$ &  $138.9^{+11.0}_{-22.0}$ &  $133.9^{+23.4}_{-6.5}$ &   3\\
 496   &   2.079 &    $-12.942 \pm   0.001$  &     $45.560 \pm 0.001$   & $70.8^{+72.7}_{-24.7}$ &  $89.4^{+10.5}_{-33.5}$ &   1 \\
 499  &    2.327 &  $-13.563 \pm 0.003$ &  $45.058 \pm  0.003$  & $163.5^{+37.0}_{-26.1}$ &  $87.0^{+32.0}_{-49.2}$ &   2 \\
  506  &    1.753 &   $-13.245\pm  0.003$  & $45.075 \pm 0.003$  & $21.8^{+7.2}_{-7.9}$ & $51.7^{+9.2}_{-9.8}$ &   1 \\
 527  &  1.651 & $-13.468 \pm  0.003$ &  $44.788 \pm 0.003$ & $47.3^{+13.3}_{-27.0}$ & $46.7^{+6.5}_{-24.4}$ &   5\\
 549   &   2.277 &  $-13.229 \pm 0.002$ &  $45.369 \pm  0.002$  & $68.9^{+31.6}_{-8.8}$ & $69.9^{+7.8}_{-6.5}$ &   4 \\
  554  &    1.707  &  $-12.719 \pm 0.002$ &  $45.573 \pm 0.002$  & $191.0^{+33.9}_{-25.6}$ &  $205.5^{+21.7}_{-16.3}$ &   3\\
 562   &   2.773 &    $-12.504 \pm   0.001$  & $46.302 \pm  0.001$ & $170.2^{+10.0}_{-27.3}$ &  $12.0^{+39.8}_{-56.3}$ &   2\\
686   &   2.130 &   $-13.084 \pm 0.002$ &  $45.444 \pm 0.002$  & $52.2^{+57.5}_{-49.0}$ &  $64.0^{+6.9}_{-6.5}$ &   2  \\
 689   &   2.007 &    $-13.241 \pm 0.003$ & $45.223 \pm 0.003$   & $105.5^{+43.8}_{-59.3}$ & $40.2^{+9.1}_{-4.2}$ &   2\\
  734  &    2.324  &    $-13.090 \pm  0.001$  &  $45.530 \pm 0.001$  & $68.0^{+38.2}_{-23.1}$ & $86.6^{+14.2}_{-16.7}$ &   5 \\
 827  &    1.966  &   $-13.443 \pm  0.006$ & $44.999 \pm 0.006$  &  $12.9^{+24.7}_{-24.5}$ & $27.6^{+1.3}_{-5.9}$ & 3 \\
\bottomrule
\label{table_CIV_SDSS}
\end{longtable}}

\section{\mq\ data}
\label{sec:MgIIdata}
\FloatBarrier
\onecolumn
\addtolength{\tabcolsep}{0pt}
\LTcapwidth=\linewidth
\setlength{\tabcolsep}{7mm}{
\begin{longtable}{lcccc}
\caption{69 \mii\ measurements with the first 68 from \protect\cite{martinezAldama2020} and the last from \protect\cite{zajacek2021}. This is an updated version of the sample in \protect\cite{Khadkaetal_2021a}. From left to right the columns list: object name, redshift, flux density at 3000\,\AA, luminosity at 3000\,\AA\ for the flat $\Lambda$CDM model ($\Omega_{m0}=0.3$, $H_0=70\,{\rm km\,s^{-1}\,Mpc^{-3}}$), and rest-frame \mii\ time lag (in days). Source redshifts corrected for peculiar velocities are indicated with an asterisk.}
\label{tab:MgII69}\\
\toprule
Object &  $z$ &  $\log \left(F_{3000}/{\rm erg}\,{\rm s^{-1}}{\rm cm^{-2}}\right)$  & $\log \left(L_{3000}/{\rm erg}\,{\rm s^{-1}}\right)$  &  $\tau$ (days)\\
\midrule
\endfirsthead
\caption{continued.}\\
\toprule
Object &  $z$ &  $\log \left(F_{3000}/{\rm erg}\,{\rm s^{-1}}{\rm cm^{-2}}\right)$  &  $\log \left(L_{3000}/{\rm erg}\,{\rm s^{-1}}\right)$ & $\tau$ (days)\\
\midrule
\endhead
\bottomrule
\endfoot
018 & 0.848 & $-13.1412\pm0.0009$ &   $44.4000 \pm    0.0009$ & $125.9^{+6.8}_{-7}$\\
028 & 1.392 & $-12.4734\pm0.0004$ & $ 45.6000  \pm   0.0004 $ & $65.7^{+24.8}_{-14.2}$\\
038 & 1.383 & $-12.3664\pm0.0003$ & $ 45.7000  \pm   0.0003 $ &$120.7^{+27.9}_{-28.7}$\\
044 & 1.233 & $-13.0431\pm0.0013$ & $44.9000  \pm   0.0013$ & $65.8^{+18.8}_{-4.8}$\\
102 & 0.861 & $-12.5575\pm0.0005$ & $45.0000 \pm  0.0005$  & $86.9^{+16.2}_{-13.3}$\\
114 & 1.226 & $-11.8369\pm0.0003$ & $46.1000 \pm    0.0003$ & $186.6^{+20.3}_{-15.4}$\\
118 & 0.715 & $-12.2592\pm0.0006$ & $45.1000 \pm  0.0006 $  & $102.2^{+27}_{-19.5}$\\
123 & 0.891 & $-12.8942\pm0.0009$ & $ 44.7000 \pm    0.0009 $ & $81.6^{+28}_{-26.6}$\\
135 & 1.315 & $-12.8122\pm0.0005$ & $45.2000 \pm    0.0005$ & $93^{+9.6}_{-9.8}$\\
158 & 1.478 & $-13.2376\pm0.0012$ & $44.9000 \pm    0.0012$ & $119.1^{+4}_{-11.8}$\\
159 & 1.587 & $-12.7139\pm0.0006$ & $45.5000 \pm    0.0006$ &  $324.2^{+25.3}_{-19.4}$\\
160 & 0.36 & $-12.8441\pm0.0013$ & $43.8000 \pm    0.0013$ & $106.5^{+18.2}_{-16.6}$\\
170 & 1.163 & $-12.6802\pm0.0005$ & $45.2000 \pm    0.0005$  & $98.5^{+6.7}_{-17.7}$\\
185 & 0.987 & $-12.8039\pm0.0094$ & $44.9000 \pm    0.0094$ & $387.9^{+3.3}_{-3}$\\
191 & 0.442 & $-13.0544\pm0.0012$ & $ 43.8000  \pm   0.0012 $ & $93.9^{+24.3}_{-29.1}$\\
228 & 1.264 & $-13.2697\pm0.0011$ & $ 44.7000 \pm    0.0011  $ & $37.9^{+14.4}_{-9.1}$\\
232 & 0.808 & $-13.1895\pm0.0014$ & $ 44.3000  \pm   0.0014$ & $273.8^{+5.1}_{-4.1}$\\
240 & 0.762 & $-13.327\pm0.0021$ & $ 44.1000 \pm    0.0021 $ & $17.2^{+3.5}_{-2.8}$\\
260 & 0.995 & $-12.4126\pm0.0004$ & $ 45.3000  \pm   0.0004  $ & $94.9^{+18.7}_{-17.2}$\\
280 & 1.366 & $-12.5531\pm0.0003$ & $ 45.5000 \pm    0.0003 $ & $99.1^{+3.3}_{-9.5}$\\
285 & 1.034 & $-13.2539\pm0.002$ & $ 44.5000 \pm    0.0020 $ & $138.5^{+15.2}_{-21.1}$\\
291 & 0.532 & $-13.2471\pm0.0016$ & $ 43.8000 \pm    0.0016$ & $39.7^{+4.2}_{-2.6}$\\
294 & 1.215 & $-12.4272\pm0.0004$ & $ 45.5000 \pm    0.0004 $ & $71.8^{+17.8}_{-9.5}$\\
301 & 0.548 & $-12.8782\pm0.0011$ & $44.2000 \pm    0.0011$ & $136.3^{+17}_{-16.9}$\\
303 & 0.821 & $-13.3066\pm0.0013$ & $44.2000 \pm  0.0013$ & $57.7^{+10.5}_{-8.3}$\\
329 & 0.721 & $-11.968\pm0.0007$ & $45.4000 \pm    0.0007$ & $87.5^{+23.8}_{-14}$\\
338 & 0.418 & $-12.9969\pm0.0013$ & $43.8000 \pm     0.0013 $ & $22.1^{+8.8}_{-6.2}$\\
419 & 1.272 & $-12.9765\pm0.0011$ & $45.0000  \pm   0.0011 $ & $95.5^{+15.2}_{-15.5}$\\
422 & 1.074 & $-13.0946\pm0.0011$ &  $44.7000  \pm   0.0011$ & $109.3^{+25.4}_{-29.6}$\\
440 & 0.754 & $-12.5157\pm0.0004$ & $ 44.9000  \pm   0.0004$ & $114.6^{+7.4}_{-10.8}$\\
441 & 1.397 & $-12.5772\pm0.0004$ & $45.5000 \pm    0.0004 $ & $127.7^{+5.7}_{-7.3}$\\
449 & 1.218 & $-12.9299\pm0.0013$ & $ 45.0000  \pm   0.0013 $ & $119.8^{+14.7}_{-24.4}$\\
457 & 0.604 & $-13.4805\pm0.0029$ & $43.7000 \pm    0.0029 $ & $20.5^{+7.7}_{-5.3}$\\
459 & 1.156 & $-12.8737\pm0.0011$ & $45.0000 \pm  0.0011 $ & $122.8^{+5.1}_{-5.7}$\\
469 & 1.004 & $-12.1222\pm0.0002$ & $ 45.6000 \pm    0.0002 $ &  $224.1^{+27.9}_{-74.3}$\\
492 & 0.964 & $-12.3786\pm0.0004$ & $45.3000  \pm   0.0004 $ & $92^{+16.3}_{-12.7}$\\
493 & 1.592 & $-12.2173\pm0.0004$ & $ 46.0000 \pm    0.0004  $ & $315.6^{+30.7}_{-35.7}$\\
501 & 1.155 & $-12.9728\pm0.0009$ & $ 44.9000 \pm    0.0009 $ & $44.9^{+11.7}_{-10.4}$\\
505 & 1.144 & $-13.0625\pm0.0011$ & $44.8000 \pm  0.0011$ & $94.7^{+10.8}_{-16.7}$\\
522 & 1.384 & $-12.9671\pm0.0006$ & $45.1000 \pm    0.0006 $ & $115.8^{+11.3}_{-16}$\\
556 & 1.494 & $-12.6492\pm0.0005$ & $45.5000 \pm  0.0005$ & $98.7^{+13.9}_{-10.8}$\\
588 & 0.998 & $-12.1158\pm0.0002$ & $45.6000 \pm  0.0002$ & $74.3^{+23}_{-18.2}$\\
593 & 0.992 & $-12.7093\pm0.0006$ & $45.0000  \pm   0.0006 $ & $80.1^{+21.4}_{-20.8}$\\
622 & 0.572 & $-12.6232\pm0.0005$ & $ 44.5000 \pm    0.0005 $ & $61.7^{+6}_{-4.3}$\\
645 & 0.474 & $-12.7268\pm0.0009$ & $44.2000  \pm   0.0009$ & $30.2^{+26.8}_{-8.9}$\\
649 & 0.85 & $-13.0437\pm0.0013$ & $44.5000 \pm    0.0013 $ & $165.5^{+22.2}_{-25.1}$\\
651 & 1.486 & $-12.9434\pm0.0011$ & $ 45.2000  \pm   0.0011 $ & $76.5^{+18}_{-15.6}$\\
675 & 0.919 & $-12.5273\pm0.0005$ & $ 45.1000  \pm   0.0005$ & $139.8^{+12}_{-22.6}$\\
678 & 1.463 & $-12.8267\pm0.0007$ & $45.3000 \pm    0.0007$ & $82.9^{+11.9}_{-10.2}$\\
709 & 1.251 & $-12.9586\pm0.001$ & $ 45.0000  \pm   0.0010 $ & $85.4^{+17.7}_{-19.3}$\\
714 & 0.921 & $-12.8296\pm0.0012$ & $44.8000  \pm   0.0012$ & $320.1^{+11.3}_{-11.2}$\\
756 & 0.852 & $-13.1462\pm0.0023$ & $ 44.4000 \pm    0.0023 $ & $315.3^{+20.5}_{-16.4}$\\
761 & 0.771 & $-12.6395\pm0.0024$ & $44.8000  \pm   0.0024   $ & $102.1^{+8.2}_{-7.4}$\\
771 & 1.492 & $-12.4477\pm0.0004$ & $45.7000 \pm    0.0004 $ & $31.3^{+8.1}_{-4.6}$\\
774 & 1.686 & $-12.5786\pm0.0004$ & $ 45.7000 \pm    0.0004 $ & $58.9^{+13.7}_{-10.1}$\\
792 & 0.526 & $-13.5353\pm0.003$ & $43.5000 \pm    0.0030 $ & $111.4^{+29.5}_{-20}$\\
848 & 0.757 & $-13.3199\pm0.0015$ & $44.1000 \pm    0.0015  $ & $65.1^{+29.4}_{-16.3}$\\
J141214 & 0.4581 & $-12.2526\pm0.00043$ & $ 44.6388  \pm   0.0004 $ & $36.7^{+10.4}_{-4.8}$\\
J141018 & 0.4696 & $-13.1883\pm0.00506$ & $  43.7288  \pm   0.0051$ & $32.3^{+12.9}_{-5.3}$\\
J141417 & 0.6037 & $-13.4926\pm0.0029$ & $ 43.6873 \pm    0.0029$ & $29.1^{+3.6}_{-8.8}$\\
J142049 & 0.751 & $-12.7205\pm0.0009$ & $44.6909 \pm    0.0009$ & $34^{+6.7}_{-12}$\\
J141650 & 0.5266 & $-13.2586\pm0.00198$ &  $43.7779  \pm   0.0020$ &  $25.1^{+2}_{-2.6}$\\
J141644 & 0.4253 & $-12.8667\pm0.00105$ & $43.9480 \pm    0.0011 $ & $17.2^{+2.7}_{-2.7}$\\
CTS252 & 1.89 & $-11.6068\pm0.09142$ & $ 46.7936 \pm    0.0935 $ & $190^{+114}_{-59}$\\
NGC 4151 & 0.004143* & $-9.5484\pm0.18206$ & $43.0304 \pm    0.2006 $ & $6.8^{+1.7}_{-2.1}$\\
NGC 4151 & 0.004143* & $-9.5484\pm0.18206$ & $ 43.0304  \pm   0.2006$ & $5.3^{+1.9}_{-1.8}$\\
CTS C30.10 & 0.90052 & $-11.5825\pm0.026$ & $46.0230 \pm    0.0260$ & $275.5^{+12.4}_{-19.5}$\\
HE0413-4031 & 1.3765 & $-11.3203\pm0.0434$ & $46.7409  \pm   0.0436$ & $302.9^{+23.7}_{-19.1}$\\
HE0435-4312 & 1.2231 & $-11.5754\pm0.036$ & $ 46.3589  \pm   0.0361$ & $296^{+13}_{-14}$\\
\end{longtable}}

\addtolength{\tabcolsep}{0pt}
\LTcapwidth=\linewidth
\setlength{\tabcolsep}{4.9mm}{
\begin{longtable}{lcccc}
\caption{25 \mii\ QSOs from \protect\cite{Yuetal2023}. See Table \ref{tab:MgII69} title for notation.}
\label{tab:MgII25}\\
\toprule
Object &  $z$ &  $\log \left(F_{3000}/{\rm erg}\,{\rm s^{-1}}{\rm cm^{-2}}\right)$  & $\log \left(L_{3000}/{\rm erg}\,{\rm s^{-1}}\right)$ &  $\tau$ (days)\\
\midrule
\endhead
\bottomrule
\endfoot
DES J024340.09+001749.40 & 1.4356 & $-12.70 \pm 0.02$ &  $45.41 \pm 0.02$ & $327.23^{+8.21}_{-22.17}$\\
DES J025254.18+001119.70 &1.6408 & $-13.13 \pm 0.01$ & $45.12 \pm 0.01$ & $159.04^{+16.66}_{-10.22}$\\
DES J024831.08+005025.60 & 0.8870 & $-12.70 \pm 0.02$ & $44.89 \pm 0.02$ & $180.71^{+45.05}_{-54.58}$\\
DES J024723.54+002536.50 & 1.8641 & $-12.50 \pm 0.01$ &  $45.89 \pm 0.01$ & $307.25^{+10.47}_{-15.36}$\\
DES J024944.09+003317.50 & 1.4800 & $-12.64 \pm 0.01$ & $45.50 \pm 0.01$ & $162.10^{+15.73}_{-26.61}$\\
DES J024455.45-011500.40 & 1.5293 & $-12.95 \pm  0.02$ &  $45.22 \pm 0.02$ &  $83.03^{+15.42}_{-35.19}$\\
DES J025225.52+003405.90 & 1.6242 & $-12.65 \pm 0.01$ &  $45.59 \pm   0.01$ & $192.44^{+14.86}_{-16.39}$\\
DES J022716.52-050008.30 & 1.6424 & $-12.81 \pm 0.01$ & $45.44 \pm  0.01$ & $187.33^{+18.54}_{-19.68}$\\
DES J022751.50-044252.70 & 1.7946 & $-12.94 \pm 0.02$ &  $45.41 \pm 0.02$ & $199.67^{+11.81}_{-11.81}$\\
DES J022208.15-065550.50 & 1.6624 & $-13.06 \pm 0.01$ & $45.20 \pm 0.01$ & $158.88^{+15.02}_{-13.90}$\\
DES J033836.19-295113.50 & 1.1480 & $-13.30 \pm 0.02$ & $44.57 \pm  0.02$ & $130.35^{+22.81}_{-39.57}$\\
DES J033903.66-293326.50 & 1.6840 & $-13.27 \pm 0.01$ & $45.01 \pm 0.01$ & $105.81^{+11.18}_{-32.04}$\\
DES J033328.93-275641.21 & 0.8386 & $-12.83 \pm 0.04$ & $44.70 \pm  0.04$ & $98.99^{+37.52}_{-19.04}$\\
DES J022436.64-063255.90 & 1.4233 & $-13.22 \pm 0.02$ & $44.88 \pm 0.02$ & $60.66^{+18.16}_{-19.81}$\\
DES J033211.42-284323.99 & 1.2369 & $-13.67 \pm  0.04$ &  $44.28 \pm 0.04$ & $61.25^{+34.87}_{-30.85}$\\
DES J033213.36-283620.99 & 1.4920 & $-12.78 \pm  0.01$ & $45.37 \pm 0.01$ & $77.05^{+18.06}_{-12.44}$\\
DES J003710.86-444048.11 & 1.0670 & $-12.12 \pm  0.01$ & $45.67 \pm 0.01$ & $208.03^{+16.93}_{-14.03}$\\
DES J003922.97-430230.41 & 1.3687 & $-12.71 \pm  0.01$ & $45.35 \pm 0.01$ & $268.92^{+17.31}_{-7.18}$\\
DES J002933.85-435240.69 & 0.9955 & $-11.85 \pm  0.01$ & $45.86 \pm 0.01$ & $270.61^{+16.04}_{-16.04}$\\
DES J003207.44-433049.00 &  1.5328 & $-12.59 \pm  0.01$ &  $45.59 \pm 0.01$ & $142.92^{+13.42}_{-8.29}$\\
DES J003015.00-430333.52 & 1.6498 & $-12.72 \pm  0.01$ & $45.54 \pm 0.01$ & $176.24^{+7.17}_{-10.57}$\\
DES J003052.76-430301.08 & 1.4275 & $-12.60 \pm 0.01$ &  $45.50 \pm 0.01$ & $159.84^{+15.24}_{-14.42}$\\
DES J003232.61-433302.99 & 1.4920 & $-12.84 \pm  0.01$ &  $45.31 \pm  0.01$ & $214.69^{+22.87}_{-17.26}$\\
DES J003234.33-431937.81 & 1.6406 & $-12.58 \pm  0.01$ &  $45.67 \pm 0.01$ & $245.78^{+6.06}_{-4.92}$\\
DES J003206.50-425325.22 & 1.7496 & $-12.75 \pm  0.01$ & $45.57 \pm 0.01$ & $169.48^{+9.46}_{-9.46}$\\
\end{longtable}}

\end{appendix}


\bsp	
\label{lastpage}
\end{document}